\documentclass[12pt]{iopart}

\usepackage{latexsym}
\usepackage{amssymb}
\usepackage{amsthm} 
\usepackage{amsfonts} 
\usepackage{graphicx,color}
\usepackage{bm}      
\usepackage[hypertex]{hyperref}

\def\re#1{(\ref{#1})}

\begin{document}

\title[]
{Topological insulators and superconductors: 
ten-fold way and dimensional hierarchy}

\author{Shinsei Ryu$^1$, Andreas P Schnyder$^{2,3}$, Akira Furusaki$^4$ and Andreas W W Ludwig$^5$}

\address{$^1$ Department of Physics, University of California, Berkeley, California 94720, USA}

\address{$^2$ Kavli Institute for Theoretical Physics, University of California, Santa Barbara, California 93106, USA}

\address{$^3$ Max-Planck-Institut f\"ur Festk\"orperforschung, Heisenbergstrasse 1, D-70569 Stuttgart, Germany} 

\address{$^4$ Condensed Matter Theory Laboratory, RIKEN, Wako, Saitama 351-0198, Japan} 

\address{$^5$ Department of Physics, University of California, Santa Barbara, California 93106, USA}

\ead{\mailto{sryu@berkeley.edu}, \mailto{schnyder@kitp.ucsb.edu}}

\date{\today}
 
\begin{abstract}
It has recently been shown that in every spatial dimension there exist
precisely five distinct classes of topological insulators or superconductors.
Within a given class, the different topological sectors 
can be distinguished, depending on the case,
by a $\mathbb{Z}$ or a $\mathbb{Z}_2$ topological invariant.
This is an exhaustive classification.
Here we construct representatives of topological insulators and superconductors
for all five classes and in arbitrary spatial dimension $d$,
in terms of Dirac Hamiltonians. 
Using these representatives we demonstrate
how topological insulators (superconductors)
in different dimensions and different
classes can be related via ``dimensional reduction'' by compactifying
one or more spatial dimensions (in ``Kaluza-Klein''-like fashion).
For $\mathbb{Z}$-topological insulators (superconductors)  
this proceeds by descending by one dimension at a time
into a different class.
The $\mathbb{Z}_2$-topological insulators (superconductors),
on the other hand, are shown to be lower-dimensional descendants of
parent $\mathbb{Z}$-topological insulators in the same class,
from which they inherit their topological properties.
The $8$-fold periodicity in dimension $d$
that exists for topological insulators (superconductors)
with Hamiltonians satisfying at least one reality condition
(arising from time-reversal or charge-conjugation/particle-hole symmetries)
is a reflection of
the $8$-fold  periodicity of the spinor representations
of the orthogonal groups SO($N$) (a form of Bott periodicity).
Furthermore, we derive for general spatial dimensions a relation
between the topological invariant that characterizes  
topological insulators and superconductors with chiral symmetry
(i.e., the winding number) and the Chern-Simons invariant.
For lower dimensional cases, this formula 
relates the winding number to
the electric polarization ($d=1$ spatial dimensions),
or to the magnetoelectric polarizability ($d=3$ spatial dimensions). 
Finally,  we also discuss topological field theories describing 
the space time theory of linear responses in
topological insulators (superconductors),
and study how the presence of inversion symmetry modifies
the classification of topological insulators (superconductors).
\end{abstract}

\maketitle

\tableofcontents
\clearpage 

\section{Introduction}

\begin{figure}[tb]
\begin{center}
    \includegraphics[height=.27\textwidth]{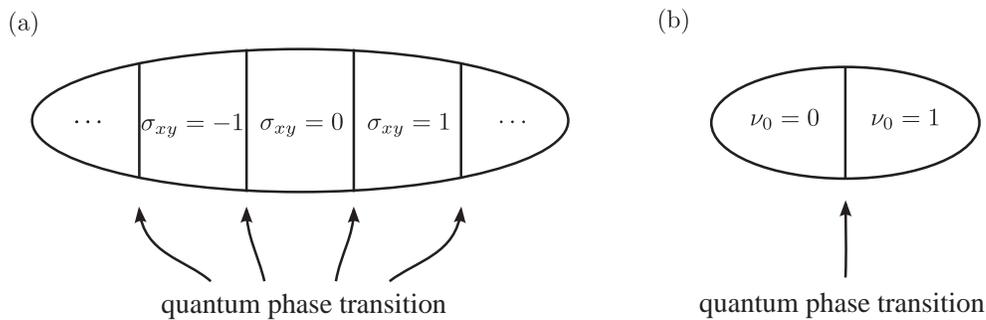}
  \caption{
Topological distinction among quantum ground states. 
(a) $\mathbb{Z}$ classification
and 
(b) $\mathbb{Z}_2$ classification.
}
\label{top_insulators}
\end{center}
\end{figure}

Topological insulators (superconductors) are gapped phases of non-interacting 
fermions which exhibit topologically protected boundary modes.
These boundary states are gapless, with extended wavefunctions,
and protected against arbitrary deformations (or perturbations)
of the Hamiltonian, as long
as the generic symmetries (such as e.g.\ time-reversal symmetry)
of the Hamiltonian are preserved and the bulk gap is not 
closed \footnote{In the presence of disorder (= lack of translational symmetry)
a sufficient condition is that all bulk states near the Fermi energy
are localized~\cite{bellissard1994}.
}.
The different phases within a given topological insulator (superconductor)
are characterized by a topological invariant, 
which is either an integer Chern/winding number or a $\mathbb{Z}_2$ quantity.
The  integer quantum Hall effect  (IQHE) is
the best known example of such a phase.
In the IQHE, the transverse (Hall) electrical 
conductance $\sigma_{xy}$ carried by
edge states is quantized 
and proportional to the topologically invariant Chern number 
\cite{Thouless82,Kohmoto}\footnote{
A profound implication of a non-vanishing Chern number is that
exponentially localized Wannier wave functions cannot be constructed
\cite{Panati,Brouder,Hastings}.
Yet another manifestation of the non-trivial nature of the 
band structure when $\sigma_{xy}\neq 0$ appears in
the scaling of the entanglement entropy 
(and the entanglement entropy spectrum) \cite{IQHEentropy}.
}.
As a consequence of the quantization of 
the Hall conductance $\sigma_{xy}$,
two different quantum Hall states with different
values of $\sigma_{xy}$ cannot be
adiabatically connected without closing the energy gap in the bulk.
Therefore they represent distinct ``topologically ordered'' phases
that are separated by a quantum phase transition (figure~\ref{top_insulators}a).

Other examples of gapped topological phases
are the (spinless) chiral $p_x+\mathrm{i} p_y$ superconductor \cite{Read00},
which breaks time-reversal symmetry ($\mathcal{T}$),
and the quantum spin Hall (QSH) 
states \cite{KaneMele,Roy06,Bernevig05,Moore06,Roy3d,Fu06_3Da,Fu06_3Db},
which are time-reversal invariant.
In the former case, the topological features in question are those
of the fermionic quasiparticle excitations deep in the superconducting phase,
whose dynamics is described by the Bogoliubov-de Gennes
(BdG) Hamiltonian \footnote{Physically,
the quasiparticles are responsible for heat (energy) transport,
and not for the electrical transport properties of the
superconductor.}. 
The BdG Hamiltonian of any superconductor has, 
by construction, a ``built-in''
charge-conjugation
 (or: particle-hole) symmetry ($\mathcal{C}$).
In analogy to the IQHE, the different topological phases of the 
$p_x+\mathrm{i} p_y$ superconductor can be labeled by an integer ($\mathbb{Z} $)
topological invariant \cite{VolovikBook}. 
The corresponding quantized conductance is the transverse (Hall) thermal, 
or ``Leduc-Righi'' conductance (divided by temperature).
The quantum spin Hall effect (QSHE), on the other hand, which occurs
in two-dimensional and three-dimensional time-reversal invariant insulators,
is characterized by a $\mathbb{Z}_2$ topological number $\nu_0$ 
\cite{KaneMele, Moore06,Roy3d,Fu06_3Da}.
This binary quantity distinguishes 
the non-trivial state ($\nu_0 =1$), whose boundary modes consist of an odd
number of Dirac fermion modes (i.e., odd number of Kramers' pairs), 
from the trivial state ($\nu_0 =0$), which is characterized by an even number
of Kramers' pairs of 
boundary (surface or edge) states (figure\ \ref{top_insulators}b).
A physical realization of the QSHE 
was theoretically predicted \cite{Bernevig-Taylor-Zhang06}
and experimentally observed
\cite{Konig06,Konig06b} in HgTe/(Hg,Ce)Te semiconductor quantum wells.
Subsequently, based both on theoretical considerations \cite{Fu06_3Db,Zhang09}
and experimental measurements \cite{Hsieh08,Hsieh09,Xia09,Hsieh09b,Chen09},
the three-dimensional $\mathbb{Z}_2$ topological insulator phase 
was shown to occur in Bismuth-related materials,
such as BiSb alloys, Bi$_2$Se$_3$ and Bi$_2$Te$_3$ .

The key difference among the IQHE, 
the QSHE and three-dimensional $\mathbb{Z}_2$ topological insulator, 
and the chiral $p_x+\rmi p_y$ superconductor lies in their 
generic symmetry properties (to be discussed in more detail below).
All these states are topological in the sense that 
they cannot be continuously deformed into a trivial insulating state
(i.e., a state without any gapless boundary modes)
while keeping the generic symmetries of the system intact and without closing
the bulk energy gap.
The key property of the QSH states, for example,
is the spin-orbit interaction in combination with
time-reversal symmetry ($\mathcal{T}$).
The latter protects the boundary states 
from acquiring a gap or becoming localized in the presence of
disorder (= terms added to the Hamiltonian which
break translational symmetry).
The chiral $p_x+\rmi p_y$ superconductor, on the other hand,
is characterized by $\mathcal{C}$ 
but the lack of $\mathcal{T}$.

The aforementioned examples of  ``symmetry protected'' topological states
are part of a larger scheme\footnote{
A complete classification of all topological insulators (superconductors)
in dimensionalities up to $d=3$ was given in 
Ref.\ \cite{SchnyderRyuFurusakiLudwig2008}.
A systematic regularity  (periodicity) of the classification as
the dimensionality is varied  was discovered  by Kitaev
in Ref.\ \cite{KitaevLandau100Proceedings}  
for all dimensions through the use of K-Theory.
A certain systematic pattern as the dimensionality is varied
was discovered for some of the topological insulators by 
Qi, Hughes and Zhang in Ref.\  \cite{Qi_Taylor_Zhang}
using a Chern-Simons action in extended space.  
Below we will use a version of the ideas of ``dimensional reduction'' 
employed in Ref.\ \cite{Qi_Taylor_Zhang}. 
}
that fully classifies 
\cite{SchnyderRyuFurusakiLudwig2008,KitaevLandau100Proceedings}
all topological insulators (superconductors)
in terms of symmetry and spatial dimension.
This classification scheme is summarized in table~\ref{periodic table}.
The first column in this table provides
the complete list of all possible ``symmetry classes'' of
single-particle Hamiltonians.
There are precisely ten such ``symmetry classes'',
a fundamental result due to Zirnbauer,
and Altland and Zirnbauer \cite{Zirnbauer96,Altland97,Huckleberry2005}.
These authors recognized
that there is a  one-to-one correspondence between
single-particle Hamiltonians and
the set of (``large'') symmetric spaces, of which there are precisely ten,
a classic result obtained in 1926 by the mathematician \'Elie Cartan.
The work of Altland and Zirnbauer extends, indeed completes,
the much earlier work by Wigner and Dyson most
well known from the context of random matrix theory
as the three ``unitary, orthogonal, and symplectic''
symmetry classes \cite{DysonThreeFold}. 
Before discussing in more detail the result of the classification 
in table~\ref{periodic table}, we will first briefly explain the notion of
``symmetry class'' and the ten-fold list of Hamiltonians,
which is the framework within  which the classification scheme in
table~\ref{periodic table} is formulated.
The reader familiar with this topic may skip this subsection.

\subsection{Review: Classification of generic Hamiltonians -- ``The Ten-Fold Way''}
\label{Review-Classification-Hamiltonian}

Consider the gapped first-quantized
Hamiltonian describing the topological insulator (superconductor)
in the bulk ($d$ spatial dimensions). The topological features of this
Hamiltonian, which we are interested in characterizing, are  (by definition)
not changed
if we modify (deform) the Hamiltonian by adding terms
(or perturbations) to it which break any translational symmetry
that may be present.
Thus the topological features must be  properties
of a translationally {\it not} invariant Hamiltonian. Seeking a classification
scheme, one must have some framework within which to classify such
translationally {\it  not} invariant Hamiltonians.
Such a classification cannot involve the notion of ordinary symmetries,
i.e., of {\it unitary} operators which commute with the Hamiltonian.
Consider, e.g.,
(Pauli-) spin rotation symmetry, ubiquitous in condensed matter systems.
The notion of spin-rotation invariance can be eliminated by writing
the Hamiltonian in block form, and by focussing attention on 
a block of the Hamiltonian. Such a block decomposition can be performed
for any symmetry which commutes with the Hamiltonian. The notion
of ``symmetry classes'' mentioned above refers to the properties
of these blocks of symmetry-less ``irreducible'' Hamiltonians:
as it turns out, there are only ten of them.
The basic idea why this is the case  is simple to understand.
The only properties that the blocks can satisfy are certain reality conditions
that follow from the ``extremely generic symmetries'' of time-reversal and
charge-conjugation (or: particle-hole symmetry). These are not
symmetries in the above sense, since they are both
represented by {\it anti-}unitary operators,
when acting on the single-particle Hilbert space.
Specifically, consider a general system of
non-interacting fermions described by a ``second-quantized''
Hamiltonian $H$. 
For a non-superconducting system, e.g.,  this reads
\begin{eqnarray}
H
= \sum_{A,B} \,  {\psi}^\dagger_{A} \, \mathcal{H}^{\ }_{A,B} \, {\psi}^{\ }_{B},
\end{eqnarray}
with fermion creation and annihilation operators
satisfying canonical anti-commutation relations
\begin{eqnarray}
\{{\psi}^{\ }_{A},{\psi}^\dagger_{B} \} = \delta_{A,B}. 
\end{eqnarray}
Here we imagine for convenience of notation that we have ``regularized''
the system on a lattice, and $A, B$ are combined labels for the lattice
sites $i,j$, and if relevant, of additional quantum numbers such as e.g.,
a Pauli-spin quantum number  
(e.g., $A=(i,a)$ with $a, b = \pm 1/2$).
Then
${\cal H}_{A,B}$ is an $N \times N$ matrix, the ``first quantized''
Hamiltonian. (Similarly, a superconducting system is described
by a BdG Hamiltonian 
for which we use the Nambu-spinor
instead of
complex fermion operators $\psi^{\ }_{A},{\psi}^\dagger_{A}$
and whose first quantized form is again
a matrix ${\cal H}$ when discretized on a lattice.)

Now, time reversal symmetry can be expressed in terms of ${\cal H}$:
the system is invariant under time reversal symmetry
if and only if the complex conjugate  of the
first quantized Hamiltonian
${\cal H}^*$ is equal to ${\cal H}$
up to a unitary rotation $U_{T}$, i.e.
\begin{equation}
\label{DEFTimeReversal}
\mathcal{T}: 
\quad 
U_{T}^\dagger
 \ {\cal  H}^{*} \  
U_{T} = + {\cal H  }.
\end{equation}
Moreover, the system is
invariant under charge-conjugation (or: particle-hole) symmetry
if and only if the complex conjugate   
of the Hamiltonian
${\cal H}^*={\cal H}^T$
is equal
 to {\it minus} ${\cal H}$ 
up to a unitary rotation $U_{C}$, i.e.
\begin{equation}
\label{DEFChargeConjugation}
\mathcal{C}: 
\quad 
U_{C}^\dagger
 \ {\cal  H}^{*} \  
U_{C} = - {\cal  H}.
\end{equation}
(This property may be less familiar, but it is easy to 
check~\cite{SchnyderRyuFurusakiLudwig2008}
that it is a characterization 
of charge-conjugation (particle-hole) symmetry
for non-interacting systems of fermions). 
A look at equations (\ref{DEFTimeReversal},\ref{DEFChargeConjugation})
reveals that 
$\mathcal{T}$
and
$\mathcal{C}$,
when acting on the single particle Hilbert space, 
are not unitary symmetries,
but rather reality conditions on the Hamiltonian ${\cal H}$
{\it modulo} unitary rotations
\footnote{
In second-quantized language,
time-reversal and particle-hole operations
can be written in terms of their action on the canonical
fermion creation and annihilation operators,
\begin{eqnarray}
\mathcal{T} {\psi}_A \mathcal{T}^{-1} 
=
\sum_B \   (U_T)_{A,B} \ {\psi}_B ,
\quad 
\mathcal{C} {\psi}_A \mathcal{C}^{-1} 
=
\sum_B \  (U^*_C)_{A,B} \   {\psi}^\dagger_B  .
\end{eqnarray}
While the particle-hole transformation is unitary, 
the time-reversal operation is anti-unitary, 
$\mathcal{T}\mathrm{i} \mathcal{T}^{-1} = -\mathrm{i}$. 
The system is time-reversal invariant (particle-hole symmetric), 
if and only if
$\mathcal{T}H\mathcal{T}^{-1}=H$
($\mathcal{C}H\mathcal{C}^{-1}=H$).
This leads directly to the conditions
(\ref{DEFTimeReversal}) and (\ref{DEFChargeConjugation})
for the first quantized Hamiltonian.
Note that 
$\mathcal{T}H\mathcal{T}^{-1}=H$ implies
$\mathcal{T}{\psi}_A(t)\mathcal{T}^{-1}=
\mathcal{T}e^{+\mathrm{i}Ht}{\psi}_Ae^{-\mathrm{i}Ht}\mathcal{T}^{-1}=
\sum_B \ (U_T)_{A,B}\  \psi_B(-t)
$. 
Iterating $\mathcal{T}$ and $\mathcal{C}$ twice,
one obtains
$\mathcal{T}^2 {\psi}_A \mathcal{T}^{-2} 
=
\sum_B \ (U^*_T U_T)_{A,B} \ {\psi}_B$,
and 
$
\mathcal{C}^2 {\psi}_A \mathcal{C}^{-2} 
=
\sum_B \ (U^*_C U_C)_{A,B} \ {\psi}_B$.
When acting on the first quantized Hamiltonian this reads
$(U^*_T U_T)^\dagger {\cal H} (U^*_T U_T)$$={\cal H}$
and
$(U^*_C U_C)^\dagger {\cal H} (U^*_C U_C)$$={\cal H}$,
respectively. The first quantized Hamiltonians ${\cal H}$
are seen (below) to run over an irreducible representation space,
and thus 
$(U^*_T U_T)$ and
$(U^*_C U_C)$ are both multiples of the identity matrix $I_{N}$
(by Schur's lemma).
Since
$U_{T}$ and $U_C$ are unitary matrices, there are only two possibilities
for each, i.e.,
$U^*_T U_T = \pm I_{N}$
and $U^*_C U_C = \pm I_{N}$. 
The time-reversal operation $\mathcal{T}$ and 
the particle-hole transformation $\mathcal{C}$
can then each square to plus or to minus the identity,
$\mathcal{T}^2=\pm 1$, and $\mathcal{C}^2=\pm 1$.
}
\footnote{
It may also be worth noting that we may assume without loss of generality 
that there is only a {\it single} time-reversal operator 
$\mathcal{T}$ and a {\it single} 
charge-conjugation operator $\mathcal{C}$.
If the (first quantized) Hamiltonian ${\cal H}$ was invariant under,
say, {\it two} charge-conjugation operations 
$\mathcal{C}_1$ and $\mathcal{C}_2$, 
then the composition $\mathcal{C}_1 \cdot \mathcal{C}_2$ of these
two symmetry operations would be a {\it unitary} symmetry
when acting on the first quantized Hamiltonian ${\cal H}$, 
i.e., the product $U_{C_1} \cdot U^*_{C_2}$ would {\it commute} with ${\cal H}$. 
By bringing the Hamiltonian ${\cal H}$ in block form, 
$U_{C_1} \cdot U^*_{C_2}$ would then be constant on each block.
Thus, on each block $U_{C_1}$ and $U_{C_2}$ would then be trivially related
to each other, and
it would suffice to consider one of the two charge conjugation operations. 
--  On the other hand, note that the product
$\mathcal{T} \cdot \mathcal{C}$
corresponds to a {\it unitary} symmetry operation when acting on
the first quantized Hamiltonian ${\cal H}$.
But in this case the unitary matrix $U_{T} \cdot U^*_{C}$
does not commute but 
{\it anti-}commutes with ${\cal H}$.
Therefore, $\mathcal{T} \cdot \mathcal{C}$ does not correspond to an
an ``ordinary'' symmetry of ${\cal H}$.
It is for this reason that we need to consider the product
$\mathcal{T} \cdot \mathcal{C}$ (called ``chiral'', or ``sublattice'' symmetry 
$\mathcal{S}$ below)
as an additional essential ingredient for the classification of the blocks,
besides time-reversal $\mathcal{T}$ and charge-conjugation
(particle-hole) symmetries $\mathcal{C}$.
}.

Now it is easy to see that there are only ten possible
ways for a system
to respond to time-reversal and charge-conjugation
(particle-hole symmetry) operations.
As for time-reversal symmetry 
($\mathcal{T}$), the Hamiltonian
can either be (i): not time-reversal invariant,
in which case we write $\mathrm{T}=0$,
or 
(ii): it may be time-reversal invariant and the anti-unitary
time-reversal symmetry operator  $\mathcal{T}$
squares to plus the identity operator, in which
case we write $\mathrm{T}=+1$, or (iii):
it may be time-reversal invariant and the anti-unitary
time-reversal symmetry operator 
$\mathcal{T}$
squares to minus the identity, in which
case we write $\mathrm{T}=-1$.
Similarly, there are three possible ways for the Hamiltonian ${\cal H}$
to respond 
to charge-conjugation (particle-hole symmetry) $\mathcal{C}$
(again, $\mathcal{C}$
may square to plus or minus the identity operator).
For these three possibilities we write $\mathrm{C}=0, +1, -1$.
Hence, there are $3 \times 3 = 9$ possible ways for a Hamiltonian
${\cal H}$ to respond to time-reversal and charge-conjugation
(particle-hole transformation).
These are not yet all ten cases because it is also necessary to consider
the behavior of the Hamiltonian under the product 
$\mathcal{S} = \mathcal{T} \cdot \mathcal{C}$
which is a unitary operation.
A moment's thought (see table~\ref{TableSymmetryClassesTwo}) shows that for
8 of the 9 possibilities the behavior  of the Hamiltonian under the product
$\mathcal{S} = \mathcal{T} \cdot \mathcal{C}$
is uniquely fixed
\footnote{
The symmetry operation $\mathcal{S}$ is sometimes called
``sublattice symmetry'', 
hence the notation $\mathcal{S}$.  However, in many instances  
$\mathcal{S}$ is not realized as a sublattice symmetry,
but is just simply the product of $\mathcal{T}$ and $\mathcal{C}$.
Therefore the term ``chiral symmetry'' to describe the symmetry 
operation $\mathcal{S}$ is sometimes more appropriate.
}.
(We write $\mathrm{S}=0$ if 
the operation $\mathcal{S}$
is not
a symmetry of the Hamiltonian, and $\mathrm{S}=1$ if it is.)
The only case when the behavior under the combined transformation
$\mathcal{S}=\mathcal{T}\cdot\mathcal{C}$
is not determined by the behavior
under $\mathcal{T}$ and $\mathcal{C}$ is the case where $\mathrm{T}=0$
and $\mathrm{C}=0$.
In this case either $\mathrm{S}=0$ or $\mathrm{S}=1$ is possible. This then
yields
$(3\times3 -1) +2 = 10$ possible behaviors of the Hamiltonian.

The list of ten possible behaviors of the first quantized
Hamiltonian under 
$\mathcal{T}$, $\mathcal{C}$ and $\mathcal{S}$
is listed in table~\ref{TableSymmetryClassesTwo}.
These are the ten generic
symmetry classes (the ``ten-fold way'') which are the
framework within which the classification scheme of topological
insulators (superconductors) is formulated.

Let us first point out a very general structure seen in 
table~\ref{TableSymmetryClassesTwo}.
This is listed in the column entitled ``Hamiltonian''.
When the first quantized Hamiltonian ${\cal H}$ is
``regularized'' (or: ``put on'')
a finite lattice, it becomes an $N\times N$ 
matrix (as discussed above). 
The entries in the column ``Hamiltonian'' specify the type of $N \times N$
matrix that the quantum mechanical time-evolution operator
$\exp( \mathrm{i} t {\cal H} )$ is. For example, for systems which
have no time-reversal or charge-conjugation symmetry properties
at all, i.e., for which $\mathrm{T}=0$,  $\mathrm{C}=0$, $\mathrm{S}=0$, which
are listed in the first row of the table, there are no
constraints on the Hamiltonian except for Hermiticity.
Thus ${\cal H}$ is a generic Hermitian matrix and the time-evolution
operator is a generic unitary matrix, so that
$\exp( \mathrm{i} t {\cal H} )$ is an element of the unitary group 
$\mathrm{U}(N)$ of unitary $N\times N$ matrices.
By imposing time-reversal symmetry
(for a system that has, e.g., no other degree of freedom
such as, e.g., spin), there exists a basis
in which ${\cal H}$ is represented by a real symmetric
$N\times N$ matrix. This, in turn, can be expressed
as saying that the time-evolution operator is
an element of the coset of groups, $\mathrm{U}(N)/\mathrm{O}(N)$.
All other entries of the column ``Hamiltonian''
can be obtained from analogous 
considerations
\footnote{
Possible realizations of
the chiral symmetry classes AIII, BDI, CII 
possessing time-evolution operators in table
\ref{TableSymmetryClassesTwo}  with $N\not = M$
are tight-binding models on bipartite graphs whose two  (disjoint)
subgraphs contain $N$ and $M$ lattice sites.}.
What is interesting about this column is that
its entries run precisely over what is known
as the complete set of ten (``large'') symmetric spaces %
\footnote[1]{
A symmetric space is a finite-dimensional Riemannian manifold of constant
curvature (its Riemann curvature tensor is covariantly constant)
which has only one parameter, its radius of curvature. 
There are also so-called exceptional
symmetric spaces which, however, are not relevant for the
problem at hand, because for them the number $N$ would
be a fixed finite number, which would prevent us from
being able to take the thermodynamic (infinite-volume) limit
of interest for all the physical systems under consideration.
},
classified in 1926 in fundamental work by the mathematician \'Elie Cartan.
Thus, as the first quantized Hamiltonian runs over all
ten possible symmetry classes, the corresponding 
quantum mechanical time-evolution operator runs over
all ten symmetric spaces.
Thus, the appearance of the Cartan symmetric spaces is
a reflection of fundamental aspects of (single-particle) quantum mechanics.
We will discuss the last column entitled ``$G/H$ (ferm.\ NL$\sigma$M)'' 
in the following subsection.

\begin{table}
\begin{center}
\begin{small}
\begin{tabular}{c||c|c|c||c||c}
\hline
{\small  Cartan label} & $\mathrm{T}$&$\mathrm{C}$ &$\mathrm{S}$ & {\small Hamiltonian}    & $G/H$ (ferm. NL$\sigma$M)
\\\hline\hline\hline
 A {\small(unitary)}    &$0$ &$0$&$0$&$\mathrm{U}(N)$&$\mathrm{U}(2n)/\mathrm{U}(n)\times \mathrm{U}(n)$ \\ \hline
 AI {\small(orthogonal)} &$+1$&$0$&$0$&$\mathrm{U}(N)/\mathrm{O}(N)$&
 $\mathrm{Sp}(2n)/\mathrm{Sp}(n)\times \mathrm{Sp}(n)$ \\ \hline
 AII {\small(symplectic)} &$-1$&$0$&$0$&$\mathrm{U}(2N)/\mathrm{Sp}(2N)$&
$\mathrm{O}(2n)/\mathrm{O}(n)\times \mathrm{O}(n)$\\\hline\hline\hline
 AIII {\small(ch. unit.)} &$0$&$0$&$1$&${\mathrm{U}(N+M)/\mathrm{U}(N)\times \mathrm{U}(M)}$ &$\mathrm{U}(n)$\\ \hline
 BDI {\small(ch. orth.)} &$+1$&$+1$&$1$&$\mathrm{O}(N+M)/\mathrm{O}(N)\times \mathrm{O}(M)$& $\mathrm{U}(2n)/\mathrm{Sp}(2n)$
 \\ \hline
 CII {\small(ch. sympl.)} &$-1$&$-1$&$1$&$\mathrm{Sp}(N+M)/\mathrm{Sp}(N)\times \mathrm{Sp}(M)$& $\mathrm{U}(2n)/\mathrm{O}(2n)$ \\\hline\hline\hline
 D ({\small BdG})                    &$0$ &$+1$&$0$ &$\mathrm{SO}(2N)$      &$\mathrm{O}(2n)/\mathrm{U}(n)$\\ \hline
 C ({\small BdG})                 &$0$ &$-1$&$0$ &$\mathrm{Sp}(2N)$    & $\mathrm{Sp}(2n)/\mathrm{U}(n)$ \\ \hline
 DIII ({\small BdG})                   &$-1$&$+1$&$1$ &$\mathrm{SO}(2N)/\mathrm{U}(N)$    & $\mathrm{O}(2n)$ \\ \hline
 CI ({\small BdG})                &$+1$&$-1$&$1$ &$\mathrm{Sp}(2N)/\mathrm{U}(N)$       & $\mathrm{Sp}(2n)$\\\hline\hline\hline
\end{tabular}
\end{small}
\caption{ \label{TableSymmetryClassesTwo}
Listed are the ten generic symmetry classes of single-particle
Hamiltonians ${\cal H}$, 
classified according to their behavior under time-reversal symmetry 
($\mathcal{T}$),
charge-conjugation (or: particle-hole) symmetry 
($\mathcal{C}$), as well as ``sublattice''
(or: ``chiral'') symmetry ($\mathcal{S}$).
The labels T, C and S, represent 
the presence/absence of time-reversal, particle-hole, and chiral symmetries,
respectively, 
as well as the types of these symmetries.
The column entitled ``Hamiltonian'' 
lists, for each of the ten symmetry classes, the symmetric
space of which
the quantum mechanical time-evolution operator $\exp( \mathrm{i} t {\cal H} )$
is an element.
The column ``Cartan label'' is the name given to the
corresponding symmetric space listed in the column 
``Hamiltonian''
in \'Elie Cartan's classification scheme (dating back to the year 1926).
The last column entitled ``$G/H$ (ferm.\ NL$\sigma$M)'' lists the
(compact sectors of the) target 
space of the NL$\sigma$M describing Anderson localization physics
at long wavelength in this given symmetry class.
}
\end{center}
\end{table}

\subsection{Review: Classification of Topological Insulators (Superconductors)}
\label{Review-Classification-TopIns}

The approach used in Ref.\ \cite{SchnyderRyuFurusakiLudwig2008}
to classify all possible topological insulators (superconductors)
rested on the existence of protected extended degrees of
freedom at the system's  boundary.
These are, in particular,
also protected in the presence of arbitrarily strong perturbations
at the boundary which break translational symmetry 
(commonly referred to as ``random'', or ``disordered'').
The existence of extended, gapless degrees of freedom
even in strongly random  fermionic systems is highly unusual, because
of the phenomenon of Anderson localization
\footnote{This is the phenomenon
that,  at least for sufficiently
strong disorder potentials, spatially extended (= delocalized)
eigenstates of the Hamiltonian tend to become localized (i.e., exponentially
decaying in space)
\cite{AndersonOriginalPaper,LeeRamakrishnanRPM,Efetov97,EversMirlinRMP}.
}. 
Thus, the degrees of
freedom at the boundary of topological insulators (superconductors)
must be of a very special kind, in that they entirely evade the phenomenon
of Anderson localization.
Our approach consists in classifying precisely
all those problems of (non-interacting) fermionic systems which
completely evade the phenomenon of Anderson localization.
We have completed this task in Ref.\ \cite{SchnyderRyuFurusakiLudwig2008}.
Thus, we have reduced the problem of classifying all topological insulators (superconductors)
in $d$ spatial bulk dimensions to a classification problem of Anderson localization
at the $(d-1)$ dimensional boundary. By solving the mentioned problem
of Anderson localization, we have thereby solved the problem of classifying
all topological insulators (superconductors).

We will now focus attention on the above-mentioned
 problem of Anderson localization
at the $(d-1)$ dimensional boundary of a $d$ dimensional topological insulator 
(superconductor).
Specifically, we will now review a solution of this problem
\cite{SchnyderRyuFurusakiLudwig2008,RefToFootnoteInLandau100}
which allows us to see 
directly the dependence
on dimensionality $d$ of the classification of  topological insulators (superconductors).
In the later,  main part of this article, we will present another point of view of
this dependence on dimensionality $d$ (using ``dimensional reduction'').

The theoretical description
of problems of Anderson localization is well known
to be very systematic and geometrical 
\cite{Wegner,Efetov97}.
A problem of Anderson localization is in general described by a random 
Hamiltonian (i.e., one that lacks translational symmetry). 
That Hamiltonian will be  in one of the ten
symmetry classes listed in table~\ref{TableSymmetryClassesTwo}
and we are currently focusing 
on Hamiltonians describing the boundary
of the topological insulator
(superconductor).  Now, as it turns out, at long length scales (much
larger than the ``mean free path'') a description in terms of 
a ``non-linear-sigma-model'' (NL$\sigma$M) emerges.
A NL$\sigma$M
is a system like that describing the classical statistical mechanics
of a Heisenberg magnet. The only difference is that while the magnet 
is formulated in terms of unit vector spins, pointing to the surface of
a two-dimensional
sphere, for a general NL$\sigma$M that spin is 
replaced \footnote{The  two-dimensional sphere $S^2$
is a particularly simple example of
a symmetric space, namely it can be written as the space 
$S^2=$ $\mathrm{U}(2)/\mathrm{U}(1)\times \mathrm{U}(1)$,
i.e., the first row in the last column of
table~\ref{TableSymmetryClassesTwo}, when $n=1$.}
by an element of one of the ten symmetric spaces listed in the last column of 
table~\ref{TableSymmetryClassesTwo}, called the ``target space'' of the
NL$\sigma$M, denoted by $G/H$
\footnote{Both $G$ and $H$ are classical Lie groups,
and $H$ is a maximal subgroup of $G$.}.
For a given symmetry class of the original Hamiltonian, 
whose time-evolution operator is characterized by the penultimate column,
the specific ``target space''
that needs to be used for the 
NL$\sigma$M describing Anderson localization in this symmetry class
is listed in the last column of
table~\ref{TableSymmetryClassesTwo}
(see also appendix \ref{subsec:RMT}). 

Now, the NL$\sigma$M on the $(d-1)$ dimensional
boundary of the $d$ dimensional topological insulator (superconductor)
completely evades Anderson localization if a certain extra 
{\it term of topological origin} can be added to the action of the 
NL$\sigma$M  which has {\it no adjustable parameter}
\cite{SchnyderRyuFurusakiLudwig2008,RefToFootnoteInLandau100}.
Whether such an extra term is allowed depends on
(i): the ``target space'' of the
NL$\sigma$M  in the symmetry class in 
question\footnote{The relevant ``target space''
is, as already mentioned above, listed in the last column of 
table~\ref{TableSymmetryClassesTwo}.},
and
(ii): the dimensionality $\bar{d}:= (d-1)$
 of the boundary on which the
NL$\sigma$M  is  defined.
There are only three {\it terms of topological origin}
which can
possibly
be added to the action of the 
NL$\sigma$M:
these are a $\theta$-term (Pruisken term) 
\cite{Pruisken84},
a $\mathbb{Z}_2$ topological term
\cite{Fendley01},
or a Wess-Zumino-Witten (WZW) term 
\cite{WittenNPB-223-1983-422,AbanovWiegmannNPB-570-2000-685}.
It is the homotopy groups of the NL$\sigma$M target spaces $G/H$
which determine whether it is possible to add such a topological term 
to a given NL$\sigma$M action (see table \ref{homotopy}).
Specifically, a $\theta$-term (Pruisken term) can
appear when $\pi_{\bar{d}}(G/H)=\mathbb{Z}$.
Similarly, a $\mathbb{Z}_2$ topological term is allowed
when $\pi_{\bar{d}}(G/H)=\mathbb{Z}_2$, and
a WZW term can be included 
\footnote{Observe that, while the homotopy group determines if
a term of topological origin for a given NL$\sigma$M action is
allowed in principle,
it depends on the specific disorder model considered,
whether such a term is actually present. In any case, if a 
term of topological origin is possible in a specific dimension
and symmetry class, then a corresponding topological insulator (superconductor)
can exist in this symmetry class (in one dimension higher).}
when $\pi_{\bar{d}+1}(G/H)=\mathbb{Z}$. 
Note that, on the one hand, one obtains, 
upon addition of a $\theta$-term (Pruisken term),
a one-parameter family of theories (on the boundary)
depending on the value of $\theta$, all of which 
reside in the same symmetry class. 
On the other hand, however,  
it is known that only for a special value
of the parameter $\theta$ Anderson localization is  avoided;
for generic values of $\theta$, this is not the case. It is for this reason
that the ability to add a $\theta$-term (Pruisken term) 
is not of interest to the question we are asking.
One is therefore left only  with a $\mathbb{Z}_2$ topological term
or a WZW term 
as the only terms of topological origin which
have no freely adjustable parameter,
and which are thus of relevance here.

The homotopy groups for all ten NL$\sigma$M 
target spaces $G/H$ listed in the last column of
table \ref{TableSymmetryClassesTwo}
are well known  from the literature (see, e.g.,
Ref.~\cite{lundell_90} for a summary) 
and this information
is summarized in table \ref{homotopy}
for the convenience of the reader.
In order to make a certain regular structure of this table apparent,
the rows of 
table~\ref{TableSymmetryClassesTwo}
have been re-ordered in a specific way \cite{KitaevPeriodicity}.
Part of this re-ordering is a subdivision of all symmetry classes into what
is called ``{\it complex case}'' and 
``{\it real case}'' in table~\ref{homotopy}. The physical origin of this subdivision
is simple to understand.  In the category  ``{\it complex case}''
appear precisely those symmetry classes in which there is
no reality condition
($\mathcal{T}$ or $\mathcal{C}$)
whatsoever imposed on the Hamiltonian.
We see from table~\ref{TableSymmetryClassesTwo}
that these are the classes which carry Cartan label A and AIII.
The Hamiltonians  in these symmetry classes are therefore ``complex''
(in this sense).
The other category
called ``{\it real case}''  in  table \ref{homotopy}
consists precisely of all the other eight symmetry classes
which have the property that there is at least one
reality condition 
[$\mathcal{T}$ or $\mathcal{C}$
-- cf.\ equations
(\ref{DEFTimeReversal},\ref{DEFChargeConjugation})]
imposed on the Hamiltonian
(see table~\ref{TableSymmetryClassesTwo}).
In this sense the Hamiltonians in 
these symmetry classes are therefore ``real''.

\begin{table}[h]
\paragraph{complex case:}
\begin{center}
\begin{tabular}{cccccccccccc}\hline
 AZ & $G/H$ &  & $\bar{d}=0$ &  $\bar{d}=1$ & $\bar{d}=2$ & $\bar{d}=3$ & $\bar{d}=4$ & $\bar{d}=5$ & $\bar{d}=6$ & $\bar{d}=7$  & \\ \hline\hline
 A & $\scriptstyle \mathrm{U}(N+M)/\mathrm{U}(N)\times \mathrm{U}(M)$ &
 \multicolumn{2}{c}{
  $\leftarrow$ $\mathbb{Z}$ \thinspace   }& 
	  \multicolumn{2}{c}{
\textcolor{blue}{\bf 0} 
\thinspace   $\longleftarrow$  \;  $\mathbb{Z}$} & \multicolumn{2}{c}{ 
\textcolor{blue}{\bf 0}
 \thinspace   $\longleftarrow$  \;  $\mathbb{Z}$} & \multicolumn{2}{c}{ 
\textcolor{blue}{\bf 0}
 \thinspace   $\longleftarrow$  \;  $\mathbb{Z}$} &   
\multicolumn{2}{c}{
\thinspace \textcolor{blue}{\bf 0} 
$\leftarrow$
}
\\\hline
{\bf AIII} & $\scriptstyle \mathrm{U}(N)$ &  & \multicolumn{2}{c}{ 
\textcolor{blue}{\bf 0}
  \thinspace  $\longleftarrow$  \;   $ \mathbb{Z}$} & 
 \multicolumn{2}{c}{ 
\textcolor{blue}{\bf 0}
 \thinspace   $\longleftarrow$  \;  $\mathbb{Z}$ } &   \multicolumn{2}{c}{ 
\textcolor{blue}{\bf 0}
  \thinspace  $\longleftarrow$  \;   $ \mathbb{Z}$} &  \multicolumn{2}{c}{ 
\textcolor{blue}{\bf 0}
  \thinspace  $\longleftarrow$  \;   $ \mathbb{Z}$}  &
  \\\hline
\end{tabular}
\end{center}
\paragraph{real case:}
\begin{center}
\begin{tabular}{cccccccccccc}\hline
  AZ   & $G/H$  &  & $\bar{d}=0$ & $\bar{d}=1$ & $\bar{d}=2$ & $\bar{d}=3$  & $\bar{d}=4 $ & $\bar{d}=5 $ & $\bar{d}=6 $ & $\bar{d}=7 $ & \\ \hline\hline
 AI & \hspace{-0.45cm} $\scriptstyle \mathrm{Sp}(N+M)/\mathrm{Sp}(N)\times \mathrm{Sp}(M)$ \hspace{-0.45cm}  & 
 \multicolumn{2}{c}{ 	$\leftarrow$  $\mathbb{Z}$  \thinspace }
 & 0& 0 & \multicolumn{2}{c}{ 
\textcolor{blue}{\bf 0}
 \thinspace   $\longleftarrow$  \; $\mathbb{Z}$} &  
\textcolor{blue}{\bf  $\mathbb{Z}_2$}
 &  
\textcolor{blue}{\bf  $\mathbb{Z}_2$}
 &  
\multicolumn{2}{c}{
\thinspace \textcolor{blue}{\bf 0} 
$\leftarrow$
}
  \\\hline
 {\bf BDI} & $\scriptstyle \mathrm{U}(2N)/\mathrm{Sp}(2N)$ & & \multicolumn{2}{c}{ 
\textcolor{blue}{\bf 0} 
\thinspace   $\longleftarrow$  \; $\mathbb{Z}$} & 0 & 0 &  \multicolumn{2}{c}{ 
\textcolor{blue}{\bf 0} 
\thinspace   $\longleftarrow$  \; $\mathbb{Z}$} &   
\textcolor{blue}{\bf  $\mathbb{Z}_2$} &  \textcolor{blue}{\bf  $\mathbb{Z}_2$} &
 \\\hline
 D & $\scriptstyle \mathrm{O}(2N)/\mathrm{U}(N)$ & & \textcolor{blue}{\bf $\mathbb{Z}_2$} & \multicolumn{2}{c}{
\textcolor{blue}{\bf 0}
 \thinspace   $\longleftarrow$  \; $\mathbb{Z}$} & 0 & 0 &   \multicolumn{2}{c}{ 
\textcolor{blue}{\bf 0}
 \thinspace   $\longleftarrow$  \; $\mathbb{Z}$} &  \textcolor{blue}{\bf  $\mathbb{Z}_2$} &
  \\\hline
{\bf DIII} & $\scriptstyle \mathrm{O}(N)$ & &  \textcolor{blue}{\bf  $\mathbb{Z}_2$} &  
\textcolor{blue}{\bf  $\mathbb{Z}_2$} & \multicolumn{2}{c}{ 
\textcolor{blue}{\bf 0}
 \thinspace   $\longleftarrow$  \; $\mathbb{Z}$} & 0 & 0 &  \multicolumn{2}{c}{ 
\textcolor{blue}{\bf 0}
 \thinspace   $\longleftarrow$  \; $\mathbb{Z}$} &
 \\\hline
 AII & $\scriptstyle \mathrm{O}(N+M)/\mathrm{O}(N)\times \mathrm{O}(M)$ & 
  \multicolumn{2}{c}{ 	$\leftarrow$  $\mathbb{Z}$  \thinspace }
 &  \textcolor{blue}{\bf $\mathbb{Z}_2$} &  \textcolor{blue}{\bf $\mathbb{Z}_2$} & \multicolumn{2}{c}{
 \textcolor{blue}{\bf 0}
 \thinspace   $\longleftarrow$  \;  $\mathbb{Z}$}   & 0 & 0 &  
\multicolumn{2}{c}{
\thinspace \textcolor{blue}{\bf 0} 
$\leftarrow$
}
 \\\hline
 {\bf CII} & $\scriptstyle \mathrm{U}(N)/\mathrm{O}(N)$   & & \multicolumn{2}{c}{ \textcolor{blue}{\bf 0} \thinspace   $\longleftarrow$  \; $\mathbb{Z}$} &		  \textcolor{blue}{\bf $\mathbb{Z}_2$} &  \textcolor{blue}{\bf $\mathbb{Z}_2$} &  \multicolumn{2}{c}{ 
\textcolor{blue}{\bf 0}
 \thinspace   $\longleftarrow$  \;  $\mathbb{Z}$}  & 0 & 0 &
 \\\hline 
 C & $\scriptstyle \mathrm{Sp}(2N)/\mathrm{U}(N)$  & & 0 & \multicolumn{2}{c}{ 
\textcolor{blue}{\bf 0}
 \thinspace   $\longleftarrow$  \; $\mathbb{Z}$} &  \textcolor{blue}{\bf $\mathbb{Z}_2$} &   \textcolor{blue}{\bf  $\mathbb{Z}_2$}  &  \multicolumn{2}{c}{ 
\textcolor{blue}{\bf 0}
 \thinspace   $\longleftarrow$  \;  $\mathbb{Z}$}  & 0 &
 \\\hline
{\bf CI} & $\scriptstyle \mathrm{Sp}(2N)$       &         & 0 & 0 &  \multicolumn{2}{c}{ 
\textcolor{blue}{\bf 0}
 \thinspace   $\longleftarrow$  \; $\mathbb{Z}$} &   \textcolor{blue}{\bf $\mathbb{Z}_2$}& \textcolor{blue}{\bf $\mathbb{Z}_2$} & \multicolumn{2}{c}{ 
\textcolor{blue}{\bf 0}
 \thinspace   $\longleftarrow$  \;  $\mathbb{Z}$}  
  \\\hline
\end{tabular}
\end{center}
\caption{
Table of homotopy groups $\pi_{\bar{d}}(G/H)$ for symmetric spaces $G/H$,
taken from the standard mathematical literature 
(see e.g.\ Ref.\ \cite{lundell_90} for a summary).
[Here, $N$ must be sufficiently large for a given $\bar{d}$.
Cartan labels of those symmetry classes invariant under the
chiral symmetry operation 
$\mathcal{S}=\mathcal{T} \cdot \mathcal{C}$ 
from table \ref{TableSymmetryClassesTwo}
are indicated by bold face letters.].
The pattern continues for higher $\bar{d}$,
with periodicity 2 for the complex case,
and 8 for the real case.
The entries corresponding to $\mathbb{Z}$
topological insulators (superconductors)
in $(\bar{d}+1)$ dimensions
(see e.g.\ table~\ref{periodic table})
are indicated by 
blue color boldface symbols.
There is a ($\bar{d}+1$)-dimensional $\mathbb{Z}_2$ topological insulator, 
whenever $\pi_{\bar{d}} (G / H) = \mathbb{Z}_2$ (also indicated in blue).
A ($\bar{d}+1$)-dimensional $\mathbb{Z}$ topological insulator,
on the other hand,
can be realized, whenever $\pi_{\bar{d}+1} (G/H) = \mathbb{Z}$.
-- This is a way \cite{SchnyderRyuFurusakiLudwig2008,RefToFootnoteInLandau100}
to directly relate  the classification of topological insulators (superconductors)
to the table of homotopy groups.
\label{homotopy}
}
\end{table}

Table \ref{homotopy} now tells us directly in which symmetry class there exist
topological insulators
or superconductors
(and of which type, $\mathbb{Z}_2$ or $\mathbb{Z}$): 
according to the above discussion we know that a  topological insulator
(superconductor)
exists in a given symmetry class in $d= \bar{d}+1$ spatial dimensions
if and only if the target space of the NL$\sigma$M on the $\bar{d}$-dimensional 
boundary allows either (a): 
for $\mathbb{Z}_2$ topological term, which  is the case when
$\pi_{\bar{d}}(G/H)=\pi_{d-1}(G/H)=\mathbb{Z}_2$,
or (b): allows for a WZW term, which
is the case when
$\pi_{d}(G/H)=\pi_{\bar{d}+1}(G/H)=\mathbb{Z}$.
By using this rule in conjunction with table \ref{homotopy} of homotopy groups, 
we arrive with the help of table \ref{TableSymmetryClassesTwo} at the table
\ref{periodic table} of topological 
insulators and superconductors
\footnote{To be explicit,
one has to move all the entries $\mathbb{Z}$ in table
 \ref{homotopy}  into the locations indicated by the arrows, and has
to replace the column label $\bar{d}$ by $d=\bar{d}+1$. 
The result is table \ref{periodic table}.}.

A look at table \ref{periodic table} reveals that 
in each spatial dimension there exist five distinct classes
of topological insulators (superconductors),
three of which are characterized by an integral ($\mathbb{Z}$) topological
number, while the remaining two possess a binary ($\mathbb{Z}_2$)
topological quantity \footnote{Note that while $d=0,1,2,3$-dimensional systems are of 
direct physical relevance,
higher dimensional topological states might be of interest indirectly,
because, for example, some of the additional components of momentum 
in a higher dimensional space may be interpreted as adiabatic parameters (external parameters
on which the Hamiltonian depends, and which can be changed adiabatically, traversing
closed paths in parameter space -- sometimes referred to as 
adiabatic ``pumping processes'').
}.

\begin{table}[t!]
\paragraph{complex case:}
\begin{center}
\begin{tabular}{cccccccccccccc}\hline
Cartan$\backslash d$ & 0  & 1 & 2 & 3 & 4 & 5 & 6 & 7 & 8 & 9 & 10 & 11 & $\cdots$ \\ \hline\hline
A   & $\mathbb{Z}$ & 0 & $\mathbb{Z}$ & 0 
    & $\mathbb{Z}$ & 0 & $\mathbb{Z}$ & 0
    & $\mathbb{Z}$ & 0 & $\mathbb{Z}$ & 0 
    & $\cdots$
\\\hline
AIII  & 0 & $\mathbb{Z}$ & 0 & $\mathbb{Z}$
      & 0 & $\mathbb{Z}$ & 0 & $\mathbb{Z}$
      & 0 & $\mathbb{Z}$ & 0 & $\mathbb{Z}$
      & $\cdots$
\\\hline
\end{tabular}
\end{center}
\paragraph{real case:}
\begin{center}
\begin{tabular}{cccccccccccccc}\hline
Cartan$\backslash d$ & 0  & 1 & 2 & 3 & 4 & 5 & 6 & 7 & 8 & 9 & 10 & 11 & $\cdots$ \\ \hline\hline
AI  & $\mathbb{Z}$ & 0 & 0 
    & 0 & $2\mathbb{Z}$ & 0 
    & $\mathbb{Z}_2$ & $\mathbb{Z}_2$ & $\mathbb{Z}$ 
    & 0 & 0 & 0 & $\cdots$ \\ \hline
BDI & $\mathbb{Z}_2$ & $\mathbb{Z}$ & 0 & 0 
    & 0 & $2\mathbb{Z}$ & 0 
    & $\mathbb{Z}_2$ & $\mathbb{Z}_2$ & $\mathbb{Z}$ 
    & 0 & 0 &  $\cdots$ \\ \hline
D   & $\mathbb{Z}_2$ & $\mathbb{Z}_2$ & $\mathbb{Z}$ 
    & 0 & 0 & 0 
    & $2\mathbb{Z}$  & 0 & $\mathbb{Z}_2$ 
    & $\mathbb{Z}_2$ & $\mathbb{Z}$  & 0 & $\cdots$ \\ \hline
DIII& 0 & $\mathbb{Z}_2$ & $\mathbb{Z}_2$ & $\mathbb{Z}$ 
    & 0 & 0 & 0 
    & $2\mathbb{Z}$  & 0 & $\mathbb{Z}_2$ 
    & $\mathbb{Z}_2$ & $\mathbb{Z}$  &  $\cdots$ \\ \hline
AII & $2\mathbb{Z}$  & 0 & $\mathbb{Z}_2$ 
    & $\mathbb{Z}_2$ & $\mathbb{Z}$ & 0 
    & 0 & 0 & $2\mathbb{Z}$ 
    & 0 & $\mathbb{Z}_2$ & $\mathbb{Z}_2$&  $\cdots$\\ \hline
CII & 0 & $2\mathbb{Z}$  & 0 & $\mathbb{Z}_2$ 
    & $\mathbb{Z}_2$ & $\mathbb{Z}$ & 0 
    & 0 & 0 & $2\mathbb{Z}$ 
    & 0 & $\mathbb{Z}_2$ &  $\cdots$\\ \hline
C   & 0  & 0 & $2\mathbb{Z}$  
    & 0 & $\mathbb{Z}_2$  & $\mathbb{Z}_2$ 
    & $\mathbb{Z}$ & 0 & 0 
    & 0 & $2\mathbb{Z}$ & 0 & $\cdots$ \\ \hline
CI  & 0 & 0  & 0 & $2\mathbb{Z}$  
    & 0 & $\mathbb{Z}_2$  & $\mathbb{Z}_2$ 
    & $\mathbb{Z}$ & 0 & 0 
    & 0 & $2\mathbb{Z}$  & $\cdots$ \\ \hline
\end{tabular}
\end{center}
\caption{
Classification of topological insulators and 
superconductors as a function of spatial dimension $d$
and symmetry class, indicated by the ``Cartan label'' (first column).
The definition of the ten 
generic
symmetry classes
of single particle Hamiltonians
(due to Altland and Zirnbauer\cite{Zirnbauer96,Altland97}%
)
is given in 
table~ \ref{TableSymmetryClassesTwo}.
The symmetry
classes are grouped in two separate lists,
complex and real cases, depending on whether
the Hamiltonian is complex, or whether one
(or more) reality conditions (arising from
time-reversal or charge-conjugation symmetries)
are imposed on it;
the symmetry classes are ordered in such a way
that a periodic pattern in dimensionality
becomes visible \cite{KitaevLandau100Proceedings}.
(See also the discussion in subsection
\ref{Review-Classification-Hamiltonian} and table \ref{homotopy}.)
The symbols $\mathbb{Z}$ and $\mathbb{Z}_2$ indicate that
the topologically distinct phases within a given 
symmetry class of topological insulators (superconductors) are
characterized by an integer invariant ($\mathbb{Z}$),
or a $\mathbb{Z}_2$ quantity, respectively. 
The symbol ``$0$'' denotes the case when
there exists no
topological insulator (superconductor), i.e., when
all quantum ground states are topologically equivalent to the trivial state.
}
\label{periodic table}
\end{table}

The topological insulators (superconductors) appear in table \ref{periodic table}
along diagonal lines. 
That is, as the spatial dimension is increased
by one, locations where the topological insulators (superconductors) appear in the table
shift down by one column. This entire pattern descends
directly from an analogous
one appearing in the table \ref{homotopy} of homotopy groups.

In the main part of the present paper we will give another
derivation of this periodicity structure of 
table \ref{periodic table}, 
by ``dimensional reduction'' and
by considering the Dirac Hamiltonian
representatives in each
of the symmetry classes
in which there exist topological insulators (superconductors) in a given spatial dimension.
By running the argument backwards, this can then,
in turn, also be viewed as another route to obtain  the table
\ref{homotopy} of homotopy groups.

Let us pause for a moment to point out a few interesting
aspects relating to table \ref{periodic table}.
First, it is interesting to note, that the 
periodic structure of table~\ref{periodic table} is
quite analogous to the
dimensional hierarchy of anomalies (``anomaly ladder'')
known in gauge theories,
e.g.,
the (3+1)-dimensional Abelian anomaly,
the  (2+1)-dimensional parity anomaly,
and
the (1+1)-dimensional non-Abelian anomaly \cite{Zumino,Nakahara03,Fujikawa}.
This may not appear to be completely unexpected, considering that
the IQHE is known as 
a condensed matter realization of the parity ``anomaly'' %
\footnote{
The IQHE is not quite an anomaly in the sense that 
the parity and 
$\mathcal{T}$
is explicitly broken. 
However, the quantization of $\sigma_{xy}$ in the IQHE
can be viewed as essentially the same phenomenon as 
the appearance of the Chern-Simons term in
QED in $D=2+1$-dimensional space-time, where
massless Dirac fermions are coupled with the electromagnetic
U(1) gauge field. 
}.

Second, we want to draw the reader's attention to the fact
that for a given symmetry class in table~\ref{periodic table}  
any $\mathbb{Z}_2$ topological insulator (superconductor) always appears
as part of a ``triplet'' that consists of a $d$-dimensional $\mathbb{Z}$
topological insulator (superconductor) and two $\mathbb{Z}_2$ topological
insulators (superconductors) in $(d-1)$ and $(d-2)$ dimensions.
This suggests that the topological characteristics of the members of
such a ``triplet'' are closely related and indeed,
it was shown in Ref.~\cite{Qi_Taylor_Zhang} that the $\mathbb{Z}_2$ classification
in symmetry class AII in $d=2$ and $3$ can be derived from
the four-dimensional $\mathbb{Z}$ topological insulator
by
a process of ``dimensional reduction''.
The same procedure has been applied \cite{Qi_Taylor_Zhang}
to derive $\mathbb{Z}_2$ classifications in those symmetry classes that 
do not possess a form of chiral symmetry
(i.e., for which $\mathrm{S}=0$) 
-- there are five of them.
In section~\ref{sec:Z2_dimreduction} we will extend this approach to
the
$\mathbb{Z}_2$ topological insulators (superconductors) in
the remaining five symmetry classes which possess 
a form of chiral symmetry
(i.e., for which $\mathrm{S}=1$).

Let us now  discuss
explicit forms of the
``terms of topological origin'' which can be added to 
the action 
of the NL$\sigma$M 
with
``target space'' $G/H$,
at the boundary of the topological
insulator (superconductor).

\subsubsection{Terms of WZW type.} \label{WZWterms}

Here there is a pattern, alternating in the dimension $\bar{d}$ of
the boundary, for the way these topological terms can be constructed. 
(i): For a $\bar{d}=\mbox{odd}$-dimensional boundary, 
and for the NL$\sigma$Ms which describe the Anderson-delocalized 
boundary of the topological bulk,
the field (denoted by $Q\in G/H$ below~%
\footnote[1]{While looking similar, the field $Q$ in the NL$\sigma$M has 
nothing to do with the spectral projector in momentum space 
defined in equation (\ref{P and Q}).} 
 -- 
cf.\ last column of  table \ref{TableSymmetryClassesTwo})
of the NL$\sigma$M field theory
is a Hermitean matrix field which satisfies certain constraints. 
The WZW term responsible for the lack of Anderson localization takes on 
the form of the $\bar{d}$-dimensional integral over
the boundary of the $\bar{d}$-dimensional Chern-Simons form
(compare equation~\re{IntegralOfChernSimonsForm} 
of the main text below).
Alternatively, it can be written also as an integral
over a $\bar{d}+1$-dimensional region whose boundary coincides with 
the physical $\bar{d}$-dimensional space, by smoothly interpolating
the NL$\sigma$M field configuration $Q$ into one dimension higher.
The fact that
$\pi_{\bar{d}+1}(G/H)=$ $\mathbb{Z}$ guarantees that 
different ways of interpolations do not matter, and 
the action depends only on the physical field configurations
in $\bar{d}$-dimensional space.
(ii): For a $\bar{d}=\mbox{even}$-dimensional boundary, 
and for the NL$\sigma$Ms which describe
the Anderson-delocalized boundary of the topological bulk,
the field in the NL$\sigma$M field theory is
either a group element or
a unitary matrix field (denoted by $g \in G/H$ below --
cf.\ last column of  table \ref{TableSymmetryClassesTwo})
which may, depending on the case, be subject to certain constraints.
The WZW term responsible for 
the lack of Anderson localization takes on
the form of a $\bar{d}+1$-dimensional integral of the winding number density,
as defined in equation~\re{winding no} of the main text below.

To illustrate these terms, we will now give explicit examples for them
in low dimensionalities:
\begin{itemize}

\item{\underbar{$d=2$:}} For the $d=2$-dimensional
topological insulator of the IQHE,
the $\bar{d}=1$-dimensional NL$\sigma$M describing 
the edge states
is the one on $\mathrm{U}(2n)/\mathrm{U}(n)\times \mathrm{U}(n)$.  
The field $Q$ of the NL$\sigma$M field theory can be 
parameterized 
as
$Q=U^{\dag}\Lambda U \in \mathrm{U}(2n)/\mathrm{U}(n)\times \mathrm{U}(n)$
where $U\in  \mathrm{U}(2n)$ and 
$\Lambda = \mathrm{diag}\, \left(I_n, -I_n\right)$.
The term 
of relevance for the absence of Anderson localization 
at the edges of the
integer quantum Hall insulator is 
$\propto \sigma_{xy} \int dx\, \mathrm{tr}[A_x]$
where $A_x := U^{\dag} \partial_x U$;
this is a $\bar{d}=1$ dimensional analogue of 
the three-dimensional Chern-Simons term.
This term can also be rewritten as a WZW type two-dimensional integral
$\propto \sigma_{xy}  \int_D dx du\, 
\epsilon^{\mu\nu} 
\mathrm{tr}\, 
\left[Q \partial_\mu Q \partial_\nu Q \right]$
($\mu,\nu=x,u$). 
Here,  we have extended the original $\bar{d}=1$ dimensional space
to a two-dimensional region $D$
by adding a fictitious space direction, parameterized by $u \in [0,1]$. 
The boundary $\partial D$ of the two-dimensional region 
coincides with the original $\bar{d}=1$ dimensional space. 
Accordingly, the original NL$\sigma$M field $Q(x)$ is smoothly extended
to $Q(x,u)$ such that it coincides with $Q(x)$ when $u=0$. 
Since 
$\pi_{2}\left[\mathrm{U}(2n)/\mathrm{U}(n)\times \mathrm{U}(n)\right]=\mathbb{Z}$
the two-dimensional integral turns out to depend only on the field configuration
on $\partial D$. 
Note that when we specialize to the case of $n=1$, all these 
expressions
are well-known from
the coherent state path integral of an SU(2) quantum spin 
whose path integral is described by a $d=1$ dimensional
NL$\sigma$M on the  target space
$\mathrm{O}(3) = \mathrm{U}(2)/\mathrm{U}(1)\times \mathrm{U}(1)$.

\item{\underbar{$d=3$}:}
    For $d=3$ topological insulators (superconductors) in classes AIII, DIII and CI,
    the relevant NL$\sigma$Ms at their $\bar{d}=2$-dimensional boundary (surface) are
    the ones with  a group manifold as target manifold
    [$\mathrm{U}(n)$, $\mathrm{O}(n)$, $\mathrm{Sp}(n)$
     for classes AIII, DIII, and CI, respectively]. 
    The relevant WZW type term can be written as a three-dimensional integral, 
    $\int_D d^2x\, du\, \epsilon^{\mu\nu\lambda}\mathrm{tr}\, \left[
    (g^{-1} \partial_{\mu} g)
    (g^{-1} \partial_{\nu} g)
    (g^{-1} \partial_{\lambda} g)
    \right]$,
    where $g\in$$\mathrm{U}(n)$, $\mathrm{O}(n)$, $\mathrm{Sp}(n)$
    \cite{WittenNPB-223-1983-422}. 
    Here again,  the original $\bar{d}=2$ dimensional space 
    is smoothly extended to the three-dimensional region $D$,
    in such a way that $\partial D$ coincides with the physical
    $\bar{d}=2$ dimensional space.
    Accordingly, the field configuration $g(x,y)$ defined on the $\bar{d}=2$ 
dimensional space
    is smoothly extended to the three-dimensional one $g(x,y,u)$ such that 
     $g(x,y,u=0)=g(x,y)$.

\item{\underbar{$d=4$}:}
     For the $d=4$ topological insulators in classes A, AII, and AI,
     the relevant $\bar{d}=3$ NL$\sigma$Ms
     have target spaces $G/H=$ $\mathrm{G}(2n)/\mathrm{G}(n)\times \mathrm{G}(n)$ 
     with 
     $\mathrm{G} = \mathrm{U}$, $\mathrm{O}$, and $\mathrm{Sp}$ respectively. 
     The field of the NL$\sigma$M can be 
     parameterized by 
     $Q=U^{\dag}\Lambda U $
     where $U\in  \mathrm{G}(2n)$. 
     The relevant 
    ``term of topological origin'' is the Chern-Simons term 
     $\propto \int d^3x\, \epsilon^{\mu\nu\lambda}
     \mathrm{tr}\,\left[  A_{\mu}\partial_{\nu}A_{\lambda} + 
     (2/3) A_{\mu}A_{\nu}A_{\lambda}  \right]$
     where $A_{\mu}= U^{\dag}\partial_{\mu}U$ ($\mu=x,y,z$). 
     The appearance of this term at the boundary
     of the $d=4$ topological insulators in class AII
     was discussed in Ref.\ \cite{NomuraKoshinoRyu}.

\item{\underbar{$d=5$}:}
For the $d=5$ topological insulators in  symmetry classes
AIII, BDI, CII, the
$\bar{d}=4$ dimensional NL$\sigma$Ms 
have target spaces
with a matrix field $g$ which is an element of
$G/H=$ $\mathrm{U}(n)$, $\mathrm{U}(2n)/\mathrm{Sp}(2n)$, 
$\mathrm{U}(2n)/\mathrm{O}(2n)$,
respectively. 
For these spaces the WZW term takes 
the form
of the $\bar{d}+1=5$ dimensional integral 
(with boundary) 
of the winding number density 
defined in equation \re{winding no density} of the main text below.
\end{itemize}

Note that those entries in table \ref{homotopy}
of homotopy groups which are integers,
$\pi_{\bar{d}}(G/H) =$ $\mathbb{Z}$, have a periodicity
in $\bar{d}$ equal to four. Therefore, the forms of the ``terms of topological origin''
of WZW type
listed in the above low-dimensional cases will repeat, with the appropriate replacement
of $\bar{d}$.

\vskip .5cm

\subsubsection{Terms of topological origin of $\mathbb{Z}_2$ type.}
The $\mathbb{Z}_2$ topological term for
the NL$\sigma$M  at the $\bar{d}=2$ dimensional surface of the
$d=3$-dimensional $\mathbb{Z}_2$ topological insulators in symmetry class AII
was discussed in Refs.\ \cite{NomuraKoshinoRyu,RyuMudryObuseFurusaki,Ostrovsky}. 
Similarly, 
at the $\bar{d}=2$ dimensional surface of the
$d=3$-dimensional $\mathbb{Z}_2$ topological insulators in symmetry class CII, 
the $\mathbb{Z}_2$ topological term is added to the NL$\sigma$M 
\cite{RyuMudryFurusakiLudwig}.

\vskip .5cm

\subsubsection{Classifying space.}
There is a third way in which
the set of Cartan symmetric spaces listed
in table \ref{TableSymmetryClassesTwo}
appear in the context of the classification of topological
insulators (superconductors).
This arises from
Fermi statistics which  allows for a distinction between those
states of the (first quantized) Hamiltonian ${\cal H}$
with energies
which lie above, and those that lie below the Fermi energy $E_F$.
In a topological insulator (superconductor) there is always an
energy gap between the Fermi energy $E_F$ and the eigenstates
of ${\cal H}$ lying above as well as those lying below $E_F$.
Therefore one may, {\it without closing the bulk gap}, 
{\it continuously deform}  the Hamiltonian ${\cal H}$
into one in which all eigenstates  below the Fermi energy $E_F$
have the same energy (say) $E=-1$, and all 
eigenstates above the Fermi energy have the same energy (say) $E=+1$.
By construction, the resulting 
``simplified Hamiltonian'' \footnote{This ``simplified Hamiltonian''
$Q$ is often referred to in Ref.\
\cite{SchnyderRyuFurusakiLudwig2008} and in the present article
as a ``projector'', since it is trivially related to the projection
operator $P$ onto all filled (with $E<E_F$) eigenstates of ${\cal H}$
by $Q= 1 - 2 P$. (The matrix in subsection \ref{WZWterms} that is denoted by the same letter $Q$ is an
entirely different object, not to be confused with $Q=1-2P$ appearing here.)},
which we often denote by $Q$, has the same topological
properties as the original Hamiltonian.
The structure of $Q$ for all symmetry classes is listed
in table III (second column)
 of the first article of Ref.\ \cite{SchnyderRyuFurusakiLudwig2008}  
for systems with translational
symmetry. ($Q$ will also be used again in the present article.)
If we were to  consider a ``zero-dimensional'' topological insulator (superconductor)
where ``space consisted of a single point'', 
then for each of the ten symmetry classes
we 
obtain\footnote{
From table III of the first article of
Ref.\ \cite{SchnyderRyuFurusakiLudwig2008},
by setting the wavevector $k\to 0$.}
a matrix $Q$ of a certain kind.
It turns out that the so-obtained matrix $Q$ for each symmetry class is
again
an element of
one of the ten symmetric spaces listed in table~\ref{TableSymmetryClassesTwo},
except that the order in which  these spaces  appear is different
from the order in the
two columns of table \ref{TableSymmetryClassesTwo}.
The resulting assignment is 
listed\footnote{The last row entitled ``classifying space'' 
in table~\ref{RMT, NLsM}, is nothing
but a list of the matrices $Q$, and this list follows from the second 
column of table III of the first article of Ref.\ \cite{SchnyderRyuFurusakiLudwig2008} by letting $k=0$.}
in the last column in table \ref{RMT, NLsM}.
The symmetric spaces to which the matrices $Q$ belong are what is called the 
``classifying space''
in the classification scheme 
of Ref.~\cite{KitaevLandau100Proceedings} 
employing K-Theory \footnote{The list of ``classifying spaces'' was obtained 
in Ref.\ \cite{KitaevLandau100Proceedings}
through the use of Clifford algebras.}.
In appendix \ref{subsec:RMT}  we collect a number of interesting relationships
between these three lists of the ten symmetric spaces:
(i) the unitary time evolution operator, (ii) the 
NL$\sigma$M target space, and (iii) the classifying space.

We end this subsection by commenting on additional information that
can be read off from
our main results, listed in Table \ref{periodic table}.
This pertains to the existence of so-called {\it weak}
topological insulators (superconductors),
as well as of zero-energy or extended 
modes localized on topological defects in 
topological insulators (superconductors).

\subsubsection{Weak topological insulators and superconductors.} 
Table \ref{periodic table} classifies
topological features
of gapped free fermion Hamiltonians that do not depend
on the presence of 
translational symmetries
of a crystal lattice
\footnote{
As we will explain below, 
in order to classify such so-called ``strong topological insulators"
it is sufficient to only consider \emph{continuous} Hamiltonians 
with momentum space $S^d$, the $d$-dimensional sphere. 
}.
In particular, these properties are not destroyed when translation
symmetry is broken, e.g., by the introduction of positional disorder.
Systems which exhibit such robust topological properties
are often also referred to as {\it strong} topological
insulators (superconductors).
This is to contrast them with so-called {\it weak} topological
insulators (superconductors) which only possess topological features
when translational symmetry is present. As soon as translational
symmetry is broken, such {\it weak} topological features are no longer
guaranteed to exist, and the system is allowed to become topologically trivial.
For example, systems defined on a $d$-dimensional lattice whose momentum space 
is the $d$-dimensional torus $T^d$,
allow for weak topological insulators which are not
strong topological insulators.
Such systems 
are topologically equivalent to parallel stacks of lower dimensional 
strong topological insulators (superconductors).
Specifically, 
the three-dimensional ``weak'' 
integer quantum Hall insulator
on a lattice, discussed in Ref.~\cite{KohmotoHalperionWu},
is essentially a layered version of the two-dimensional
integer quantum Hall insulator. Hence, it
is characterized by a triplet of Chern numbers, each describing
the winding of a map from the two-dimensional torus $T^2$,
which is a subspace of the three-dimensional momentum space $T^3$, 
onto the complex Grassmannian, $G_{m,m+n} ( \mathbb{C} )$.
Similarly, in symmetry class AII, there exists a three-dimensional 
weak topological insulator \cite{Moore06, Roy3d,Fu06_3Da,Fu06_3Db},
which consists of layered
two-dimensional $\mathbb{Z}_2$ topological quantum states, 
and whose weak topological features are described by a triplet of
$\mathbb{Z}_2$ invariants.

The three-dimensional ``weak" 
integer quantum Hall insulators, as well as
the weak topological insulators of Refs.~\cite{Moore06,Roy3d,Fu06_3Da,Fu06_3Db}
are both examples of  $d=3$-dimensional 
weak topological insulators of ``codimension" one, i.e., they can be viewed as
one-dimensional arrays of $d=2$-dimensional strong topological insulators. 
The existence of these states can be read off from the $d=2$-column
and the rows of Table \ref{periodic table} labeled A and AII, respectively.
In general, $d$-dimensional weak 
topological states can be of any ``codimension" $k$, with $0 < k \leq d$.
The existence of weak topological insulators and superconductors
in $d$ spatial dimensions and for a given symmetry class can be readily
inferred from table~\ref{periodic table}.
That is, $d$-dimensional weak topological insulators (superconductors) of
``codimension" $k$ can occur whenever there
exists a strong topological state in the same symmetry class,
but in $d-k$ dimensions.  
Specifically, the topological order  (characterized by elements in
$\mathbb{Z}$ or in $\mathbb{Z}_2$)
can be specified for each of the $d \choose k$  ``orientations'' of the 
submanifold.
Moreover, it is also possible that in addition ``strong" topological order, 
i.e., in the full space dimension $d$, exists.
In the K-Theory description of Ref.~\cite{KitaevLandau100Proceedings}
the presence of weak topological features is described by additional
summands in the Abelian group of topological invariants 
(i.e., $\mathbb{Z}$ or $\mathbb{Z}_2$)  which appears when
homotopy classes of maps
from the sphere (strong topological insulators) are replaced by maps
from the torus $T^d$
(weak and strong topological insulators).

\subsubsection{ Zero modes localized on topological defects.}  
Another interesting application of table~\ref{periodic table} is to determine
whether $r$-dimensional topological defects in topological insulators
(superconductors) can support  localized (isolated) zero modes.
For example, let us first consider a point-like defect 
(i.e., $r=0$), which is embedded in a $d$-dimensional system ($d\ge 1$)
of any symmetry class.
Furthermore, let us assume that
at the defect time-reversal symmetry
is broken ($\mathrm{T}=0$)
but particle-hole symmetry is preserved with $\mathrm{C}=+1$,
i.e., we are dealing with 
symmetry class D.
From table \ref{periodic table}
we infer that in symmetry class D in $d=r+1=1$ dimensions there is 
a $\mathbb{Z}_2$ classification, which implies that the ($r=0$)-dimensional
boundary can support gapless states, and, therefore, a point-like defect
with the symmetries of class D can bind zero modes.
This situation is realized, e.g., for a vortex core in a  
($p \pm \mathrm{i} p$)-wave superconductor, 
which can bind isolated Majorana modes.
In general, whether it is possible for a $r$-dimensional topological defect
of a given symmetry class to support
gapless states or not is determined  by the entry of
table~\ref{periodic table} at 
the intersection of  column $d=r+1$
and the row of the given symmetry class.
An application of this general statement
can for example be found in the work of Ref.~\cite{ran_vishwanath_09},
where a  dislocation line ($r=1$) in a $d=3$-dimensional lattice
hosting a weak topological insulator in symmetry class AII was found to bind 
an extended gapless zero mode. (This corresponds in table \ref{periodic table}
to the intersection of the row denoting  symmetry class AII 
with the column $d=r+1=1+1=2$, the latter representing
the $d=3$-dimensional weak topological insulator, which corresponds to a strong
topological insulator in ``codimension" $k=1$.)
To illustrate the use of table \ref{periodic table}, we can for example predict that
a similar binding of extended zero modes to lattice dislocation lines ($r=1$)
can also occur in weak $d=3$-dimensional topological insulators of
``codimension" $k=1$
in symmetry classes A, D, DIII,  and C.

\subsection{Outline of the present article}

In this paper we discuss in detail the mechanism behind the dimensional
periodicity and shift property appearing in table~\ref{periodic table}.
We first demonstrate that there exists, for all five symmetry classes
of topological insulators or superconductors, 
and in all dimensions $d$,   a  representative of the Hamiltonian
in this class which has the form of a Dirac
Hamiltonian, by constructing explicitly such a representative.
We use these Dirac Hamiltonian representatives in the five
symmetry classes with topological insulators (superconductors) to obtain
relationships between these 
five symmetry classes in different dimensions.
First we
construct dimensional hierarchies (``dimensional ladders'') relating
$\mathbb{Z}$ topological 
insulators (superconductors) in different dimensions and symmetry classes
(see sections~\ref{sec: dimensional ladder: complex case}
and \ref{sec: dimensional ladder: real case}).
This is done by using a process of ``dimensional reduction'' in which
spatial dimensions are compactified (in ``Kaluza-Klein like'' fashion),
thus relating higher to lower dimensional theories.
By employing the same dimensional reduction process,
we derive in section~\ref{sec:Z2_dimreduction}
the topological classification of the $\mathbb{Z}_2$ topological
insulators (superconductors) from their higher-dimensional parent
$\mathbb{Z}$ topological insulators (superconductors)
in the same symmetry class.
Sections~\ref{sec: dimensional ladder: complex case},
\ref{sec: dimensional ladder: real case},~\ref{sec:Z2_dimreduction}
taken together provide a complete, independent derivation of table
\ref{periodic table}
of topological insulators (superconductors).

In section \ref{class AIII gen diss}
we establish a connection between 
the topological invariant (winding number) defined 
for topological insulators and superconductors with chiral symmetry
and yet another topological invariant, the Chern-Simons invariant,
for general dimensions
[equation (\ref{eq: master formula winding num and Wilson loop})].
For lower dimensional cases, the formula 
(\ref{eq: master formula winding num and Wilson loop})
relates the winding number to
the electric polarization ($d=1$ spatial dimensions),
and 
to the 
the magnetoelectric polarizability ($d=3$ spatial dimensions).

In section \ref{field_theory} we list,
for topological insulators in symmetry classes A and AIII the
topological field theory
describing the space time theory of linear responses
in all space time dimensions $D=d+1$.
For symmetry class A
 this is the Chern-Simons action, and in class
AIII it is the theta term with theta angle $\theta=\pi$.
(Generalizations to topological singlet superconductors
with $\mathrm{SU}(2)$ spin-rotation symmetry are mentioned in
section \ref{Discussion}).

Finally, as already mentioned in subsection
\ref{Review-Classification-TopIns},
we have discussed for all dimensionalities
the  ``terms of topological origin'' of Wess-Zumino-Witten  type
which can be added to  NL$\sigma$M field
theories on  symmetric spaces.
In particular, in odd dimensions, these are the Chern-Simons terms.

In appendix~\ref{subsec:RMT} we collect a number of interesting relationships
between the Cartan classification of generic Hamiltonians
due to Altland and Zirnbauer,
the NL$\sigma$M target space of Anderson localization,
and the classifying space appearing in the
K-theory approach to topological insulators.
The representation theory of the
spinor representations of the
orthogonal groups $\mathrm{SO}(N)$
is briefly reviewed in appendix~\ref{app: spinors}.
Furthermore, we study in appendix~\ref{secParity} the influence
of inversion symmetry on the classification of topological insulators 
(superconductors), 
combined with either time-reversal or  charge-conjugation (particle-hole) symmetry.

\section{Dimensional hierarchy: complex case ($\mathrm{A}\to \mathrm{AIII}$)}
\label{sec: dimensional ladder: complex case}

In this section, we consider the relationship 
between $2n$-dimensional class A
topological insulator and 
$(2n-1)$-dimensional class AIII
topological insulator. 
(Recall from table \ref{TableSymmetryClassesTwo} that Hamiltonians
in symmetry classes A and AIII are not invariant under time-reversal
and charge-conjugation symmetries 
($\mathcal{T}$ and $\mathcal{C}$), 
but Hamiltonians in class AIII
possess ``chiral symmetry'' called 
$\mathcal{S}$ in the table, 
i.e., they {\it anti-}commute with a unitary operator.)
First we present a general discussion about
class A and AIII topological insulators
(sections \ref{class A gen discuss} and \ref{class AIII gen diss}) 
and then we will focus on 
Dirac Hamiltonian representatives in these
symmetry classes 
(sections \ref{dirac insul and dim reduc}, \ref{example eins}, and \ref{example zwei}).
The insights gained from the study
of these Dirac Hamiltonian examples
are applicable to a wider class of insulators.
Finally, in section \ref{field_theory}  we discuss 
effective topological field theories describing 
topological response functions of class A and AIII topological insulators.

\subsection{Class A in $d=2n+2$ dimensions}
\label{class A gen discuss}

The class of problems that 
we will consider below 
is non-interacting fermionic systems
with translation invariance. 
For such systems, the eigenvalue problem 
at each momentum $k$ in the Brillouin zone (BZ) 
is described by
\begin{eqnarray}
\mathcal{H}(k)
|u^{\ }_{a}(k) \rangle
=
E_{a} (k)|u^{\ }_{a}(k) \rangle,
\quad
a = 1,\ldots, N_{\mathrm{tot}}. 
\end{eqnarray}
Here, $\mathcal{H}(k)$ is a $N_{\mathrm{tot}}\times N_{\mathrm{tot}}$
single-particle Hamiltonian in momentum space,
and $|u^{\ }_{a}(k) \rangle$ is the $a$-th Bloch
wavefunction with energy $E_{a} (k)$.
We assume that
there is a finite gap at the Fermi level, 
and therefore, we obtain 
a unique ground state by filling all states below
the Fermi level. 
(In this paper, we always adjust $E_a(k)$ in such a way 
that the Fermi level is at zero energy.) 

We assume there are $N_-$ ($N_+$)
occupied (unoccupied) Bloch wavefunctions 
for each $k$
with $N_{+} + N_{-} = N_{\mathrm{tot}}$.
We call the set of filled Bloch wavefunctions 
$
\{
|u^{-}_{\hat a}(k) \rangle
\}
$,
where 
hatted indices $\hat{a}=1,\ldots, N_-$ 
labels the occupied bands only.
We introduce the spectral projector 
onto the filled Bloch states
and the ``$Q$-matrix'' by
\begin{eqnarray}
P(k)
=
\sum_{\hat{a}}
|u^{-}_{\hat a}(k) \rangle
\langle u^{-}_{\hat a}(k) |, 
\qquad
Q(k) = 1 - 2P(k).
\label{P and Q}
\end{eqnarray}

The ground state (filled Fermi sea)
is characterized at each $k$ by 
the set of normalized vectors
$
\{
|u^{-}_{\hat a}(k) \rangle
\}
$,
which is a member of $\mathrm{U}(N_\mathrm{tot})$,
or, equivalently,
by $Q(k)$,
up to basis
transformations within occupied and unoccupied bands.
Thus $Q(k)$
can be viewed as an element of
the complex Grassmannian
$G_{N_+,N_++N_-}(\mathbb{C})=
\mathrm{U}(N_++N_- )/[\mathrm{U}(N_+)\times \mathrm{U}(N_-)]$. 
For a given system,
$Q(k)$ defines
a map from BZ into 
the complex Grassmannian,
and hence 
classifying topological classes of
band insulators 
is equivalent to
counting 
how many distinct classes there are
for the space of all such mappings.
The answer to this question is given by the homotopy group
$\pi_d(G_{m,m+n}(\mathbb{C}))$, which is nontrivial
in even dimensions $d=2n+2$.

Define for the occupied bands,
the non-Abelian Berry connection
\cite{Wilczek and Zee}
\begin{eqnarray}
\fl\qquad
\mathcal{A}^{\hat a \hat b}(k)
=
A^{\hat a \hat b}_{\mu}(k)\rmd k_{\mu} 
=
\langle u^{-}_{\hat a}(k) |\rmd u^{-}_{\hat b}(k) \rangle,
\quad 
\mu=1,\cdots,d, 
\quad 
\hat a, \hat b
=1,\cdots, N_-,
\end{eqnarray}
where 
$A^{\hat a \hat b}_{\mu}
=
-(A^{\hat b \hat a}_{\mu})^*$. 
The Berry curvature is defined by 
\begin{eqnarray}
\mathcal{F}^{\hat a\hat b} (k) 
=
\rmd\mathcal{A}^{\hat a\hat b}
+
\left(
\mathcal{A}^2
\right)^{\hat a\hat b}
=
\frac{1}{2} 
F^{\hat a\hat b}_{\mu\nu}(k) \rmd k_{\mu} \wedge
 \rmd k_{\nu}. 
\label{eq: suN gauge}
\end{eqnarray}
Class A insulators 
in even spatial dimensions
$d=2n+2$
($n=0,1,2,\ldots$)
can be 
characterized by the Chern form
of the Berry connection
in momentum space. 
The $(n+1)$-th Chern character is
\begin{eqnarray}
\mathrm{ch}_{n+1} 
(\mathcal{F})
=
\frac{1}{(n+1)!}
\tr
\left(
\frac{ \rmi\mathcal{F}}{2\pi}
\right)^{n+1}.
\end{eqnarray}
The integral of the Chern character in $d=2n+2$ dimensions is
an integer, 
the $(n+1)$st Chern number, 
\begin{eqnarray} 
\label{Chern_number}
\mathrm{Ch}_{n+1}[\mathcal{F}]
=
\int_{\mathrm{BZ}^{d=2n+2}}
\mathrm{ch}_{n+1} 
(\mathcal{F})
=
\int_{\mathrm{BZ}^{d=2n+2}}
\frac{1}{(n+1)!}
\tr
\left(
\frac{ \rmi\mathcal{F}}{2\pi}
\right)^{n+1}
\in 
\mathbb{Z}.
\end{eqnarray}
Here, $\int_{\mathrm{BZ}^d}$ denotes
the integration over $d$-dimensional $k$-space.
For lattice models, it can be taken as 
a Wigner-Seitz cell in the reciprocal lattice space.
On the other hand, 
for continuum models, it can be
taken as $\mathbb{R}^d$.
Assuming that the asymptotic behavior of the
Bloch wavefunctions 
approach a $k$-independent value
as $|k|\to\infty$, 
the domain of integration 
can be regarded as $S^{d}$
\footnote{
For Dirac insulators with linear dispersion, discussed below, 
this assumption is not entirely correct.
If the $k$-linear behavior of the Dirac spectrum persists to 
infinitely large momentum,
the behavior of the wavefunctions at $|k|\to \infty$ is not trivial:
the one-point compactification thus does not work.
In this case, 
the domain of integration becomes effectively a half of $S^d$, and
accordingly, integer topological numbers become half integers.
However, the Dirac spectrum can be properly regularized,
in such a way that the behavior of the wavefunctions becomes
trivial at larger $k$,
see for example equation (\ref{eq: regularized mass}).
This is also the case when the Dirac Hamiltonians are 
formulated on a lattice [see equation (\ref{tight-binding})]. 
}.

When, $d=2$ ($n=0$),
$\mathrm{Ch}_{1}[\mathcal{F}]$
is the TKNN integer \cite{Thouless82}, 
\begin{eqnarray}
\mathrm{Ch}_{1}[\mathcal{F}]
=
\frac{\rmi}{2\pi}
\int_{\mathrm{BZ}^{d=2}}
\tr
\left(
\mathcal{F}
\right)
=
\frac{\rmi}{2\pi}
\int
\rmd^2 k 
\tr
\left(
F_{1 2}
\right) ,
\end{eqnarray}
which is nothing but the
quantized Hall conductance
$\sigma_{xy}$ in units of $(e^2/h)$. 
The second Chern-number ($n=1$),
$\mathrm{Ch}_{2}[\mathcal{F}]$,
\begin{eqnarray}
\mathrm{Ch}_{2}[\mathcal{F}]
=
\frac{-1}{8\pi^2}
\int_{\mathrm{BZ}^{d=4}}
\tr
\left(
\mathcal{F}^2
\right)
=
\frac{-1}{32\pi^2}
\int \rmd^4 k
\epsilon^{\kappa\lambda \mu\nu}
\tr
\left(
F_{\kappa\lambda}
F_{\mu\nu} 
\right),
\end{eqnarray}
can be used to describe 
a topological insulator in $d=4$,
or, alternatively,
a certain adiabatic ``pumping process''
in lower spatial dimension \cite{Qi_Taylor_Zhang}.

The Chern character $\mathrm{ch}_{n+1}$ can be 
written in terms of its Chern-Simons form,
\begin{eqnarray}
\mathrm{ch}_{n+1}(\mathcal{F})
=
\rmd Q_{2(n+1)-1}(\mathcal{A},\mathcal{F}).
\end{eqnarray}
Here the Chern-Simons form is defined as
\begin{eqnarray}
Q_{2n+1}(\mathcal{A},\mathcal{F})
:=
\frac{1}{n!}
\left(
\frac{\rmi}{2 \pi}
\right)^{n+1}
\int^1_0 \rmd t\,  
\tr
\left(
\mathcal{A} \mathcal{F}^{n}_t 
\right),
\label{eq: def CS form}
\end{eqnarray}
where
\begin{eqnarray}
\mathcal{F}_t
=
t \rmd\mathcal{A} + t^2 \mathcal{A}^2 
=
t \mathcal{F} 
+ (t^2-t) \mathcal{A}^2.
\end{eqnarray}
For example, 
\begin{eqnarray}
Q_1\left(\mathcal{A},\mathcal{F}\right)
=
\frac{\rmi}{2\pi} \tr
\mathcal{A},
\quad
Q_3\left(\mathcal{A},\mathcal{F}\right)
=
\frac{ -1}{8\pi^2} 
\tr
\left(
\mathcal{A} \rmd \mathcal{A}
+
\frac{2}{3} \mathcal{A}^3
\right).
\end{eqnarray}

\subsection{Class AIII in $d=2n+1$ dimensions}
\label{class AIII gen diss}

\subsubsection{Winding number.} \label{sec: winding no}

We now discuss 
band insulators in symmetry class AIII.
By definition, for all class AIII band insulators,
we can find a unitary matrix $\Gamma$ which anticommutes
with the Hamiltonians, 
\begin{eqnarray}
\left\{
\mathcal{H}(k),  \Gamma
\right\}
= 
0,
\quad
\Gamma^2=1.
\label{chiral symmetry}
\end{eqnarray}
It follows that the spectrum 
is symmetric with respect to zero energy
and $N_{+}=N_{-}=:N$. 
As a consequence of the chiral symmetry (\ref{chiral symmetry}), 
all class AIII Hamiltonians, 
as well as their $Q$-matrix (\ref{P and Q}),
can be brought into block off-diagonal form, 
\begin{eqnarray}
Q(k)
=
\left(
\begin{array}{cc}
0  & q(k) \\
q^{\dag}(k) & 0 
\end{array}
\right),
\quad
q \in \mathrm{U}(N), 
\label{eq: off diagonal Q}
\end{eqnarray}
in the basis  in which $\Gamma$ is diagonal. 
The off-diagonal component $q(k)$ defines a map from BZ
onto U($N$), and classifying class AIII topological insulators reduces
to considering the homotopy group $\pi_d(\mathrm{U}(N))$.
The homotopy group is nontrivial in odd spatial dimensions $d=2n+1$,
while it is trivial in even spatial dimensions, i.e.,
there is no non-trivial topological insulator
in class AIII
in even spatial dimensions.

In odd spatial dimensions $d=2n+1$,
class AIII topological insulators are characterized
by the winding number \cite{SchnyderRyuFurusakiLudwig2008}%
\begin{eqnarray} 
\label{winding no}
\nu_{2n+1}[q]
: =
\int_{\mathrm{BZ}^{d=2n+1}}
\omega_{2n+1}[q],
\end{eqnarray}
where the winding number density is given by
\begin{eqnarray} \label{winding no density}
\omega_{2n+1}[q]
&:=&
\frac{(-1)^n n!}{(2n+1)!}
\left(\frac{\rmi}{2\pi}\right)^{n+1}
\tr
\left[(q^{-1} \rmd q)^{2n+1}\right]
\\  
&=&
\frac{(-1)^n  n!}{(2n+1)!}
\left(\frac{\rmi}{2\pi}\right)^{n+1}
\epsilon^{\alpha_1\alpha_2\cdots}
\mathrm{tr} \left[
 q^{-1} \partial_{\alpha_1} q \cdot
 q^{-1} \partial_{\alpha_2} q
\cdots\right]
\rmd^{2n+1} k. 
\nonumber
\end{eqnarray}

\subsubsection{Chern-Simons invariant.}

While
derived from the Chern form,
the Chern-Simons forms,
introduced 
in equation (\ref{eq: def CS form}), 
themselves define 
a characteristic class for 
an odd-dimensional manifold.
This suggests that the Chern-Simons forms can be
used to characterize also AIII topological insulators
in $d=2n+1$ dimensions.
Integrating the Chern-Simons form over the BZ, 
we introduce
\begin{eqnarray}
\label{IntegralOfChernSimonsForm}
\mathrm{CS}_{2n+1} [\mathcal{A},\mathcal{F}] 
:=
\int_{\mathrm{BZ}^{2n+1}}
Q_{2n+1}(\mathcal{A},\mathcal{F}). 
\end{eqnarray}
Unlike the winding number,
$\mathrm{CS}_{2n+1} [\mathcal{A},\mathcal{F}]$ 
is defined without assuming chiral symmetry
and can be used for non-chiral topological insulators (superconductors).

Before discussing how useful Chern-Simons forms are, 
however, 
observe that they
are not gauge invariant. 
Neither are
the integral of the Chern-Simons forms
over the BZ. 
However, 
for two different choices of gauge
$\mathcal{A}$ and $\mathcal{A}'$,
which are connected by a gauge transformation
\begin{eqnarray}
\mathcal{A}'
= g^{-1}\mathcal{A} g + g^{-1} \rmd g,
\quad
\mathcal{F}'
= g^{-1}\mathcal{F} g,
\end{eqnarray}
we have 
\begin{eqnarray}
Q_{2n+1}(\mathcal{A}', \mathcal{F}')
-
Q_{2n+1}(\mathcal{A}, \mathcal{F})
=
Q_{2n+1}(g^{-1}\rmd g, 0)
+
\rmd\alpha_{2n},
\label{eq: Cartan homotopy}
\end{eqnarray}
where $\alpha_{2n}$ is some $2n$ form \cite{Nakahara03}.
We note that
\begin{eqnarray}
Q_{2n+1}\left(g^{-1}\rmd g, 0 \right)
&=
\frac{1}{n!}
\left(
\frac{\rmi}{2 \pi}
\right)^{n+1}
\int^1_0 \rmd t \,
\tr
\left(
\mathcal{A} (t^2-t)^n \mathcal{A}^{2n} 
\right)
\nonumber \\
&=
\frac{1}{n!}
\left(
\frac{\rmi}{2 \pi}
\right)^{n+1}
\tr
\left[
(g^{-1} \rmd g)^{2n+1} 
\right]
\int^1_0 \rmd t\, 
(t^2-t)^n 
\nonumber \\
&=
\omega_{2n+1}[g]. 
\label{eq: CS form when pure gauge}
\end{eqnarray}
This is nothing but the winding number density,
and its integral
\begin{eqnarray}
\nu_{2n+1}[g]
:=
\int_{\mathrm{BZ}^{2n+1}}
Q_{2n+1}\left(g^{-1}\rmd g, 0 \right)
=
\int_{\mathrm{BZ}^{2n+1}}
\omega_{2n+1}[g]
\end{eqnarray}
is an integer,
which counts the non-trivial winding
of the map
$g(k): \mathrm{BZ}^{2n+1} \to \mathrm{U}(N)$.
Note that
$\pi_{2n+1} \left[
\mathrm{U}(N)
\right]=\mathbb{Z}$
(for large enough $N$). 
We thus conclude
\begin{eqnarray}
\mathrm{CS}_{2n+1}\left[\mathcal{A}',\mathcal{F}'\right]
-
\mathrm{CS}_{2n+1}\left[\mathcal{A},\mathcal{F}\right]
=
\mbox{integer}, 
\label{eq: CS invariant upto integer}
\end{eqnarray}
and hence the exponential 
\begin{eqnarray}
W_{2n+1}
: =
\exp\{
 2\pi \rmi
\mathrm{CS}_{2n+1}
\left[\mathcal{A},\mathcal{F}\right]
\}
\end{eqnarray}
is a well-defined, gauge invariant quantity, 
although it is not necessarily quantized.

The discussion so far has been general. In particular,
it is {\it not} restricted to {\it chiral}
topological insulators.
We now compute the Chern-Simons invariant
for class AIII topological insulators
in $d=2n+1$. 
We first explicitly write down
the Berry connection 
for chiral symmetric Hamiltonians.
To this end, we observe that 
for unoccupied and occupied bands,
the Bloch wavefucntions satisfy
\begin{eqnarray}
Q (k) |u^+_{\hat{a}}(k)\rangle
= 
+|u^+_{\hat{a}}(k)\rangle,
\qquad
Q (k) |u^-_{\hat{a}}(k)\rangle
= 
-|u^-_{\hat{a}}(k)\rangle,
\label{eigenequations}
\end{eqnarray}
respectively.
Introducing
\begin{eqnarray}
|u^\epsilon_{\hat{a}}(k)\rangle
=\frac{1}{\sqrt2}
\left(
\begin{array}{c}
\chi^\epsilon_{\hat{a}}(k)\\
\eta^\epsilon_{\hat{a}}(k)
\end{array}
\right),
\qquad
\epsilon=\pm,
\end{eqnarray}
we rewrite (\ref{eigenequations}) as
\begin{eqnarray}
\left(
\begin{array}{cc}
0  & q(k) \\
q^{\dag}(k) & 0 
\end{array}
\right)
\left(
\begin{array}{c}
\chi^{\pm}_{\hat{a}}(k)\\
\eta^{\pm}_{\hat{a}}(k)
\end{array}
\right)
=
\pm 
\left(
\begin{array}{c}
\chi^{\pm}_{\hat{a}}(k)\\
\eta^{\pm}_{\hat{a}}(k)
\end{array}
\right).
\end{eqnarray}
We can then construct a set of eigen Bloch functions
as 
\begin{eqnarray}
|u^{\epsilon}_{\hat{a}}(k)\rangle
=
\frac{1}{\sqrt{2}}
\left(
\begin{array}{c}
 \chi^{\epsilon}_{\hat{a}} \\
\epsilon q^{\dag}(k)\chi^{\epsilon}_{\hat{a}}
\end{array}
\right).
\end{eqnarray}
The $N$-dimesional space spanned by 
the occupied states 
$\{|u^{-}_{\hat{a}}(k)\rangle\}$
can be obtained by first choosing 
$N$ independent orthonormal vectors
$n^{\epsilon}_{\hat{a}}$
which are $k$-independent,
and then from $n^{\epsilon}_{\hat{a}}$, 
\begin{eqnarray}
|u^{\epsilon}_{\hat{a}}(k)\rangle_{\mathrm{N}}
=
\frac{1}{\sqrt{2}}
\left(
\begin{array}{c}
n^{\epsilon}_{\hat{a}} \\
\epsilon q^{\dag}(k) n^{\epsilon}_{\hat{a}}
\end{array}
\right).
\label{u_N}
\end{eqnarray}
From these Bloch functions,
the Berry connection is computed as
\begin{eqnarray}
{ }_{\mathrm{N}}\langle u^{\epsilon}_{\hat{a}}(k) |
\partial_{\mu} u^{\epsilon}_{\hat{b}} (k)
\rangle_{\mathrm{N}}
\rmd k_{\mu} 
&=&
\frac{1}{2}
\left[
\langle n^{\epsilon}_{\hat{a}} |
\partial_{\mu} |n^{\epsilon}_{\hat{b}} \rangle 
+
\epsilon^{2}
\langle n^{\epsilon}_{\hat{a}} q(k)|
\partial_{\mu} |q^{\dag}(k) n^{\epsilon}_{\hat{b}} \rangle 
\right]
\rmd k_{\mu} 
\nonumber \\ 
&=&
\frac{1}{2}
\left[
0
+
\epsilon^{2}
\langle n^{\epsilon}_{\hat{a}} q(k)|
\partial_{\mu} |q^{\dag}(k) n^{\epsilon}_{\hat{b}} \rangle 
\right]
\rmd k_{\mu} 
\nonumber \\ 
&=&
\frac{1}{2}
[q(k)\partial_{\mu} q^{\dag}(k) ]_{\hat{a}\hat{b}}
\rmd k_{\mu} 
=:
\mathcal{A}^{\mathrm{N}}_{\hat{a}\hat{b}},
\end{eqnarray}
where
we have made a convenient choice
$
(n^{\epsilon}_{\hat{a}})_{\hat{b}}
=
\delta_{\hat{a}\hat{b}}
$.
While this looks \emph{almost}
like a pure gauge, it is not exactly so because of
the factor $1/2$.
Note also that this calculation
shows that the Berry connection for the 
occupied and unoccupied bands are identical.
With equations (\ref{eigenequations}) through (\ref{u_N}), 
we have succeeded in
constructing eigen Bloch wavefuctions
of the  ``$Q$-matrix'', equation
(\ref{eq: off diagonal Q}), 
in block-off diagonal basis, 
that are free from any singularity. I.e., 
we have explicitly demonstrated that
there is no obstruction
to constructing eigen wavefuctions globally. 
We emphasize that this applies to  
all symmetry classes with chiral symmetry
(AIII, BDI, DIII, CII, CI)
for any spatial dimension
($d=2n$ as well as $d=2n+1$).

Alternatively, 
one can construct another set of eigen Bloch functions
\begin{eqnarray}
|u^{\epsilon}_{\hat{a}}(k)\rangle_{\mathrm{S}}
=
\frac{1}{\sqrt{2}}
\left(
\begin{array}{c}
\epsilon q(k) n^{\epsilon}_{\hat{a}} \\
n^{\epsilon}_{\hat{a}}
\end{array}
\right).
\label{u_S}
\end{eqnarray}
For these Bloch functions,
the Berry connection is computed as
\begin{eqnarray}
{ }_{\mathrm{S}} \langle u^{\epsilon}_{\hat{a}}(k) |
\partial_{\mu} u^{\epsilon}_{\hat{b}} (k)\rangle_{\mathrm{S}}
\rmd k_{\mu} 
&=&
\frac{1}{2}
\left[
\epsilon^2
\langle n^{\epsilon}_{\hat{a}} q^{\dag}(k)|
\partial_{\mu} | q(k) n^{\epsilon}_{\hat{b}} \rangle 
+
\langle n^{\epsilon}_{\hat{a}} |
\partial_{\mu} | n^{\epsilon}_{\hat{b}} \rangle 
\right]
\rmd k_{\mu} 
\nonumber \\ 
&=&
\frac{1}{2}
\left[
\epsilon^{2}
\langle n^{\epsilon}_{\hat{a}} q^{\dag}(k)|
\partial_{\mu} |q(k) n^{\epsilon}_{\hat{b}} \rangle 
+
0
\right] 
\rmd k_{\mu} 
\nonumber \\ 
&=&
\frac{1}{2}
[q^{\dag}(k)\partial_{\mu} q(k) ]_{\hat{a}\hat{b}}
\rmd k_{\mu} 
=:
\mathcal{A}^{\mathrm{S}}_{\hat{a}\hat{b}},
\end{eqnarray}
where
we have made a convenient choice
$
(n^{\epsilon}_{\hat{a}})_{\hat{b}}
=
\delta_{\hat{a}\hat{b}}
$.
The two gauges
are related to each other by
\begin{eqnarray}
\mathcal{A}^{\mathrm{S}}
=
g^{-1} \mathcal{A}^{\mathrm{N}} g
+
g^{-1} \rmd g,
\end{eqnarray}
where
the transition function $g$ is the block off-diagonal projector,
$g=q$:
\begin{eqnarray}
\fl
q^{\dag}(k) 
\left[
\frac{1}{2}q(k) \partial_{\mu} q^{\dag}(k)
\right]
q(k) 
+
q^{\dag}(k) 
\partial_{\mu} 
q(k) 
&=
\frac{1}{2} 
\left[\partial_{\mu} q^{\dag}(k)\right] q(k) 
+
q^{\dag}(k) 
\partial_{\mu} 
q(k) 
\nonumber \\ 
&=
-\frac{1}{2} 
q^{\dag}(k) \partial_{\mu} q(k) 
+
q^{\dag}(k) 
\partial_{\mu} 
q(k)
\nonumber\\
&=
\frac{1}{2} 
q^{\dag}(k) 
\partial_{\mu} 
q(k). 
\end{eqnarray}

We now 
compute the
Chern-Simons invariant
for class AIII topological insulators
in $d=2n+1$.
In the gauge where
$\mathcal{A}=\mathcal{A}^\mathrm{S}= (1/2) (q^{-1} \rmd q)$, 
we have
\begin{eqnarray}
\rmd\mathcal{A}
&=
\frac{1}{2} 
\rmd(q^{-1} \rmd q)
=
-\frac{1}{2}
(q^{-1} \rmd q) (q^{-1} \rmd q),
\nonumber \\ 
\mathcal{F}_t
&=
t\, \rmd\mathcal{A}
+
t^2 \mathcal{A}^2
=
\left(
-\frac{t}{2}
+\frac{t^2}{4}
\right)
(q^{-1} \rmd q)^2.
\end{eqnarray}
Then,
\begin{eqnarray}
Q_{2n+1}(\mathcal{A},\mathcal{F})
&=
\frac{1}{n!}
\left(
\frac{\rmi}{2 \pi}
\right)^{n+1}
\int^1_0 \rmd t\,  
\tr
\left(
\mathcal{A} \mathcal{F}^{n}_t 
\right)
\nonumber \\ 
&=
\frac{1}{n!}
\left(
\frac{\rmi}{2 \pi}
\right)^{n+1}
\int^1_0 \left(
-\frac{t}{2}
+\frac{t^2}{4}
\right)^n 
\frac{\rmd t}{2}
\tr
\left[
 (q^{-1} \rmd q)^{2n+1}
\right]
\nonumber \\
&=
\frac{1}{n!}
\left(
\frac{\rmi}{2 \pi}
\right)^{n+1}
\frac{1}{2}\int^{1}_0 
u^n(u-1)^n
 \rmd u\,  
\tr
\left[
 (q^{-1} \rmd q)^{2n+1}
\right],
\end{eqnarray}
which is a half of the winding number density, 
\begin{eqnarray}
Q_{2n+1}(\mathcal{A}^{\mathrm{S}},
\mathcal{F}^{\mathrm{S}})
=
\frac{1}{2}
\omega_{2n+1}[q].
\end{eqnarray}
Hence we conclude 
\begin{eqnarray}
\mathrm{CS}_{2n+1}
\left[\mathcal{A}^{\mathrm{S}},
\mathcal{F}^{\mathrm{S}}\right]
=
\int_{\mathrm{BZ}^{d=2n+1}}
Q_{2n+1}(\mathcal{A}^{\mathrm{S}},\mathcal{F}^{\mathrm{S}})
=
\frac{1}{2} \nu_{2n+1} [q].
\end{eqnarray}
As a corollary, 
\begin{eqnarray}
W_{2n+1}
 =
\exp\{
 2\pi \rmi \,
\mathrm{CS}_{2n+1} \left[\mathcal{A},\mathcal{F}\right]
\}
=
\exp\{ \pi  \rmi \, \nu_{2n+1}[q]\}.
\label{eq: master formula winding num and Wilson loop}
\end{eqnarray}
Thus, while in general 
the quantity $W_{2n+1}$ is not 
quantized, 
for class AIII Hamiltonians in $d=2n+1$, 
$W_{2n+1}$ can take only two values, $W_{2n+1}=\pm 1$.

Physically,
$\mathrm{CS}_{2n+1}\left[\mathcal{A},\mathcal{F}\right]$ 
in $d=1$ ($n=0$) spatial dimension
takes the form of the U(1) Wilson loop
defined for $\mathrm{BZ}^{d=1}\simeq S^1$.
It is quantized for chiral symmetric systems
\cite{Ryu02}.
Also, the logarithm of $W_1$ 
represents
the electric polarization
\cite{Kingsmith93,Resta94}.
For a lattice system,
the non-invariance of 
$\mathrm{CS}_{1}\left[\mathcal{A},\mathcal{F}\right]$
[i.e., it can change under a gauge transformation,
$\mathrm{CS}_{1}\left[\mathcal{A},\mathcal{F}\right]$
$\to$
$\mathrm{CS}_{1}\left[\mathcal{A},\mathcal{F}\right]
+ 
(\mathrm{integer})
$]
has a clear meaning:
since the system is periodic, 
the displacement of electron coordinates
has a meaning only within a unit cell,
and two electron 
coordinates
that differ by an integer multiple of the
lattice constant should be identified.

In $d=3$ ($n=1$) spatial dimensions 
$\mathrm{CS}_{3}$ 
represents
the magnetoelectric polarizability
\cite{Qi_Taylor_Zhang, Essin08,Xiao07}.
While this was discussed originally for
three-dimensional $\mathbb{Z}_2$ topological insulators
in symplectic symmetry (class AII),
the magnetoelectric polarizability is also quantized for
class AIII.

\subsection{Dirac insulators and dimensional reduction}
\label{dirac insul and dim reduc}

We start from a $d=2n+3$-dimensional \textit{gapless} Dirac Hamiltonian
defined in momentum space,
\begin{eqnarray}
\mathcal{H}^{d=2n+3}_{(2n+3)}(k)
= 
\sum_{a=1}^{d=2n+3} k_a \Gamma^a_{(2n+3)}.
\end{eqnarray}
Here,
$k_{a=1,\ldots, d}$
are $d$-dimensional momenta, 
and 
$\Gamma^{a=1,\ldots, 2n+3}_{(2n+3)}$ 
are $(2^{n+1}\times 2^{n+1})$-dimensional Hermitian matrices that satisfy
$
\{\Gamma^a_{(2n+3)},\Gamma^b_{(2n+3)}\}=2\delta_{a,b}.
$
Some results for 
the gamma matrices, on which the following discussion is based,
are summarized in appendix~\ref{app: spinors}.
The massless Dirac Hamiltonian $\mathcal{H}^{d=2n+3}_{(2n+3)}(k)$
cannot be realized on a lattice in a naive way, 
because of the fermion doubling problem.

By replacing $k_{2n+3}$ by a mass term,
we obtain a $d=2n+2$ dimensional 
class A topological Dirac insulator, 
\begin{eqnarray}
\mathcal{H}^{d=2n+2}_{(2n+3)}(k,m)
= 
\sum_{a=1}^{d=2n+2} k_a \Gamma^a_{(2n+3)}
+m \Gamma^{2n+3}_{(2n+3)}.
\label{eq: class A massive dirac}
\end{eqnarray}
The topological character of this Dirac insulator will be further
discussed below. 

By setting in addition $k_{2n+2}=0$,
we obtain an insulator in one dimension lower, 
\begin{eqnarray}
\mathcal{H}^{d=2n+1}_{(2n+3)}(k,m)
= 
\sum_{a=1}^{d=2n+1} k_a \Gamma^a_{(2n+3)}
+m \Gamma^{2n+3}_{(2n+3)}.
\label{eq: class AIII massive dirac}
\end{eqnarray}
By construction, 
$\mathcal{H}^{d=2n+1}_{(2n+3)}(k,m)$
anticommutes with $\Gamma^{2n+2}_{(2n+3)}$, 
\begin{eqnarray}
\left\{
\mathcal{H}^{d=2n+1}_{(2n+3)}(k,m),\, 
\Gamma^{2n+2}_{(2n+3)}
\right\}
=0, 
\end{eqnarray}
and hence it is a member of class AIII.

This construction of the lower-dimensional models
from their higher-dimensional ``parent''
is an example of the Kaluza-Klein dimensional reduction.
To obtain 
$\mathcal{H}^{d=2n+1}_{(2n+3)}(k,m)$
from 
$\mathcal{H}^{d=2n+2}_{(2n+3)}(k,m)$, 
one can first compactify the $(2n+2)$-th spatial direction
into a circle $S^1$.
The $(2n+2)$-th component of the momentum is 
then quantized,
$k_{2n+2}= 2\pi N_{2n+2}/ \ell$,
where $N_{2n+2}\in \mathbb{Z}$,
and
$\ell$ is the radius of $S^1$. 
The energy eigenvalues now carry the integral label $N_{2n+2}$,
in addition to the continuous label, $k_{i=1,\ldots, 2n+1}$.
By making the radius of the circle very small,
all levels with $N_{2n+2}\neq 0$ 
have very large energies while the levels with $N_{2n+2}= 0$ 
are not affected by the small radius limit.
In this way, in the limit $\ell\to 0$, there is a separation
of energy scales, and we only
keep the states with $N_{2n+2}= 0$,
while neglecting all states with $N_{2n+2}\neq 0$
(the Kaluza-Klein modes).

Similarly, 
$\mathcal{H}^{d=2n+2}_{(2n+3)}(k,m)$
is naturally obtained 
from
$\mathcal{H}^{d=2n+3}_{(2n+3)}(k)$
by dimensional reduction, 
if we also include a U(1) gauge field. 
We can introduce a (external, electromagnetic) U(1) gauge field 
$a_{i=1,\cdots,2n+3}$
by  minimal coupling,
$\mathcal{H}^{d=2n+3}_{(2n+3)}(k)
\to
\mathcal{H}^{d=2n+3}_{(2n+3)}(k+a)
$.
In the $S^1$ compactification of the $(2n+3)$-th spatial coordinate, 
we again keep modes with $k_{2n+3}=0$ only,
while neglecting all the Kaluza-Klein modes with $k_{2n+3}\neq 0$.
The vector field (gauge field) $a_{i=1,\ldots 2n+3}$ in $d=2n+3$
dimensions can be decomposed into
a vector field ($a_{i=1,\ldots 2n+2}$) 
and a scalar field 
($a_{2n+3}\equiv \varphi$)
 in $d=2n+2$, 
leading to
\begin{eqnarray}
\mathcal{H}^{d=2n+2}_{(2n+3)}(k+a,\varphi)
= 
\sum_{i=1}^{d=2n+2} (k_i+a_i) \Gamma^i_{(2n+3)}
+
\varphi \Gamma^{2n+3}_{(2n+3)}.
\label{eq: Dirac + external gauge}
\end{eqnarray}
Switching off the 
$d=2n+2$-dimensional
gauge field,
and
assuming the scalar field is constant $\varphi=m$,
we obtain $\mathcal{H}^{d=2n+2}_{(2n+3)}(k,m)$.

For the massive Dirac insulator (\ref{eq: class A massive dirac})
in class A in $2n+2$ dimensions, 
the $n$-th Chern number $\mathrm{Ch}_{n+1}$ 
is nonvanishing \cite{Golterman92}.
Accordingly, for the massive Dirac insulator
(\ref{eq: class AIII massive dirac})
in class AIII in $2n+1$ dimensions, 
the winding number and the Chern-Simons form are
non-zero.

As the Chern-Simons form
can be derived from its higher-dimensional
parent, the Chern form,
one may wonder if
there is a connection between
$\mathrm{CS}_{2n+1}=\frac{1}{2}\nu_{2n+1}$ of
a class AIII Dirac topological insulator in $d=2n+1$
and $\mathrm{Ch}_{n+1}$ of
a class A Dirac topological insulator in $d=2n+2$.
We now try to answer this question.

Let us start from 
\begin{eqnarray}
\mathcal{H}^{d=2n+2}_{(2n+3)}(k,m)
= 
\sum_{a=1}^{d=2n+2} k_a \Gamma^a_{(2n+3)}
+m(k) \Gamma^{2n+3}_{(2n+3)}.
\end{eqnarray}
Here we have regularized 
the Dirac spectrum by making the mass term $k$
dependent,
\begin{eqnarray}
m(k) = m - C k^2,
\quad 
\mbox{with}
\quad
\mathrm{sgn} (C) = \mathrm{sgn} (m).
\label{eq: regularized mass}
\end{eqnarray}
With this regularization, as $|k|\to \infty$,
$\mathcal{H}^{d=2n+2}_{(2n+3)}(k,m)
\propto \Gamma^{2n+3}_{(2n+3)}$,
and hence the wavefunctions become $k$-independent
(i.e., the BZ is topologically $S^d$).
The sign choice of the constant $C$,
$\mathrm{sgn} (C) = \mathrm{sgn} (m)$,
makes the Chern invariant for the Dirac insulator
non-zero. 

By dropping $k_{2n+2}$, 
we get the (regularized) Dirac insulator 
in one dimension lower,
\begin{eqnarray}
\mathcal{H}^{d=2n+1}_{(2n+3)}( \tilde{k},m)
= 
\sum_{a=1}^{d=2n+1} k_a \Gamma^a_{(2n+3)}
+m(\tilde{k},0) \Gamma^{2n+3}_{(2n+3)} ,
\end{eqnarray}
where $\tilde{k}=(k_1,\ldots, k_{2n+1})$. 
As mentioned before, 
$\mathcal{H}^{d=2n+1}_{(2n+3)}( \tilde{k},m)$
is chiral symmetric, 
$
\{
\mathcal{H}^{d=2n+1}_{(2n+3)}( \tilde{k},m),
\Gamma^{2n+2}_{(2n+3)}
\}
=0
$. 
We work in the basis where $\Gamma^{2n+2}_{(2n+3)}$ is diagonal.
In this basis,
$\mathcal{H}^{d=2n+1}_{(2n+3)}$ is block-off diagonal, 
\begin{eqnarray}
\mathcal{H}^{d=2n+1}_{(2n+3)}(\tilde{k},m)
=
\left(
\begin{array}{cc}
0 & \tilde{D}(\tilde{k}) \\
\tilde{D}^{\dag}(\tilde{k}) & 0
\end{array}
\right).
\label{eq: chiral H, block off diagonal}
\end{eqnarray}
Correspondingly,
\begin{eqnarray}
\mathcal{H}^{d=2n+2}_{(2n+3)}(k)
=
\left(
\begin{array}{cc}
\Delta(k_{2n+2}) & D(k) \\
D^{\dag}(k) & -\Delta(k_{2n+2})
\end{array}
\right),
\end{eqnarray}
where 
$\Delta(k_{2n+2})=k_{2n+2}$,
and 
$D(k) =  D(\tilde{k},k_{2n+2})$ satisfies
\begin{eqnarray}
D(\tilde{k},k_{2n+2}) = \tilde{D}(\tilde{k}),
\quad
\mbox{when} 
\quad 
k_{2n+2}=0. 
\end{eqnarray}

We now look for eigenfunctions
of $\mathcal{H}^{d=2n+2}_{(2n+3)}$ with
negative eigenvalues,
\begin{eqnarray}
\left(
\begin{array}{cc}
\Delta(k_{2n+2}) & D(k) \\
D^{\dag}(k) & -\Delta(k_{2n+2}) 
\end{array}
\right)
\left(
\begin{array}{c}
\chi \\
\eta
\end{array}
\right)
=
-\lambda(k)
\left(
\begin{array}{c}
\chi \\
\eta
\end{array}
\right).
\end{eqnarray}
To this end, we first solve 
the following auxiliary eigenvalue problem
\begin{eqnarray}
\left(
\begin{array}{cc}
0 & D(k) \\
D^{\dag}(k) & 0
\end{array}
\right)
\left(
\begin{array}{c}
\chi \\
\eta
\end{array}
\right)
=
-\tilde{\lambda}(k)
\left(
\begin{array}{c}
\chi \\
\eta
\end{array}
\right),
\end{eqnarray}
where $\tilde{\lambda}=\sqrt{\tilde{k}^2+[m(k)]^2}$.
We see from (\ref{u_N}) and (\ref{u_S}) that the solutions are given by 
\begin{eqnarray}
\frac{1}{\sqrt2}
\left(\begin{array}{c}
-n_{\hat a} \\
q^\dag(k) n_{\hat a}
\end{array}
\right)
\quad
\mbox{or}
\quad
\frac{1}{\sqrt2}
\left(\begin{array}{c}
q(k) n_{\hat a} \\
-n_{\hat a}
\end{array}
\right),
\label{auxiliary_solution}
\end{eqnarray}
where $q(k)$ and $q^{\dag}(k)$ are
the off-diagonal blocks of the $Q$-matrix
derived from the auxiliary Hamiltonian.
If we specialize to the case where $k=(\tilde{k},0)$, 
the projector of 
$\mathcal{H}^{d=2n+1}_{(2n+3)}(\tilde{k})$
is given in terms of
$q(\tilde{k})$ and $q^{\dag}(\tilde{k})$. 
We then construct
two different sets of
normalized eigenfunctions
of $\mathcal{H}^{d=2n+2}_{(2n+3)}$
\begin{eqnarray}
|u^{-}_{\hat a} (k)\rangle_{\mathrm{N}}
&=&
\frac{1}{
\sqrt{ 2\lambda (\lambda+ \Delta) }
}
\left(
\begin{array}{c}
-\tilde{\lambda} n_{\hat a} \\
( \lambda+\Delta) q^{\dag} n_{\hat a}
\end{array}
\right), 
\end{eqnarray}
and
\begin{eqnarray}
|u^{-}_{\hat a} (k)\rangle_{\mathrm{S}}
&=&
\frac{1}{
\sqrt{ 2 \lambda (\lambda -\Delta) }
}
\left(
\begin{array}{c}
(\Delta-\lambda)  q n_{\hat a} \\
\tilde{\lambda} n_{\hat a}
\end{array}
\right),
\end{eqnarray}
with the eigenvalue $-\lambda=-\sqrt{\tilde{\lambda}^2+\Delta^2}$.
When specialized to 
$k=(\tilde{k},0)$, 
these wavefunctions 
yield the eigenfunctions (\ref{auxiliary_solution}) of 
the chiral Dirac Hamiltonian $\mathcal{H}^{d=2n+1}_{(2n+3)}$.

Since the Hamiltonian
$\mathcal{H}^{d=2n+2}(k)$
is characterized by the non-zero Chern number,
$\mathrm{Ch}_{n+1}\left[\mathcal{F}\right]\neq 0$,
it is not possible to define wavefunctions globally.
We thus need to split the BZ ($\mathrm{BZ}^{d=2n+2}$) into two parts, 
$\mathrm{BZ}^{d=2n+2}_{\mathrm{N}}$ 
and
$\mathrm{BZ}^{d=2n+2}_{\mathrm{S}}$.
We choose the interface of these two patches
as 
$
\partial\mathrm{BZ}^{d=2n+2}_{\mathrm{N}}
=
-\partial\mathrm{BZ}^{d=2n+2}_{\mathrm{S}}
=\mathrm{BZ}^{d=2n+1}
$,
on which a lower dimensional Hamiltonian ``lives". 
For each patch,
$\mathrm{BZ}^{d=2n+2}_{\mathrm{N},\mathrm{S}}$,
we can choose a global gauge.
Observe that the wavefunction
$|u^{-}_{\hat a} (k)\rangle_\mathrm{N}$
is well-defined for $k_{2n+2}>0$,
and has a singularity
at $(\tilde{k},k_{2n+2})=(0,-\sqrt{m/C})$.
Similarly,
$|u^{-}_{\hat a} (k)\rangle_\mathrm{S}$
is well-defined for $k_{2n+2}<0$
and singular
at $(\tilde{k},k_{2n+2})=(0,\sqrt{m/C})$.

\begin{figure}[tb]
\begin{center}
    \includegraphics[height=.26\textwidth]{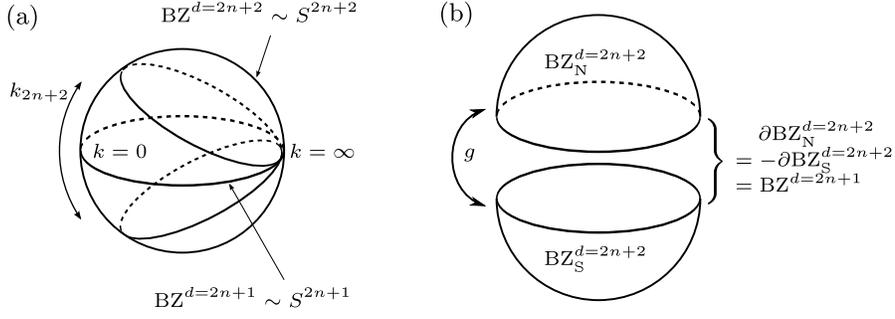}
  \caption{
(a) 
The $d=2n+2$ dimensional Brillouin zone, 
$\mathrm{BZ}^{d=2n+2}\sim S^{2n+2}= S^{2n+1}\wedge S^1$, 
and
its $d=2n+1$ dimensional descendant 
$\mathrm{BZ}^{d=2n+1}\sim S^{2n+1}$
located at the equator of $S^{2n+2}$. 
(b) Splitting the $d=2n+2$ dimensional Brillouin zone in half,
or embedding $d=2n+1$ dimensional Brillouin zone
into a higher-dimensional one.
}
\label{sphere}
\end{center}
\end{figure}

The two gauges are thus complementary 
and
at the boundary $\partial \mathrm{BZ}^{d=2n+2}_{\mathrm{N}}
=-\partial \mathrm{BZ}^{d=2n+2}_{\mathrm{S}}$
they are glued by the transition function
\begin{eqnarray}
\mathcal{A}^{\mathrm{S}} 
&=& g^{-1} 
\mathcal{A}^{\mathrm{N}} g + g^{-1} \rmd g,
\end{eqnarray}
where $g(\tilde{k})\in \mathrm{U}(N_-)$ with 
$\tilde{k} \in \partial \mathrm{BZ}^{d=2n+2}_{\mathrm{N}}$.
Then, 
\begin{eqnarray}
\mathrm{Ch}_{n+1}\left[\mathcal{F}\right]
&=&
\int_{\mathrm{BZ}^{d=2n+2}}
\mathrm{ch}_{n+1} 
(\mathcal{F})
=
\int_{\mathrm{BZ}^{d=2n+2}}
\rmd Q_{2n+1}(\mathcal{A},\mathcal{F})
\nonumber \\
&=&
\int_{\mathrm{BZ}_{\mathrm{N}}^{d=2n+2}}
\rmd Q_{2n+1}(\mathcal{A}^{\mathrm{N}},\mathcal{F})
+
\int_{\mathrm{BZ}_{\mathrm{S}}^{d=2n+2}}
\rmd Q_{2n+1}(\mathcal{A}^{\mathrm{S}},\mathcal{F})
\nonumber \\
&=&
\int_{\partial \mathrm{BZ}_{\mathrm{N}}^{d=2n+2}}
Q_{2n+1}(\mathcal{A}^{\mathrm{N}},\mathcal{F})
-
\int_{\partial \mathrm{BZ}_{\mathrm{N}}^{d=2n+2}}
Q_{2n+1}(\mathcal{A}^{\mathrm{S}},\mathcal{F}),
\end{eqnarray}
where we have used Stokes theorem. 
From formula (\ref{eq: Cartan homotopy}),
it follows that 
\begin{eqnarray}
\mathrm{Ch}_{n+1}\left[\mathcal{F}\right]
&=&
\int_{\partial \mathrm{BZ}^{d=2n+1}_{\mathrm{N}}}
Q_{2n+1}\left(g^{-1}\rmd g, 0 \right). 
\end{eqnarray}
By use of equation (\ref{eq: CS form when pure gauge})
we conclude that the RHS of the above equation
is nothing but the winding number. 
Thus, the Chern number can be expressed 
as a winding number of the transition function
$g$.
For the case we are interested in, i.e., when
a $d=2n+1$ topological insulator
``lives'' on 
$\partial \mathrm{BZ}^{d=2n+1}_{\mathrm{N}}$,
the transition function
is given by the off-diagonal block of the projector,
$g(\tilde{k})=q(\tilde{k})$.
This is how 
two topological insulators 
in $d=2n+2$
and $d=2n+1$ are
related.

\subsection{Example: $d=3\to 2\to 1$}
\label{example eins}

Consider three mutually anticommuting,
Hermitian matrices
\begin{eqnarray}
\Gamma^{a=1,2,3}_{(3)}=
\{
\sigma_{x},
\,
\sigma_{y},
\,
\sigma_{z}
\},
\end{eqnarray}
where $\sigma_{x,y,z}$ are
the $2\times 2$ Pauli matrices. 
These matrices can be used to construct
a $d=3$ gapless (Weyl) chiral fermion
\begin{eqnarray}
\mathcal{H}^{d=3}_{(3)}(k)
=
\sum_{a=1}^3
k_a \Gamma^a_{(3)}. 
\end{eqnarray}
By replacing the third component of
the momemtum by a mass term,  
$k_{a=3}\to m$,
we get a $d=2$ dimensional 
Dirac Hamiltonian, 
\begin{eqnarray}
\mathcal{H}^{d=2}_{(3)}(k,m)
&=
\sum_{a=1}^2
k_a \Gamma^a_{(3)}
+
m \Gamma^3_{(3)}
\nonumber \\
&=
k_x \sigma_x 
+
k_y \sigma_y
+
m \sigma_z.
\label{QHE Dirac Hamiltonian}
\end{eqnarray}
This is a class A Hamiltonian.
The Bloch wavefunctions are given by
[$\lambda(k) := \sqrt{ k^2 + m^2 }$]
\begin{eqnarray}
|u^+ (k) \rangle
&=
\frac{1}{\sqrt{2 \lambda \left(\lambda - m\right) }}
\left(
\begin{array}{c}
k_x - \rmi k_y  \\
\lambda - m
\end{array}
\right),
\nonumber \\
|u^- (k) \rangle
&=
\frac{1}{\sqrt{2 \lambda \left(\lambda + m\right) }}
\left(
\begin{array}{c}
-k_x + \rmi k_y  \\
\lambda + m \\
\end{array}
\right).
\label{eq: wfns}
\end{eqnarray}
We note that $|u^- (k) \rangle$
is well-defined for
all $k$ when $m>0$.
(When $m<0$, by choosing a gauge properly,
we can obtain a similar well-defined wavefunction.)
Assuming $m>0$,
we adopt $|u^- (k) \rangle$,
for which we obtain
the Berry connection
\begin{eqnarray}
A_x (k,m) =
+\frac{\mathrm{i} k_y }{2 \lambda \left(\lambda +m\right)},
\qquad
A_y (k,m) =
-\frac{ \mathrm{i} k_x }{2 \lambda \left(\lambda +m\right)},
\end{eqnarray}
and the Berry curvature
\begin{eqnarray}
F_{xy}\left(k,m \right) =
\partial_{k_x} A_y - \partial_{k_y} A_x
= -\frac{\rmi m }{ 2 \lambda^3 } .
\end{eqnarray}
The Chern number 
is non-zero, 
\begin{eqnarray}
\mathrm{Ch}_{1}
[\mathcal{F}]
=
\frac{\mathrm{i}}{2 \pi } \int \rmd^2 k F_{xy}
=
\frac{\rmi}{2 \pi} \int \rmd^2 k 
\frac{-\rmi m }{ 2 \lambda^3 } 
=
\frac{1}{2}
\frac{m}{|m|}, 
\end{eqnarray}
which is nothing but 
the Hall conductance $\sigma_{xy}$.
This model can be realized, 
at low energies,
in the honeycomb lattice model
introduced by Haldane \cite{Haldane88}.

Finally, setting $k_2=0$,
we get a $d=1$ dimensional Dirac Hamiltonian, 
\begin{eqnarray}
\mathcal{H}^{d=1}_{(3)}(k,m)
&=&
k_x \sigma_x 
+
m \sigma_z.
\label{eq: 1d AIII massive Dirac}
\end{eqnarray}
This is nothing but the chiral 
topological Dirac insulator in class AIII,
as it anticommutes with $\sigma_y$.
A topological invariant can be defined 
for the $d=1$ dimensional Dirac Hamiltonian
as follows.
We first go to the canonical form
by 
$
(\sigma_x, \sigma_y, \sigma_z)
\to
(\sigma_x, \sigma_z, -\sigma_y)
$,
$
\mathcal{H}^{d=1}_{(3)}(k,m)
\to 
k_x \sigma_x 
-
m \sigma_y.
$
This $d=1$ dimensional Hamiltonian 
describes the physics of polyacetylene
(see, for example, 
Refs.\ \cite{Su-Schrieffer, Stone}.)
The Bloch wavefunction with negative eigenvalue is
\begin{eqnarray}
|u^- (k_x) \rangle
&=&
\frac{1}{\sqrt{2 (k^2_x+m^2)  }}
\left(
\begin{array}{c}
-k_x - \rmi m  \\
\lambda  \\
\end{array}
\right).
\label{WF in d=1}
\end{eqnarray}
The off-diagonal block of the projection 
operator $Q(k)$ in this basis is 
\begin{eqnarray}
q(k) &=&
-\frac{k_x + \rmi m}{\sqrt{k^2_x + m^2}},
\end{eqnarray}
and the winding number is given by
\begin{eqnarray}
\nu_1 [q]
=
\frac{\rmi}{2\pi}
\int_{\mathrm{BZ}}
q^{-1} \rmd q
=
\frac{\rmi}{2\pi}
\int^{\infty}_{-\infty}  
\rmd k_x \,
\frac{ -\rmi m}{ k^2_x + m^2}
=
\frac{1}{2} \frac{m}{|m|}. 
\label{eq: winding number nu1 for AIII dirac}
\end{eqnarray}
From equation (\ref{WF in d=1}) the Berry connection is obtained,
\begin{eqnarray}
\mathcal{A}(k_x) 
=
\langle u^-(k_x)|\rmd u^-(k_x)\rangle 
=
\frac{1}{2}
\frac{ -\rmi m }{ k^2_x + m^2}
\rmd k_x.
\end{eqnarray}
Then, we find
\begin{eqnarray}
\mathrm{CS}_1\left[\mathcal{A},\mathcal{F}\right]
=
\frac{\rmi}{2\pi}
\int_{\mathrm{BZ}} \mathrm{tr}\,  \mathcal{A} 
=
\frac{\rmi}{2\pi}
\int^{\infty}_{-\infty} \rmd k_x
\frac{1}{2}
\frac{ -\rmi m }{ k^2_x + m^2}
=
\frac{1}{4}
\frac{m}{|m|}.
\end{eqnarray}
This is a half of the winding number
$\nu_1[q]$
as expected. 
Hence, the Wilson ``loop'' is given by 
\begin{eqnarray}
W_1 &=&
\exp \left\{
2\pi \mathrm{i} \mathrm{CS}_1\left[\mathcal{A},\mathcal{F}\right]
\right\}
=
\exp \int_{\mathrm{BZ}} \mathrm{tr}\,  \mathcal{A} 
= 
e^{ \pm \pi \rmi /2 }.
\end{eqnarray}

Here, we mention, however, that there is a subtlety in computing
the Wilson loop.
In the basis
where the Hamiltonian in momentum space is real, 
$\mathcal{H}^{d=1}_{(3)}(k,m)
=
k_x \sigma_x 
+
m \sigma_z$
(\ref{eq: 1d AIII massive Dirac}), 
the Bloch wavefunction is real
for a given $k$, and as a consequence the Berry connection 
$\mathcal{A}(k_x)$ vanishes identically.
One would then conclude $W_1=1$, not $W_1=-1$,
although $\nu_1= \frac{1}{2} \mathrm{sgn}(m)$. 
This puzzle can be solved by properly regularizing 
the Dirac insulator. 
For simplicity, let us 
replace $k_x$ by $\sin k_x$,
and $m$ by $m-1+\cos(k_x)$.
With such a regularization, the BZ is topologically $S^1$, 
and one finds a singularity in the wavefunction
at $k_x=\pi$, where the phase of the wavefunction 
jumps by $\pi$
(i.e., there is a Dirac string at $k_x=\pi$.).
If, on the other hand, we use a different basis (\ref{WF in d=1}),
we can avoid to have 
a Dirac string in the BZ, and the Wilson loop can 
be obtained by integrating the Berry connection over the BZ.

\subsection{Example: $d=5\to 4\to 3$}
\label{example zwei}

\label{subsec: A-AIII examples in d=5,4,3}

Consider five mutually anticommuting,
Hermitian matrices
\begin{eqnarray}
\Gamma^{a=1,\ldots,5}_{(5)}=
\{
\alpha_{x},
\,
\alpha_{y},
\,
\alpha_{z},
\,
\beta,
\, 
-\rmi \beta \gamma^5
\},
\end{eqnarray}
where we are using the Dirac representation, 
\begin{eqnarray}
\alpha_i
=\left(\begin{array}{cc}
0 & \sigma_i\\
\sigma_i & 0
\end{array}\right),
\qquad
\beta=\left(\begin{array}{cc}
1 & 0\\
0 & -1\end{array}\right),
\qquad
\gamma^{5}=\left(\begin{array}{cc}
0 & 1\\
1 & 0\end{array}
\right).
\end{eqnarray}
Observe that
$
\alpha_x
\alpha_y
\alpha_z
\beta 
= -\rmi \beta\gamma_5
$.
These matrices can be used to construct
a ($d=5$)-dimensional gapless chiral fermion
\begin{eqnarray}
\mathcal{H}^{d=5}_{(5)}(k)
=
\sum_{a=1}^5
k_a \Gamma^a_{(5)}. 
\end{eqnarray}
By replacing the fifth component of
the momemtum by a mass term,  
$k_{a=5}\to m_5$,
we obtain a $d=4$ dimensional 
Dirac Hamiltonian, 
\begin{eqnarray}
\mathcal{H}^{d=4}_{(5)}(k,m_5)
=
\sum_{a=1}^4
k_a \Gamma^a_{(5)}
+
m_5 \Gamma^5_{(5)}.
\end{eqnarray}
It is known that for this gapped Hamiltonian,
the second Chern number is non-zero
\cite{Golterman92,Qi_Taylor_Zhang}.

\begin{figure}
\begin{center}
\includegraphics[width=0.9\textwidth]{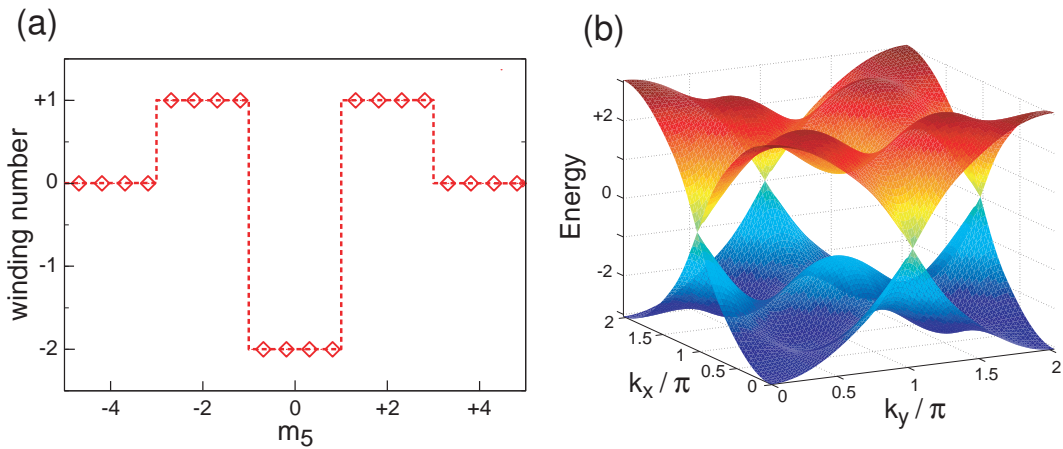}
\caption{      \label{surface_modes_AIII}
(a)
Winding number $\nu_3$ for Hamiltonian \re{matrixAIII} as a function of $m_5$.
(b)
Two-dimensional energy spectrum of the surface states of model \re{matrixAIII} with mass $m_5=+0.5$.
There are two inequivalent surface modes in agreement with the
winding number $\nu_3 (m_5 =+0.5) = -2$.
}
\end{center}
\end{figure}

Finally, setting $k_4=0$,
we get a $d=3$ dimensional Dirac Hamiltonian, 
\begin{eqnarray}
\mathcal{H}^{d=3}_{(5)}(k,m_5)
&=&
\sum_{a=1}^3
k_a \Gamma^a_{(5)}
+
m_5 \Gamma^5_{(5)}
=
\sum_{a=1}^3
k_a \alpha_a
-
m_5 \rmi \beta \gamma^5 .
\end{eqnarray}
This is nothing but 
the chiral topological Dirac insulator in class AIII
discussed in Ref.\ \cite{SchnyderRyuFurusakiLudwig2008}.
The tight-binding version of this model for a simple cubic lattice is given by
\begin{eqnarray} \label{matrixAIII}
\mathcal{H}^{d=3}_{(5)}(k,m_5)
&=&
\sum_{a=1}^3 \sin k_a  \Gamma^a_{(5)} + 
\left( m_5 + \sum_{a=1}^3 \cos k_a \right)  \Gamma^5_{(5)}.
\label{tight-binding}
\end{eqnarray}
By taking open boundary conditions in the $z$ direction and periodic
boundary conditions in the $x$ and $y$ directions, we can study the
surface states of this model.
When the winding number $\nu_3$ is non-vanishing,
there are $| \nu_3 |$ surface Dirac states,
which cross the bulk band gap (see figure~\ref{surface_modes_AIII}).

We now study the Berry connection 
of the above 
class A ($d=4$) and 
class AIII ($d=3$) Dirac insulators.
To treat these two cases
in a unified fashion, 
let us consider the following $d=4$ Dirac Hamiltonian
\begin{eqnarray}
\mathcal{H}(k,m_0) 
&=&
\sum_{a=1}^3 k_a \alpha_a
+
m_0 \beta 
+
k_w (- \rmi \beta \gamma^5).
\end{eqnarray}
Upon identification 
$k_w \to m_5$
it can be also viewed as
a $d=3$ dimensional Dirac Hamiltonian with two masses.
The four eigenvalues are 
\begin{eqnarray}
E(k) 
&=&
\pm \lambda(k),
\quad
\lambda(k)
=
\sqrt{k^2 + m^2_0}.
\end{eqnarray}
The two normalized negative energy 
eigenstates with $E(k)=-\lambda(k)$ are
\begin{eqnarray}
|u^{-}_1(k)\rangle
&=&
\frac{1}{
\sqrt{
2\lambda  (\lambda + m_0)
}
}
\left(
\begin{array}{c}
-k_x + \rmi k_y \\
\rmi k_w + k_z \\
0 \\
\lambda+ m_0
\end{array}
\right),
\nonumber \\
|u^{-}_2(k)\rangle
&=&
\frac{1}{
\sqrt{
2\lambda  (\lambda + m_0)
}
}
\left(
\begin{array}{c}
-k_z + \rmi k_w \\
-k_x - \rmi k_y \\
\lambda+m_0 \\
0 \\
\end{array}
\right).
\label{eq: wfn 4x4 occupied}
\end{eqnarray}
The two normalized positive energy 
eigenstates with $E(k)=+\lambda(k)$ are
\begin{eqnarray}
|u^{+}_1(k)\rangle
&=&
\frac{1}{
\sqrt{
2\lambda  (\lambda - m_0)
}
}
\left(
\begin{array}{c}
k_x - \rmi k_y \\
- \rmi k_w -k_z \\
0 \\
\lambda-m_0
\end{array}
\right)
,
\nonumber \\
|u^{+}_2(k)\rangle
&=&
\frac{1}{
\sqrt{
2\lambda  (\lambda - m_0)
}
}
\left(
\begin{array}{c}
-\rmi k_w+k_z \\
k_x + \rmi k_y \\
\lambda-m_0 \\
0 
\end{array}
\right).
\label{eq: wfn 4x4 unoccupied}
\end{eqnarray}
Note that if $m_0>0$, 
$|u^{+}_{1,2}(k)\rangle$ are not well-defined
at $\lambda(k) = m_0$ (i.e, $k=0$).
Observe that when $m_0>0$,
$|u^-_{1,2}(k)\rangle$ have the form of wavefunctions
expected from the general consideration 
(corresponding to the $\mathcal{A}^\mathrm{N}$ gauge choice).

When $m_0=0$,
the Hamiltonian anticommutes with $\beta$
and is block off-diagonal. 
The off-diagonal component 
of the $Q$-matrix is given by 
\begin{eqnarray}
q(k) = 
\frac{-1}{\lambda}\left(
k\cdot \sigma -\rmi m_5
\right)
\end{eqnarray}
By noting $(\mu=1,2,3)$
\begin{eqnarray}
\partial_{\mu}q 
=
\frac{k_{\mu}}{\lambda^3}
\left(k\cdot \sigma -\rmi m_5 \right)
-
\frac{1}{\lambda}\sigma_{\mu},
\quad 
q^{\dag}\partial_{\mu}q 
=
\frac{-1}{\lambda^2}
\left(
k_{\mu}-k\cdot \sigma \sigma_{\mu} -\rmi m \sigma_{\mu}
\right), 
\end{eqnarray}
and also,
\begin{eqnarray}
&
\int \rmd^3 k\, 
\epsilon^{\mu\nu\rho}
\tr\left[
q^{\dag}\partial_{\mu}q
q^{\dag}\partial_{\nu}q
q^{\dag}\partial_{\rho}q
\right]
\nonumber \\
&=
\int \rmd^3 k\, 
\frac{-\epsilon^{\mu\nu\rho}}{\lambda^6}
\tr
\left[
-3 \rmi m_5 (k\cdot \sigma) \sigma_{\mu} (k\cdot \sigma)
   \sigma_{\nu} \sigma_{\rho}
+\rmi m^3_5 \sigma_{\mu} \sigma_{\nu} \sigma_{\rho}
\right]
\nonumber \\
&=
(-3 \rmi m_5)
\epsilon_{\mu\nu\rho}
\frac{2 \rmi}{3}
\int \rmd^3 k\,
\frac{\epsilon^{\mu\nu\rho} k^2}{(k^2+m^2_5)^3}
+
\rmi m^3_5 
2\rmi\epsilon_{\mu\nu\rho}
\int \rmd^3 k\,
\frac{-\epsilon^{\mu\nu\rho}}{(k^2+m^2_5)^3}
\nonumber \\
&=
12m_5
\int \rmd^3k
\frac{1}{(k^2+m^2_5)^2}, 
\end{eqnarray}
the winding number is given by
\begin{eqnarray}
\nu_3[q] 
=
\frac{12}{24 \pi^2}
m_5
\int^{\infty}_0 \rmd k 
\frac{4\pi k^2}{(k^2+m^2_5)^2}
=
\frac{1}{2\pi^2}
4\pi
\times 
\frac{\pi}{4}
\frac{m_5}{|m_5|} 
=
\frac{1}{2}
\frac{m_5}{|m_5|}.
\end{eqnarray}

We now compute the
Chern-Simons invariant
(magnetoelectric polarizability)
for the massive Dirac Hamiltonian. 
For the lower two occupied bands, we can introduce a U(2) gauge field by 
$
A_{\mu}^{\hat{a}\hat{b}}(k)
\rmd k_{\mu} =
\langle u_{\hat{a}}^-(k)|\rmd u_{\hat{b}}^-(k)\rangle
$
($\hat{a},\hat{b}=1,2$).
The U(2) gauge field is given by,
in the matrix notation, 
\begin{eqnarray}
A_x 
&=&
\frac{\rmi}{2\lambda (\lambda +m_0 )}
\left(
\begin{array}{cc}
+k_y & k_w +\rmi k_z\\
k_w -\rmi k_z & -k_y \\
\end{array}
\right),
\nonumber \\
A_y
&=&
\frac{\rmi}{2\lambda (\lambda +m_0 )}
\left(
\begin{array}{cc}
-k_x & \rmi k_w - k_z\\
-\rmi k_w - k_z & +k_x \\
\end{array}
\right),
\nonumber \\
A_z 
&=&
\frac{\rmi}{2\lambda (\lambda +m_0 )}
\left(
\begin{array}{cc}
-k_w &  -\rmi k_x + k_y \\
\rmi k_x + k_y & +k_w \\
\end{array}
\right).
\end{eqnarray}
The gauge field can be decomposed into U(1) ($a^{0}$) and SU(2) ($a^{j=x,y,z}$)
parts as 
\begin{eqnarray}
A_{\mu}^{\ }(k)=
a_{\mu}^{0}(k)\frac{\sigma_{0}}{2}+a_{\mu}^{j}(k)\frac{\sigma_{j}}{2}.
\label{eq: decomposition of U(2) into U(1)xSU(2)}
\end{eqnarray}
The U(1) part of the Berry connection is trivial,
whereas the SU(2) part is given by 
\begin{eqnarray}
&&
a^{x}_{x} = 
\rmi\frac{k_w}{\lambda(\lambda+ m_0)},
\qquad
a^{y}_{x} =
\rmi\frac{-k_z}{\lambda(\lambda+ m_0)},
\qquad
a^{z}_{x} =
\rmi\frac{+k_y}{\lambda(\lambda+ m_0)},
\nonumber \\
&&
a^{x}_{y} = 
\rmi\frac{-k_z}{\lambda(\lambda+ m_0)},
\qquad
a^{y}_{y} = 
\rmi\frac{-k_w}{\lambda(\lambda+ m_0)},
\qquad
a^{z}_{y} =
\rmi\frac{-k_x}{\lambda(\lambda+ m_0)},
\nonumber \\
&&
a^{x}_{z} = 
\rmi\frac{+k_y}{\lambda(\lambda+ m_0)},
\qquad
a^{y}_{z} = 
\rmi\frac{+k_x}{\lambda(\lambda+ m_0)},
\qquad
a^{z}_{z} = 
\rmi\frac{-k_w}{\lambda(\lambda+ m_0)}.
\end{eqnarray} 
The Chern-Simons form can be computed as 
\begin{eqnarray}
\mathrm{CS}_3
&=&
\frac{-1}{8\pi^2}
\int \rmd^3 k
\epsilon^{\mu\nu\rho}
\mathrm{tr}\,\left(
A_{\mu} \partial_{\nu} A_{\rho}
+
\frac{2}{3}
A_{\mu} A_{\nu} A_{\rho}
\right)
\nonumber \\
&=&
\frac{1}{8\pi^2}
\int \rmd^3 k
\frac{m_5 
(2\lambda + m_0 )
}{
\lambda^3 (\lambda + m_0 )^2
}
\nonumber \\
\!\!&=&\!\! 
\frac{1}{\pi}
\frac{m_5}{|m_5|}
\mathrm{arctan}
\left[
\frac{
(m^2_5 + m^2_0)^{1/2} - m_0}
{
(m^2_5 + m^2_0)^{1/2} + m_0
}
\right]^{1/2}. 
\end{eqnarray}
When $m_0=0$,
\begin{eqnarray}
\mathrm{CS}_3
=
\frac{1}{4}\frac{m_{5}}{|m_{5}|}.
\end{eqnarray}
As the CS term is determined only modulo 1, 
the fractional part
$\frac{1}{4}\mathrm{sgn}(m_5)$ is an intrinsic property.
On a lattice, as there are two-copies of the $4\times4$ Dirac Hamiltonians,
the total Chern-Simons invariant is given by 
$\mathrm{CS}_3=\frac{1}{2}\mathrm{sgn}(m_{5})$,
and $e^{2\pi\rmi\mathrm{CS}_3}=-1$.

There is a subtlety in computing $\mathrm{CS}_3$ and $W_3$,
which is similar to the case of the Wilson loop $W_1$.
When $m_0\neq 0$ and $m_5=0$,
the Chern-Simons form is zero identically,
and one would conclude $W_3=1$, 
while 
chiral symmetry for the case of $m_0\neq 0$ and $m_5=0$
allows us to define $\nu_3$,
and 
$\nu_3= \frac{1}{2} \mathrm{sgn}(m)$.
As before, 
this puzzle can be solved by properly regularizing 
the Dirac insulator.
With such a regularization, the BZ is topologically $S^3$, 
and one finds a singularity in the wavefunctions. 
If, on the other hand, we use a different basis,
we can avoid to have a Dirac string in the BZ. 

\subsection{Description in terms of field theory for the linear responses in
$D=d+1$ space-time dimensions}
\label{field_theory}

For class A and AIII topological insulators or superconductors,
there is always 
a conserved U(1) quantity;
either electric charge or one component (the $z$-component, say)
of spin is conserved \cite{FosterLudwigAIII-Interactions}. 
Transport properties of these conserved quantities in these 
topological insulators or superconductors
in $d$ spatial dimensions
can be described in terms of 
an effective field theory for the linear responses
in $D=d+1$ space-time dimensions,
which can be obtained by
coupling an external
space-time dependent gauge field $a_\mu$ to
 the  topological insulator or superconductor (we will
choose the Dirac representative;
see the  discussion near (\ref{eq: Dirac + external gauge})),
and by integrating out
the gapped fermions. 
One can then read off the 
theory of the linear responses
from the so-obtained (classical) 
effective action for $a_\mu$.

The theory for the linear responses of a topological insulator
in symmetry class A and in $d$ spatial dimensions is then
given by the Chern-Simons action in $D=d+1$ space-time dimensions,
\begin{eqnarray}
S_{\mathrm{CS}}^{D=2n+1} [a]
=
2\pi \rmi 
k
\int_{\mathbb{R}^{D=2n+1}} Q_{2n+1}[a],
\quad
k = \mathrm{Ch}_{n}.
\label{eq: CS action}
\end{eqnarray}
Here, 
$Q[a]$ is the Chern-Simons form of the U(1) gauge field $a_{\mu}$
defined in $D=2n+1$ dimensional (Euclidean) spacetime,
and the level $k$ of the Chern-Simons term is given in terms of 
$\mathrm{Ch}_{n}$, 
which is the Chern invariant 
computed from
the Bloch wavefunctions
in $d=2n$ space dimensions.

On the other hand,
the theory of linear responses of a topological insulator in symmetry class
AIII in $d=2n-1$ spatial dimensions is given by a theta term in
$D=2n$ space-time dimensions,
\begin{eqnarray}
S_{\theta }^{D=2n}  [a]
=
\rmi \theta \int_{\mathbb{R}^{D=2n}} 
\mathrm{ch}_n[f],
\quad
\theta = \pi \nu_{2n-1},
\label{eq: theta term}
\end{eqnarray}
where $f$ is the field strength of the external gauge field $a_{\mu}$, 
and the theta angle $\theta$ is related to the winding number 
$\nu_{2n-1}$, which can be computed from the Bloch wavefunctions. 
The integration of fermions in $D=2n$ dimensions is trickier
than in  $D=2n+1$ dimensions;
while a simple derivative expansion of the fermion determinant 
would appear to yield
the Chern-Simons action (\ref{eq: CS action}), 
a more careful evaluation yields in fact
the theta term, which is obtained from the Fujikawa Jacobian
\cite{Pavan2008}.

The theta angle in the theta term (\ref{eq: theta term})
can take on a priori any value,
if there is no discrete symmetry
that pins its value. 
In class AII $\mathbb{Z}_2$ topological insulators
in $D=4n$ dimensions, 
time-reversal symmetry 
demands $\theta$ to be an integer multiple of
$\pi$.
Similarly, 
in class AIII, chiral symmetry pins 
$\theta$ to be an integer multiple of
$\pi$.
This can be seen from the symmetry 
properties of the Chern-Simons and theta terms, 
\begin{eqnarray}
\mathcal{T}: S_{\mathrm{CS}}^{D=2n+1}   \to (-1)^n 
S_{\mathrm{CS}}^{D=2n+1}, 
\quad
S_{\theta }^{D=2n}   \to (-1)^{n+1} S_{\theta }^{D=2n} ,
\nonumber \\
\mathcal{C}: 
S_{\mathrm{CS}}^{D=2n+1}   \to (-1)^{n+1} 
S_{\mathrm{CS}}^{D=2n+1},
\quad
S_{\theta }^{D=2n}   \to (-1)^n S_{\theta }^{D=2n} .
\end{eqnarray} 
Here, the 
(anti-unitary) time-reversal symmetry operation
$\mathcal{T}$ may either square to plus or minus the identity.
Similarly, the charge-conjugation symmetry operation
$\mathcal{C}$ may either square to plus or minus the identity.
Since the chiral symmetry $\mathcal{S}$ can be expressed as the product of
these two symmetry operations, i.e.,
$\mathcal{S}=\mathcal{T} \cdot \mathcal{C}$, one obtains
\begin{eqnarray}
\mathcal{S}:
S_{\theta }^{D=2n}   \to - S_{\theta }^{D=2n}.
\end{eqnarray} 
Together with the $2\pi$ periodicity of $\theta$,
it follows that
$\theta$ is either $0$ or $\pi$.

\section{Dimensional 
hierarchy for
$\mathbb{Z}$ topological insulators:
 real case}
\label{sec: dimensional ladder: real case}

\subsection{Dimensional reduction}

We now consider eight ``real'' symmetry classes. 
In particular,
we consider topological 
insulators (superconductors) 
with integer classification,
$\mathbb{Z}$ or $2\mathbb{Z}$.
($\mathbb{Z}_2$ cases will be discussed 
in section \ref{sec:Z2_dimreduction}.) 
A connection similar to 
the one between classes $\mathrm{A}$ 
and $\mathrm{AIII}$
can be established.

The massive Dirac Hamiltonians relevant to this section
are all obtained from the following 
massless Dirac Hamiltonian
\begin{eqnarray}
(0):\quad &&
\mathcal{H}^{d=2n+3}_{(2n+3)}(k)
= 
\sum_{a=1}^{d=2n+3} k_a \Gamma^a_{(2n+3)},
\label{eq: top Hamiltonian 0}
\end{eqnarray}
by replacing some momenta by a mass,
or by simply dropping them. 
Specifically, we will consider 
\begin{eqnarray}
(i):\quad &&
\mathcal{H}^{d=2n+2}_{(2n+3)}(k,m)
= 
\sum_{a=1}^{d=2n+2} k_a \Gamma^a_{(2n+3)}
+ 
m \Gamma^{2n+3}_{(2n+3)}, 
\label{eq: Dirac Hamiltonian I}
\\
(ii):\quad &&
\mathcal{H}^{d=2n+1}_{(2n+3)}(k,m)
=
\sum_{a=1}^{d=2n+1}  k_a \Gamma^{a}_{(2n+3)}
+
m \Gamma^{2n+3}_{(2n+3)}, 
\label{eq: Dirac Hamiltonian II}
 \\
(iii):\quad &&
\mathcal{H}^{d=2n}_{(2n+3)}(k,m)
=
\sum_{a=1}^{d=2n} k_a \Gamma^a_{(2n+3)}
+
 m 
\mathcal{M}, 
\label{eq: Dirac Hamiltonian III}
 \\
(iv):\quad &&
\mathcal{H}^{d=2n-1}_{(2n+3)}(k,m)
= 
\sum_{a=1}^{d=2n-1} k_a \Gamma^a_{(2n+3)}
+
 m \mathcal{M}, 
\label{eq: Dirac Hamiltonian IV}
\end{eqnarray}
where
\begin{eqnarray}
\mathcal{M}
:=
\mathrm{i}
\Gamma^{2n+3}_{(2n+3)}
\Gamma^{2n+2}_{(2n+3)}
\Gamma^{2n+1}_{(2n+3)}. 
\end{eqnarray}
The massive Dirac Hamiltonian
$\mathcal{H}^{d=2n+2}_{(2n+3)}(k,m)$
is obtained from its gapless parent
$\mathcal{H}^{d=2n+3}_{(2n+3)}(k)$
by replacing $k_{2n+3}$ with a mass $m$.
By further removing $k_{2n+2}$, 
we obtain
$\mathcal{H}^{d=2n+1}_{(2n+3)}(k,m)$.
This is essentially the same procedure as
in the previous section, where we
obtained the
class AIII topological Dirac Hamiltonian
from its parent class A Hamiltonian. 
On the other hand, 
$\mathcal{H}^{d=2n}_{(2n+3)}(k,m)$
can be obtained from
$\mathcal{H}^{d=2n+3}_{(2n+3)}(k)$
by first removing three components of
the momentum $k_{2n+3,2n+2, 2n+1}$,
and then adding $\mathcal{M}$ as a mass term. 
Finally,
$\mathcal{H}^{d=2n-1}_{(2n+3)}(k,m)$
is obtained from 
$\mathcal{H}^{d=2n}_{(2n+3)}(k,m)$
by dimensional reduction. 

\begin{table}[t!] 
\begin{center}
\begin{tabular}{cccccccccccccc}\hline
AZ$\backslash d$ & 0  & 1 & 2 & 3 & 4 & 5 & 6 & 7 & 8 & 9 & 10 & 11 & $\cdots$ \\ \hline\hline
AI  & $\textcolor{red}{i}$ &  &  &  & $\textcolor{red}{iii}$ &  & & $\textcolor{red}{0}$  & $\textcolor{red}{i}$ &  &  & & $\cdots$ \\ \hline
BDI &  & $\textcolor{blue}{ii}$ &  &  &  & $\textcolor{blue}{iv}$ &  & &  & $\textcolor{blue}{ii}$ &  & & $\cdots$ \\ \hline
D   &  & \textcolor{red}{0} & $\textcolor{red}{i}$ &  &  & &  $\textcolor{red}{iii}$ &  &  & \textcolor{red}{0} & $\textcolor{red}{i}$ & & $\cdots$\\ \hline
DIII&  &  &  & $\textcolor{blue}{ii}$ & &  &   & $\textcolor{blue}{iv}$ &  &   &  & $\textcolor{blue}{ii}$ &   $\cdots$ \\ \hline
AII & $\textcolor{red}{iii}$ &  &  & \textcolor{red}{0} & $\textcolor{red}{i}$ &     &  & & $\textcolor{red}{iii}$  &  &  & \textcolor{red}{0} & $\cdots$ \\ \hline
CII &  & $\textcolor{blue}{iv}$&  &  & & $\textcolor{blue}{ii}$ &   &  &  & $\textcolor{blue}{iv}$  &  &  & $\cdots$\\ \hline
C   &  &  & $\textcolor{red}{iii}$ &  &  & \textcolor{red}{0} & $\textcolor{red}{i}$ & &   &  & $\textcolor{red}{iii}$  &  & $\cdots$ \\ \hline
CI  &  &  &  & $\textcolor{blue}{iv}$ &  &  &   & $\textcolor{blue}{ii}$ &  &   &  & $\textcolor{blue}{iv}$  &  $\cdots$ \\ \hline
\end{tabular}
\caption{
Symmetry class of 
Dirac Hamiltonians 
$\mathcal{H}^{d=2n+3}_{(2n+3)}(k)$ (\ref{eq: top Hamiltonian 0}),
$\mathcal{H}^{d=2n+2}_{(2n+3)}(k,m)$ (\ref{eq: Dirac Hamiltonian I}),
$\mathcal{H}^{d=2n+1}_{(2n+3)}(k,m)$ (\ref{eq: Dirac Hamiltonian II}),
$\mathcal{H}^{d=2n}_{(2n+3)}(k,m)$ (\ref{eq: Dirac Hamiltonian III})
and,
$\mathcal{H}^{d=2n-1}_{(2n+3)}(k,m)$ (\ref{eq: Dirac Hamiltonian IV}),
marked by
``0'',
``$i$'',
``$ii$'',
``$iii$'',
and
``$iv$'',
respectively. 
Chiral and non-chiral symmetric classes are colored
in blue and red, respectively. 
\label{eq: symmetires of real Dirac}
}
\end{center}
\end{table}

Here,  an important difference from the dimensional reduction in
the complex case is the shifting of symmetry classes. 
The parent Hamiltonian $\mathcal{H}^{d=2n+3}_{(2n+3)}(k)$
has one (and only one) of four discrete symmetries,
$\mathcal{T}$ (squaring to plus or minus the identity),
or
$\mathcal{C}$ (squaring to plus or minus the identity),
depending on the dimensionality $d$
(see table \ref{eq: symmetires of real Dirac} and appendix~\ref{app: spinors}).
That is, it is a member of any one of AI, D, AII, and C. 

\subsubsection{From $d=2n+3$ to $d=2n+2$.}
When dimensionally reducing
$\mathcal{H}^{d=2n+3}_{(2n+3)}(k)$
to
$\mathcal{H}^{d=2n+2}_{(2n+3)}(k,m)$,
the symmetry of the parent Hamiltonian
$
\mathcal{H}^{d=2n+3}_{(2n+3)}(k)
$
is broken by the mass term: 
While the sign of the kinetic term is reversed under the discrete symmetry,
the sign of the mass term remains unchanged.
Thus, the new massive Hamiltonian is a member of a different symmetry class. 
By this procedure,
we get an even dimensional 
Hamiltonian $\mathcal{H}^{d=2n+2}_{(2n+3)}(k,m)$ with the
``shifted'' symmetry,
\begin{eqnarray}
\mathrm{AI}\to \mathrm{C},
\quad
\mathrm{C}\to \mathrm{AII},
\quad
\mathrm{AII}\to \mathrm{D},
\quad
\mathrm{D}\to \mathrm{AI}. 
\end{eqnarray}
This can be easily proved by taking an explict form
of gamma matrices
(see appendix~\ref{app: spinors}).
Observe that, 
while we have chosen to remove
$k_{2n+3}$ and hence $\Gamma^{2n+3}_{(2n+3)}$,
any other component could have been removed
instead of $k_{2n+3}$ and $\Gamma^{2n+3}_{(2n+3)}$.
However, all these different choices are unitarily equivalent,
as they are simply related by a permutation of Clifford 
generators, $\Gamma^{a}_{(2n+3)}$.

\subsubsection{From $d=2n+2$ to $d=2n+1$.}
Further dimensionally reducing 
$\mathcal{H}^{d=2n+2}_{(2n+3)}(k,m)$
to
$\mathcal{H}^{d=2n+1}_{(2n+3)}(k,m)$,
we obtain a chiral symmetric Dirac Hamiltonian.
Together with existing discrete symmetry,
this chiral symmetry yields one more discrete symmetry.
One can see the symmetry of the newly generated Hamiltonian 
is shifted, 
\begin{eqnarray}
\mathrm{D}\to \mathrm{BDI},
\quad
\mathrm{AII} \to \mathrm{DIII},
\quad 
\mathrm{C}\to \mathrm{CII},
\quad
\mathrm{AI} \to \mathrm{CI}.
\end{eqnarray}
This again can be easily proved by taking an explict form
of gamma matrices (see appendix~\ref{app: spinors}).

\subsubsection{From $d=2n+3$ to $d=2n$.}
Let us now consider
$
\mathcal{H}^{d=2n}_{(2n+3)}(k,m)
$.
One can check that this Hamiltonian 
belongs to the same symmetry class 
as its parent 
$
\mathcal{H}^{d=2n+3}_{(2n+3)}(k)
$.
To study the topological properties of 
$
\mathcal{H}^{d=2n}_{(2n+3)}(k,m)
$,
observe that,
by construction,
$\mathcal{H}^{d=2n}_{(2n+3)}(k,m)$ commutes with
a product made of any two of $\Gamma^{2n+1,2n+2, 2n+3}_{(2n+3)}$, 
\begin{eqnarray}
\left[
\mathcal{H}^{d=2n}_{(2n+3)}(k,m), 
\Gamma^{a}_{(2n+3)}
\Gamma^{b}_{(2n+3)}
\right]
=0
\end{eqnarray}
where
$
(a,b)
=
(2n+1,2n+2)$,
$(2n+2,2n+3)$,
or
$(2n+3,2n+1)$.
This suggests that
the Hamiltonian can be made block-diagonal.
We note that in the 
representation of the gamma matrices
we are using,
the Hamiltonian is block diagonal, 
\begin{eqnarray}
\mathcal{H}^{d=2n}_{(2n+3)}(k,m)
&=& 
\sum_{a=1}^{d=2n} k_a 
\Gamma^a_{(2n+1)}
\otimes \sigma_3
-
m \Gamma^{2n+1}_{(2n+1)} \otimes \sigma_0
\nonumber \\
&=&
\left(
\begin{array}{cc}
\mathcal{H}^{d=2n}_{(2n+1)}(k,-m)
& 0 \\
0 &
-\mathcal{H}^{d=2n}_{(2n+1)}(k,m)
\\
\end{array}
\right)
\end{eqnarray}
By use of
a unitary transformation
for the lower right block, 
\begin{eqnarray}
\Gamma^{2n+1}_{(2n+1)}
\left[
-\mathcal{H}^{d=2n}_{(2n+1)}(k,m)
\right]
\Gamma^{2n+1}_{(2n+1)}
&=&
\mathcal{H}^{d=2n}_{(2n+1)}(k,-m),
\end{eqnarray}
$\mathcal{H}^{d=2n}_{(2n+3)}(k,m)$
reduces to two copies of
$\mathcal{H}^{d=2n}_{(2n+1)}(k,-m)$. 
On the other hand, 
we have already seen that 
each
$
\mathcal{H}^{d=2n}_{(2n+1)}(k,-m)
=
\sum_{a=1}^{d=2n} k_a 
\Gamma^a_{(2n+1)}
-
m\Gamma^{2n+1}_{(2n+1)} 
$
has a non-trivial Chern number $\pm 1$. 
We thus conclude that the
Chern number
characterizing $\mathcal{H}^{d=2n}_{(2n+3)}(k)$
is $\pm 2$.

\subsubsection{From $d=2n$ to $d=2n-1$.}
Finally,
we discuss the topological insulators (superconductors)
$\mathcal{H}^{d=2n-1}_{(2n+3)}(k,m)$
marked by ``$iv$'' in 
table \ref{eq: symmetires of real Dirac}.
One can check that the symmetry class of this Hamiltonian
is shifted
from that of
$\mathcal{H}^{d=2n}_{(2n+3)}(k,m)$ as
\begin{eqnarray}
\mathrm{AI}\to \mathrm{CI},
\quad
\mathrm{C} \to \mathrm{CII},
\quad 
\mathrm{AII}\to \mathrm{DIII},
\quad
\mathrm{D} \to \mathrm{BDI}.
\end{eqnarray}
Just as
$\mathcal{H}^{d=2n}_{(2n+3)}(k,m)$
is characterized by a non-zero Chern invariant which is even, 
the descendant 
$\mathcal{H}^{d=2n-1}_{(2n+3)}(k,m)$
is also characterized by an even winding number.

The symmetry properties of all the Dirac Hamiltonians 
discussed here 
are summarized 
in table \ref{eq: symmetires of real Dirac}.

\subsection{Example: $d=3\to 2\to 1$}

Consider a $d=3$-dimensional gapless Dirac fermion
\begin{eqnarray}
\mathcal{H}(k)
= 
k_x \sigma_x 
+
k_y \sigma_y
+
k_z \sigma_z.
\end{eqnarray}
This Hamiltonian is a member of class AII,
since
\begin{eqnarray}
\sigma_y \mathcal{H}(k)\sigma_y
= 
\mathcal{H}^*(-k).
\end{eqnarray}
Starting from this Hamiltonian,
we dimensionally reduce this Hamiltonian
successively and 
obtain the lower dimensional Hamiltonians 
with shifted symmetry classes,
$\mathrm{AII}\to \mathrm{D} \to \mathrm{BDI}$. 
We replace $k_z$ by a mass term,
and consider a $d=2$-dimensional massive Dirac fermion
\begin{eqnarray}
\mathcal{H}(k,m)
= 
k_x \sigma_x 
+
k_y \sigma_y
+
m \sigma_z.
\label{class D Dirac Hamiltonian for d=2}
\end{eqnarray}
This Hamiltonian is now a member of class D since
\begin{eqnarray}
\sigma_x \mathcal{H}^*(k,m)\sigma_x
=
-\mathcal{H}(-k,m).
\label{class D Dirac Hamiltonian for d=2 symmetry}
\end{eqnarray}
As mentioned before,
the Chern number is non-zero for this model,
$
\mathrm{Ch}_2= \frac{1}{2} \mathrm{sgn}(m)
$.
This Hamiltonian is in the same universality class 
as the chiral $p+\rmi p$-wave topological superconductor
\cite{Read00,SenthilFisher2000}.

The single-particle Hamiltonian (\ref{class D Dirac Hamiltonian for d=2}),
of course, looks identical to (\ref{QHE Dirac Hamiltonian}).
This degeneracy is lifted once we consider more generic types
of perturbations than the mass term, 
which can in principle be spatially inhomogeneous. 
Also, because of the $\mathcal{C}$ symmetry
[see equation (\ref{class D Dirac Hamiltonian for d=2 symmetry})],
the Hamiltonian (\ref{class D Dirac Hamiltonian for d=2})
can be viewed as a single-particle Hamiltonian for real (Majorana)
fermions, 
while this is not the case for a class AIII 
single-particle Hamiltonian.
In other words, 
a class D system can be written in the second quantized form as 
\begin{eqnarray}
H =
\frac{1}{2} 
\sum_{k\in \mathrm{BZ}^2 }
\left(
a^{\dag}_k,
a^{\ }_{-k}
\right)
\mathcal{H}(k,m)
\left(
\begin{array}{c}
a^{\ }_{k} \\
a^{\dag}_{-k}
\end{array} 
\right). 
\end{eqnarray}
where $a^{\dag}_k/a^{\ }_k$ is a fermion creation/annihilation 
operator with momentum $k$.

By further switching off $k_y$, or 
by the dimensional reduction in the $y$-direction, 
\begin{eqnarray}
\mathcal{H}(k,m)
= 
k_x \sigma_x 
+
m \sigma_z.
\end{eqnarray}
This Hamiltonian is a member of class BDI since 
\begin{eqnarray}
\sigma_x \mathcal{H}^*(k,m)\sigma_x
=
-\mathcal{H}(-k,m),
\quad
\sigma_z \mathcal{H}^*(k,m)\sigma_z
=
\mathcal{H}(-k,m).
\end{eqnarray}
This model can be realized as 
a continuum limit of the 
lattice Majorana fermion model 
discussed in Ref.\ \cite{FidkowskiKitaev2009}. 
For this $d=1$ topological massive Dirac Hamiltonian, 
the winding number and the Wilson loop can be computed 
in the same way as we did for 
the $d=1$ class AIII topological Dirac insulators
(\ref{eq: 1d AIII massive Dirac}).
From (\ref{eq: winding number nu1 for AIII dirac}), 
we immediately see $\nu_1= \frac{1}{2} \mathrm{sgn}(m)$.

\subsection{Example: $d=5\to 2$}

Consider $d=2\cdot 1 + 3$ dimensions.
We can take the following five matrices as
gamma matrices, 
\begin{eqnarray} \label{mfivegam}
\Gamma^{a=1,\ldots,5}_{(5)}=
\left\{
\tau_y \otimes \sigma_{x,y,z},
\quad
\tau_x \otimes \sigma_{0},
\quad
\tau_z \otimes \sigma_{0}
\right\}
\end{eqnarray}
Note that
\begin{eqnarray}
\sigma_y 
\left(
\Gamma^{a}_{(5)}
\right)^T  \sigma_y 
=
+\Gamma^{a}_{(5)}.
\end{eqnarray}
Thus, the gapless Hamiltonian 
\begin{eqnarray}
\mathcal{H}(k)
&=&
\sum_{a=1}^5 k_a \Gamma^{a}_{(5)}
\end{eqnarray}
is a member of class C, as seen from 
\begin{eqnarray}
\sigma_y k_a 
\left(
\Gamma^{a}_{(5)}
\right)^T  \sigma_y 
=
-(-k_a) \Gamma^{a}_{(5)}.
\end{eqnarray}
We now remove three momenta, and add one mass term.
There are four possible mass terms that are 
compatible with the class C symmetry
\begin{eqnarray} \label{possMterm}
\tau_{x,z,0}\otimes \sigma_y,
\qquad
\tau_{y}\otimes \sigma_0.
\end{eqnarray}
To construct a kinetic term we pick two 
gamma matrices from the list~\re{mfivegam}.
For any choice of two gamma matrices there is only one mass term in
equation \re{possMterm} that
anticommutes with the chosen kinetic term.
In this way, we obtain, for example, 
the following massive Hamiltonian 
\begin{eqnarray}
\mathcal{H}(k,m)
=
\tau_y \otimes \sigma_x k_x
+
\tau_y \otimes \sigma_z k_z
+
m \tau_0 \otimes \sigma_y .
\end{eqnarray}
Observe that
$\rmi (\tau_y\otimes\sigma_y)(\tau_x\otimes\sigma_0)(\tau_z\otimes\sigma_0)
=\tau_0\otimes\sigma_y$.
It is easy to see that
this Hamiltonian has the doubled Chern number, 
$\mathrm{Ch}_1=\pm 2$. 
To see this,
first rotate $\tau_y\to \tau_z$,
\begin{eqnarray}
\mathcal{H}(k,m)
&\to &
\tau_z \otimes \sigma_x k_x
+
\tau_z \otimes \sigma_z k_z
+
m \tau_0 \otimes \sigma_y 
\nonumber \\
&=&
\left(
\begin{array}{cc}
\sigma_x k_x+ \sigma_z k_z+ \sigma_y m &  0 \\
0 & -\sigma_x k_x- \sigma_z k_z + \sigma_y m \\
\end{array}
\right),
\end{eqnarray}
and then apply a unitary transformation, 
\begin{eqnarray}
&&
\left(
\begin{array}{cc}
\sigma_0 &  0 \\
0 & \sigma_y 
\end{array}
\right)
\left(
\begin{array}{cc}
\sigma_x k_x+ \sigma_z k_z+ \sigma_y m &  0 \\
0 & -\sigma_x k_x- \sigma_z k_z + \sigma_y m \\
\end{array}
\right)
\left(
\begin{array}{cc}
\sigma_0 &  0 \\
0 & \sigma_y 
\end{array}
\right)
\nonumber \\
&=&
\left(
\begin{array}{cc}
\sigma_x k_x+ \sigma_z k_z+ \sigma_y m &  0 \\
0 & \sigma_x k_x + \sigma_z k_z + \sigma_y m \\
\end{array}
\right).
\end{eqnarray}

The $d=2$ dimensional Hamiltonian constructed here
has the same topological properties as 
the $d=2$ dimensional $d+\rmi d$-wave superconductor
discussed in Ref.\ \cite{SenthilMarstonFisher}.

\section{$\mathbb{Z}_2$ topological insulators and dimensional reduction}
\label{sec:Z2_dimreduction}

In the previous sections we have studied 
$\mathbb{Z}$ topological insulators (superconductors),
which are characterized by an integer -- the Chern or winding number.
In this section we discuss $\mathbb{Z}_2$ topological insulators (superconductors)
and show how they can be derived as lower dimensional descendants 
of parent $\mathbb{Z}$ topological insulators (superconductors)
(see table~\ref{table: Z2 topins dimreduct}).
Recently, Qi \textit{et al.}~\cite{Qi_Taylor_Zhang} have employed
this procedure to derive descendant $\mathbb{Z}_2$ topological 
insulators (superconductors) for those symmetry classes that
break chiral 
symmetry.
Here, we show how this approach can be generalized and applied to all eight
symmetry classes of table~\ref{table: Z2 topins dimreduct}.
We first treat chiral symmetric systems, and then review the case of
chiral symmetry breaking classes.
The reasoning follows closely the one of Ref.~\cite{Qi_Taylor_Zhang}.

\begin{table}[h]
\begin{center}
\begin{tabular}{cccccccccccccc}\hline
AZ$\backslash d$ & 0  & 1 & 2 & 3 & 4 & 5 & 6 & 7 & 8 & 9 & 10 & 11 & $\cdots$ \\ \hline\hline
AI  & \textcolor{red}{$ \mathbb{Z}  $} & 0 & 0 
    & 0 & \textcolor{red}{$2 \mathbb{Z}$} & 0 
    & \textcolor{red}{$\mathbb{Z}_2$} & \textcolor{red}{$\mathbb{Z}_2$} & \textcolor{red}{$\mathbb{Z}$}
    & 0 & 0 & 0 & $\cdots$ \\ \hline
{\bf BDI } &  \textcolor{blue}{$\mathbb{Z}_2$} &  \textcolor{blue}{$\mathbb{Z}$} & 0 & 0 
    & 0 & \textcolor{blue}{$2 \mathbb{Z}$} & 0 
    &  \textcolor{blue}{$\mathbb{Z}_2$} &  \textcolor{blue}{$\mathbb{Z}_2$} &  \textcolor{blue}{$\mathbb{Z}$}
    & 0 & 0 &  $\cdots$ \\ \hline
D   & \textcolor{red}{$\mathbb{Z}_2$} & \textcolor{red}{$\mathbb{Z}_2$} & \textcolor{red}{$\mathbb{Z}$}
    & 0 & 0 & 0 
    & \textcolor{red}{$2 \mathbb{Z}$}  & 0 & \textcolor{red}{$\mathbb{Z}_2$} 
    & \textcolor{red}{$\mathbb{Z}_2$} & \textcolor{red}{$\mathbb{Z}$}  & 0 & $\cdots$ \\ \hline
{\bf DIII}& 0 &  \textcolor{blue}{$\mathbb{Z}_2$} &  \textcolor{blue}{$\mathbb{Z}_2$} &  \textcolor{blue}{$\mathbb{Z}$}
    & 0 & 0 & 0 
    & \textcolor{blue}{$2 \mathbb{Z}$}  & 0 &  \textcolor{blue}{$\mathbb{Z}_2$}
    &  \textcolor{blue}{$\mathbb{Z}_2$} &  \textcolor{blue}{$\mathbb{Z}$}  &  $\cdots$ \\ \hline
AII & \textcolor{red}{$2 \mathbb{Z}$}  & 0 & \textcolor{red}{$\mathbb{Z}_2$} 
    & \textcolor{red}{$\mathbb{Z}_2$} & \textcolor{red}{$\mathbb{Z}$} & 0 
    & 0 & 0 & \textcolor{red}{$2 \mathbb{Z}$} 
    & 0 & \textcolor{red}{$\mathbb{Z}_2$} & \textcolor{red}{$\mathbb{Z}_2$} &  $\cdots$\\ \hline
{\bf CII } & 0 & \textcolor{blue}{$2 \mathbb{Z}$}  & 0 &  \textcolor{blue}{$\mathbb{Z}_2$}
    &  \textcolor{blue}{$\mathbb{Z}_2$} &  \textcolor{blue}{$\mathbb{Z}$} & 0 
    & 0 & 0 & \textcolor{blue}{$2 \mathbb{Z}$}
    & 0 &  \textcolor{blue}{$\mathbb{Z}_2$} &  $\cdots$\\ \hline
C   & 0  & 0 & \textcolor{red}{$2 \mathbb{Z}$}
    & 0 &  \textcolor{red}{$\mathbb{Z}_2$}  & \textcolor{red}{$\mathbb{Z}_2$}
    & \textcolor{red}{$\mathbb{Z}$} & 0 & 0 
    & 0 & \textcolor{red}{$2 \mathbb{Z}$} & 0 & $\cdots$ \\ \hline
{\bf CI}  & 0 & 0  & 0 & \textcolor{blue}{$2 \mathbb{Z}$}  
    & 0 &  \textcolor{blue}{$\mathbb{Z}_2$}  &  \textcolor{blue}{$\mathbb{Z}_2$} 
    & \textcolor{blue}{$\mathbb{Z}$} & 0 & 0 
    & 0 & \textcolor{blue}{$2 \mathbb{Z}$}  & $\cdots$ \\ \hline
\end{tabular}
\end{center}
\caption{
All $\mathbb{Z}_2$ topological insulators (superconductors) can be 
described through dimensional reduction as lower dimensional descendants
of parent $\mathbb{Z}$ topological insulators (superconductor)
in the same symmetry class,
which are characterized by non-trivial winding/Chern numbers.
Symmetry classes in which chiral (``sublattice'') symmetry is present (S=1)
are colored blue (the Cartan label marked in bold-face), whereas those
where it is absent (S=0) are colored red. 
\label{table: Z2 topins dimreduct}
}
\end{table}

\subsection{Topological insulators with chiral symmetry}
\label{sec: dimensional reduction in chiral classes}

Four among the eight symmetry classes of
table~\ref{table: Z2 topins dimreduct}
(see also table~\ref{TableSymmetryClassesTwo}) 
are invariant under chiral (``sublattice'') symmetry,
which is a combination of particle-hole and time-reversal symmetry.
These four symmetry classes are called BDI, CI, CII, and DIII.
It is possible to characterize the topological properties of
chiral symmetric $\mathbb{Z}$ topological insulators by 
a winding number (topological invariant), which is defined in terms of the block-off diagonal projector
(see section \ref{sec: winding no}).
From this topological invariant we can derive a $\mathbb{Z}_2$ classification
by imposing
the constraint of additional discrete symmetries (such as $\mathcal{C}$ and $\mathcal{T}$) 
for the lower-dimensional descendants.
For a $d$-dimensional $\mathbb{Z}$ topological insulator (superconductor) in a given symmetry class,
we distinguish between first and second descendants,
whose (reduced) dimensionality is $d-1$ and $d-2$, respectively.

\subsubsection{Symmetry properties of winding number.} 
\label{sec: symmetry properties of winding no}

In order to better understand the role played by $\mathcal{T}$ and $\mathcal{C}$,
we first study the transformation properties of the winding
number \re{winding no} and 
the winding number density \re{winding no density}
under these discrete symmetries.
First of all, we note that the winding number density is purely real, 
$ w^{\ast}_{2n+1} [q]   = w_{2n+1} [q] $,
which can be checked by direct calculation.  
Without any additional discrete symmetries, i.e., for symmetry class AIII,
the off-diagonal projector $q$, and hence the winding number density (\ref{winding no density}) are 
not subject to additional constraints.
However, for the symmetry classes
BDI, DIII, CII, and CI,  the presence of time reversal and particle-hole symmetries
relates the projector $q$
at wavevector $k$ to the one at wavevector $-k$.
As a consequence, the configurations of the winding number density $w$
in momentum space are restricted by time-reversal and particle-hole symmetries.
Let us now derive these symmetry constraints on the winding number density $w_{2n+1}[q]$
for any odd spatial dimension $d=2n+1$.

\paragraph{Classes DIII and CI:}
First we consider symmetry classes DIII and CI, where
the block off-diagonal projector satisfies~\cite{SchnyderRyuFurusakiLudwig2008}
\begin{eqnarray} \label{sym prop DIII and CI}
q^T(-k) = \epsilon q(k),
\quad
\epsilon=
\left\{
\begin{array}{cc}
-1, & \mathrm{DIII} \\
+1, & \mathrm{CI}.
\end{array}
\right.
\end{eqnarray}
We introduce the auxiliary functions
\begin{eqnarray} \label{eq: auxiliary functions}
v_{\mu}(k)
:=
\left(
q^{\dag} \frac{\partial}{\partial k_{\mu}} q
\right)(k)
\quad \textrm{and} \quad
\tilde{v}_{\mu}(k)
:=
\left(
q \frac{\partial}{\partial k_{\mu}} q^{\dag}
\right)(k).
\end{eqnarray}
From the symmetry property \re{sym prop DIII and CI} it follows that
$v^{\ }_{\mu}(k)= -\tilde{v}^*_{\mu}(-k)$,
irrespective of the value taken by $\epsilon$.
Noting that  $(\rmd q^{\dag})q = - q^{\dag}\rmd q$  we find
\begin{eqnarray}
&
\epsilon^{\alpha_1\alpha_2\cdots \alpha_{2n+1}}
\mathrm{tr}\,\left[
v_{\alpha_1}(k)
v_{\alpha_2}(k)
\cdots
v_{\alpha_{2n+1}}(k)
\right]
\nonumber \\
&\qquad =
\epsilon^{\alpha_1\alpha_2\cdots \alpha_{2n+1}}
\mathrm{tr}\,\left[
v_{\alpha_1}(-k)
v_{\alpha_2}(-k)
\cdots
v_{\alpha_{2n+1}}(-k)
\right]^*.
\end{eqnarray}
Thus,  the winding number density for symmetry class
DIII and CI   is subject to the constraint
\begin{eqnarray}
w_{2n+1} \left[ q(k) \right]
=
(-1)^{n+1} w_{2n+1}^*\left[ q (-k) \right] .
\end{eqnarray}
Consequently, 
the winding number $\nu_{4n+1} [q]$ is vanishing
in $d=4n+1$ dimensions,
i.e., there exists no non-trivial topological state
characterized by an integer winding number
in $(4n+1)$-dimensional systems belonging to symmetry class DIII or CI
(cf.\ table~\ref{table: Z2 topins dimreduct}).
Conversely, in $d=(4n+3)$ dimensions both in class DIII and CI
there are topologically non-trivial states
({\it ii} or {\it iv} in table \ref{eq: symmetires of real Dirac}),
and one would naively expect that for each of these states
there are lower dimensional descendants characterized by
$\mathbb{Z}_2$ topological invariants.
However, this is not the case, as we will explain below.

\paragraph{Classes BDI and CII:}
Next we consider symmetry classes BDI and CII, where
the projector satisfies~\cite{SchnyderRyuFurusakiLudwig2008}
\begin{eqnarray} \label{sym prop BDI and CII}
Gq^*(-k)G^{-1} = q(k),
\quad
G=
\left\{
\begin{array}{cc}
\sigma_0, & \mathrm{BDI} \\
\rmi\sigma_y, & \mathrm{CII} .
\end{array}
\right.
\end{eqnarray}
Introducing the auxiliary function $v^{\ }_{\mu}(k)$,
equation \re{eq: auxiliary functions}, as before, we find that
$
v^{\ }_{\mu}(k)
=
-G v^*_{\mu}(-k)G^{-1},
$
for both class BDI and CII.
From the cyclic property of the trace it follows that
\begin{eqnarray}
&
\epsilon^{\alpha_1\alpha_2\cdots \alpha_{2n+1}}
\mathrm{tr}\,\left[
v_{\alpha_1}(k)
v_{\alpha_2}(k)
\cdots
v_{\alpha_{2n+1}}(k)
\right]
\nonumber \\
&\qquad =
\epsilon^{\alpha_1\alpha_2\cdots \alpha_{2n+1}}
(-1)^{2n+1}
\mathrm{tr}\,\left[
v_{\alpha_1}(-k)
v_{\alpha_2}(-k)
\cdots
v_{\alpha_{2n+1}}(-k)
\right]^*.
\end{eqnarray}
Hence, the winding number density in class BDI and CII  is constrained by
\begin{eqnarray}
w_{2n+1} \left[ q (k) \right]
=
(-1)^{n}
w_{2n+1}^*\left[ q (-k) \right].
\end{eqnarray}
As a result, the winding number $\nu_{4n-1} [q]$ is vanishing
in $d=4n-1$ dimensions, i.e., there are no non-trivial
$(4n-1)$-dimensional $\mathbb{Z}$ topological insulators
belonging to symmetry class BDI or CII
(see table~\ref{table: Z2 topins dimreduct}).
On the other hand, in $(4n+1)$ dimensions there exist $\mathbb{Z}$
topological states in both class BDI and CII
({\it ii} or {\it iv} in table \ref{eq: symmetires of real Dirac}),
and one would expect, as before, that for each of these states
there are lower-dimensional topologically non-trivial descendants.
Again, this expectation turns out to be incorrect, as we will explain below.

\subsubsection{$\mathbb{Z}_2$ classification of first descendants.}
\label{sec: Z2 classification of 1st}

\begin{figure}
\begin{center}
    \includegraphics[height=.4\textwidth]{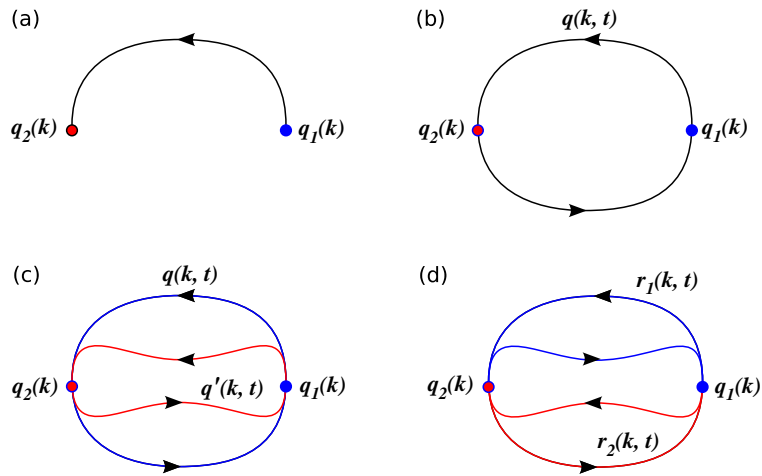}
  \caption{      \label{interpol_paths}
(a) One-parameter interpolation of two topological insulators
represented by the projectors $q_1$ and $q_2$. 
In (b),
the one-way interpolation (a) is extended 
by making use of the discrete symmetry.
(c) Two different 
interpolations, each colored red and blue, respectively. 
In (d) the original interpolations are rearranged 
into two different interpolations. 
}
\end{center}
\end{figure}

In this section, we show how a $\mathbb{Z}_2$ topological classification
is obtained for $2n$-dimensional, chiral symmetric 
insulators (supercondcutors) under the constraint of additional
discrete symmetries 
(i.e., $\mathcal{C}$  and $\mathcal{T}$).
To uncover the $\mathbb{Z}_2$ topological characteristics of chiral
symmetric systems, we study
the topological distinctions among the block off-diagonal projectors $q(k)$.
Let us consider two projectors $q_1(k)$ and $q_2(k)$,
whose momentum space configuration is restricted by
the symmetry constraint of one of the classes DIII/CI,
equation \re{sym prop DIII and CI}, or BDI/CII,
equation \re{sym prop BDI and CII}.
We introduce a continuous interpolation $q(k, t)$, $t \in [0, \pi]$ between
these two projectors (see figure~\ref{interpol_paths}a) with
\begin{eqnarray} \label{interpol for q half}
q(k, t=0)=q_1(k) 
\qquad \textrm{and} \qquad
q(k, t=\pi)=q_2(k).
\end{eqnarray}
Since the topological space of $2n$-dimensional class AIII insulators is
simply connected, 
the continuous deformation $q(k,t)$ is well defined.
In general, $q(k,t)$ does not satisfy the symmetry constraints encoded by
equation \re{sym prop DIII and CI} or equation \re{sym prop BDI and CII},
respectively.
However, by combining equation \re{interpol for q half}
with its symmetry transformed partner, we can construct an interpolation
that obeys the discrete symmetries of
the given symmetry class  (see figure~\ref{interpol_paths}b). 
In the case of symmetry class DIII/CI we define for $t \in [\pi, 2 \pi]$ 
\begin{eqnarray} \label{extended interpol DIII/CI}
q (k, t) = 
\epsilon q^T ( -k, 2\pi -t ), 
\quad
\epsilon =
\left\{ 
\begin{array}{l l}
-1, & \textrm{DIII}\\
+1, & \textrm{CI}, \\
\end{array} 
\right.
\end{eqnarray}
while, in the case of symmetry class BDI/CII we set for  $t \in [\pi, 2 \pi]$ 
\begin{eqnarray} \label{extended interpol BDI/CII}
q (k, t) = 
G q^\ast ( -k, 2 \pi -t ) G^{-1} ,
\quad
G =
\left\{ 
\begin{array}{l l}
\sigma_0, &  \textrm{BDI} \\
\rmi \sigma_y, & \textrm{CII} . \\
\end{array} 
\right.
\end{eqnarray}
Equations~(\ref{interpol for q half})--(\ref{extended interpol BDI/CII})
with the parameter $t$ replaced 
by the wavevector component $k_{2n+1}$ represent a $(2n+1)$-dimensional
projection operator 
respecting the symmetry constraints of the corresponding symmetry class.
Consequently, a winding number for $q(k,t)$, $\nu_{2n+1} [ q(k,t)]$,
can be defined  in the $(k,t)$ space.
Two different interpolations 
$q(k,t)$ and $q^{\prime} (k,t)$
generally give different winding numbers,
$\nu_{2n+1} [q (k,t) ] \ne \nu_{2n+1} [ q^{\prime}(k, t)]$.
However, we can show that symmetry constraint \re{extended interpol DIII/CI}
or \re{extended interpol BDI/CII}, respectively, leads to
\begin{eqnarray} \label{claim for nu}
\nu_{2n+1}[q(k,t)]
-
\nu_{2n+1}[q'(k,t)]
=0
\quad
\mbox{mod}
~
2 .
\end{eqnarray}
To prove equation \re{claim for nu} we introduce two new interpolations
$r_1 (k, t)$ and $r_2 (k, t)$ that transform into each 
other under the discrete symmetry operations (see figure~\ref{interpol_paths})
\begin{eqnarray}
r_1 (k,t) 
&=&
\left\{
\begin{array}{l l}
q (k, t) , 
& \qquad 
t \in \left[ 0, \pi \right] 
\cr
q^{\prime} (k, 2 \pi - t),
& \qquad
t \in \left[ \pi, 2 \pi \right],
\end{array}
\right.
\nonumber\\
r_2 (k,t) 
&=&
\left\{
\begin{array}{l l}
q^{\prime} (k, 2 \pi - t), 
& \qquad 
t \in \left[ 0, \pi \right] 
\cr
q (k, t) , 
& \qquad
t \in \left[ \pi, 2 \pi \right],
\end{array}
\right.
\end{eqnarray}
These are recombinations of the deformations $q(k,t)$ and
$q^{\prime}(k,t)$ with
\begin{eqnarray}
\nu_{2n+1}[q(k,t)]
-
\nu_{2n+1}[q'(k,t)]
=
\nu_{2n+1}[r_1(k,t)]
+
\nu_{2n+1}[r_2(k,t)] .
\end{eqnarray}
Now, we make use of the result from
section~\ref{sec: symmetry properties of winding no}.
Namely, for symmetry class BDI/CII we found
\begin{eqnarray} \label{sym prop w BDI/CII}
w_{4n+1} [ q (k, t) ]
=
w^\ast_{4n+1} [ q (-k,-t) ]
\end{eqnarray}
in $d=4n+1$ spatial dimensions, whereas
for symmetry class DIII/CI 
\begin{eqnarray}
w_{4n-1} [ q (k, t) ]
=
w^\ast_{4n-1} [ q (-k,-t) ]
\end{eqnarray}
in $d=4n-1$ spatial dimensions.
Let us first consider class BDI/CII in $d=4n+1$ dimensions.
With equation \re{winding no} and  \re{sym prop w BDI/CII} we obtain 
\begin{eqnarray}
\nu_{4n+1}[r_1]
&=&
\int  \rmd^{4n} k\, \rmd t \,  w^{\ }_{4n+1} [ r_1 (k,t) ]
\nonumber\\
&=&
\int  \rmd^{4n} k\, \rmd t \,  w^{\ast}_{4n+1} [ r_1 (-k,-t) ]
\nonumber\\
&=&
\int  \rmd^{4n} k\, \rmd t \,  w^{\ast}_{4n+1} [ r_2 (k,t) ] =
\nu_{4n+1}[r_2 ] ,
\end{eqnarray}
where we made use of the fact that $r_1$ transforms into $r_2$
under the discrete symmetry operations of class BDI/CII.
In conclusion, we have shown 
$ \nu_{4n+1}[q(k,t)]
-
\nu_{4n+1}[q'(k,t)]
=
2 \nu_{4 n+1}[r_1(k,t)]
\in 2 \mathbb{Z}$
for any two interpolations $q(k,t)$ and $q'(k,t)$ belonging to class BDI/CII.
Hence, we can define a relative invariant for the $4n$-dimensional projection operators
$q_1(k)$ and $q_2(k)$
\begin{eqnarray} \label{relaitve_invariant}
\nu_{4n} \left[ q_1(k), q_2 (k) \right]
= 
(-1)^{\nu_{4n+1} \left[ q (k,t) \right] },
\end{eqnarray}
which is independent of the particular choice of the
interpolation $q(k,t)$ between $q_1(k)$ and $q_2(k)$. 
Once we have
identified  a ``vacuum'' projection 
operator, e.g., $q_0 ( k ) \equiv q_0$, we can construct
with equation \re{relaitve_invariant} a $\mathbb{Z}_2$ invariant:  non-trivial Hamiltonians are
characterized by $\nu_{4n} [ q, q_0 ] = -1$, whereas trivial ones satisfy $\nu_{4n} [ q, q_0 ] = +1$.
Finally, we note that
the calculation leading to the relative invariant $\nu_{4n} [q_1,q_2]$, equation \re{relaitve_invariant}, can be repeated for symmetry class DIII/CI in $d=4n-1$ dimensions yielding a relative invariant $\nu_{4n-2} [ q_1, q_2]$.

For $d=4n-1$ ($d=4n+1$)
the dimensional reduction seems to be possible both
for class DIII and CI (BDI and CII). 
However, for a given $n$ the dimensional reduction
seems to be meaningful only for one of them;
while for one of them
(corresponding to {\it ii} in
table \ref{eq: symmetires of real Dirac}),
there is a decendent $\mathbb{Z}_2$ insulator
in one dimension lower,
for the other (corresponding to {\it iv} in
table \ref{eq: symmetires of real Dirac}),
there are no decendent $\mathbb{Z}_2$ insulators. 
In other words, 
for a given $n=\mathrm{odd}$ ($\mathrm{even}$),
there are lower dimensional  
$\mathbb{Z}_2$ topological insulators
for either one of class DIII and CI (BDI and CII),
but not both. 
This is because the above procedure does not apply
when the classification of topological insulator is 
not $\mathbb{Z}$, but $2\mathbb{Z}$.

\subsubsection{$\mathbb{Z}_2$ classification of second descendants.} 

The dimensional reduction procedure presented in the previous subsection can be repeated once more to obtain
a $\mathbb{Z}_2$ classification of the second descendants. 
As before, we first focus on symmetry class BDI/CII and
study the topological distinctions among the block off-diagonal projectors.
We consider two $(4n-1)$-dimensional projectors $q_1(k)$ and $q_2(k)$, which satisfy
the symmetry constraints imposed by class BDI/CII.
We define an adiabatic interpolation $q ( k , t )$, $t \in [0, 2 \pi]$
\begin{eqnarray} \label{interpol 2nd desc}
&&
q (k, t=0) = q_1(k), 
\quad \qquad  \quad
q (k, t=\pi) = q_2 (k) ,
\nonumber\\
&&
q (k, t) = 
G q^\ast ( -k,  -t ) G^{-1} ,
\quad
G =
\left\{ 
\begin{array}{l l}
\sigma_0, &  \textrm{BDI} \\
\mathrm{i} \sigma_y, & \textrm{CII} .   \\
\end{array} 
\right. .
\end{eqnarray}
We can interpret $q(k,t)$ as a $4n$-dimensional projector
belonging to symmetry class BDI/CII.
Therefore, for two deformations $q(k,t)$ and $q'(k,t)$ of the
form \re{interpol 2nd desc}
a relative invariant $\nu_{4n} [q (k,t), q'(k,t) ]$ can be defined,
as discussed in section~\ref{sec: Z2 classification of 1st}.
It turns out that due to condition \re{interpol 2nd desc}
the invariant $\nu_{4n} [q (k,t), q'(k,t) ]$ is independent of the
particular choice of interpolations, i.e.,
$\nu_{4n} [q (k,t), q'(k,t) ] = +1$
for any two deformations $q(k,t)$ and $q'(k,t)$
satisfying condition \re{interpol 2nd desc}.
In order to prove this, we consider a continuous interpolation $r(k,t,s)$
between the two deformations $q(k,t)$ and $q'(k,t)$ with
\begin{eqnarray}
&&
r ( k, t, s=0)
= q(k,t) ,
\quad \; \;
r ( k, t, s=\pi)
= q' (k,t) ,
\nonumber\\
&&
r ( k, t=0, s )
= q_1 (k) ,
\quad \quad
r ( k, t=\pi , s )
= q_2 (k ) ,
\nonumber\\
&&
r (k, t ,s ) = 
G r^\ast ( -k,  -t ,-s ) G^{-1} ,
\quad
G =
\left\{ 
\begin{array}{l l}
\sigma_0, &  \textrm{BDI} \\
\mathrm{i} \sigma_y, & \textrm{CII} . \\
\end{array} 
\right.
\end{eqnarray}
This represents a $(4n+1)$-dimensional off-diagonal projector
in symmetry class BDI/CII with the winding number
$\nu_{4n+1} \left[ r(k,t,s) \right] $.
We note that $r(k,t,s)$ is not only a deformation
between $q(k,t)$ and $q'(k,t)$, but can also be viewed as
a continuous interpolation between $r (k, 0, s) \equiv q_1 (k) $
and $r(k, \pi, s) \equiv q_2 (k) $  (for any $s \in [ 0, 2 \pi ] )$.
Therefore, 
we find 
$\nu_{4n} [q (k,t), q'(k,t) ] 
=
\nu_{4n} [r (k,t,0), r(k, t, \pi ) ] 
= 
\nu_{4n} [r (k,0,s), r(k,\pi, s ) ] 
$
Since $r (k,0,s) \equiv q_1(k)$ and $ r(k,\pi, s ) \equiv q_2(k)$
are independent of $s$, we find that
$\nu_{4n} [ r (k,0,s) ,r(k,\pi, s) ] = (-1)^{\nu_{4n+1} [r] } = +1$.
Hence, we have shown that $\nu_{4n} [q (k,t), q'(k,t) ]$ only
depends on $q_1(k)$ and $q_2(k)$.
Therefore, $\nu_{4n} [ q_0 , q (k,t) ]$ together with a reference
(``vacuum'') projector $q_0$
constitutes a well-defined $\mathbb{Z}_2$ invariant in $4n-1$ dimensions.
 
\subsubsection{Example: $ d=3 \to 2 \to 1$.}

As an example we consider a 3-dimensional topological Dirac superconductor
belonging to symmetry class DIII.
Consider a $d=3$ dimensional Dirac Hamiltonian, 
\begin{eqnarray}
\mathcal{H}^{d=3}_{(5)}(k,m)
&=&
k_x \alpha_x
+
k_y \alpha_y 
+
k_z \alpha_z
-
m \rmi \beta \gamma^5 .
\label{eq: d=3 DIII massive dirac}
\end{eqnarray}
This is nothing but 
the chiral topological Dirac superconductor in class DIII. 
This Hamiltonian is essentially identical
to the 
BdG Hamiltonian describing the Bogoliubov quasiparticles 
in the B phase of superfluid $^3$He
\cite{SchnyderRyuFurusakiLudwig2008, topSCstanford, topSCRoy}
\footnote{
In  Ref.\ \cite{Salomaa88},
a topological invariant counting the number of
gap-closing Dirac points in an extended, higher-dimensional  
parameter space was discussed 
for a four-band example (\ref{eq: d=3 DIII massive dirac}),
as opposed to $\nu_3[q]$ defined for a given topological phase. 
Moreover, the crucial role for the protection of topological properties 
arising from the combined time-reversal and charge-conjugation
symmetries of symmetry class DIII
was not discussed. 
}.
It also describes an auxiliary Majorana hopping problem
for an interacting bosonic model on the diamond lattice
\cite{RyuDiamond}. 
In this basis, 
discrete symmetries are given by 
\begin{eqnarray}
\mathcal{T}:\quad
(\sigma_y \otimes \tau_x)
\left[\mathcal{H}^{d=3}_{(5)}(-k,m) \right]^*
(\sigma_y \otimes \tau_x) 
=
\mathcal{H}^{d=3}_{(5)}(k,m),
\nonumber \\
\mathcal{C}: \quad
(\sigma_y \otimes \tau_y)
\left[\mathcal{H}^{d=3}_{(5)}(-k,m) \right]^*
(\sigma_y \otimes \tau_y) 
=
-\mathcal{H}^{d=3}_{(5)}(k,m). 
\end{eqnarray}
Combining these two,
we have chiral symmetry, which allows us to define
the winding number $\nu_3$. 
From the calculations in
Sec.\ \ref{subsec: A-AIII examples in d=5,4,3},
the winding number is non-zero,
$\nu_3=\pm 1/2$,
depending on the sign of the mass.

By dimensional reduction, we obtain the Hamiltonian in 
one-dimension lower, 
\begin{eqnarray}
\mathcal{H}^{d=2}_{(5)}(k,m)
&=&
k_x \alpha_x
+
k_y \alpha_y
-
m \rmi \beta \gamma^5 .
\label{eq: d=2 DIII massive dirac}
\end{eqnarray} 
This Hamiltonian is a 2D analogue of  $^3$He-B,
and is unitarily equivalent to the direct product of
spinless 
$p+ \rmi p$
and
$p- \rmi p$ 
wave superconductors. 
As this is obtained from dimensional reduction
of a parent topological 
superconductor (\ref{eq: d=3 DIII massive dirac})
this is a $\mathbb{Z}_2$ state.
Finally, by further reducing dimensions, 
\begin{eqnarray}
\mathcal{H}^{d=1}_{(5)}(k,m)
&=&
k_x \alpha_x
-
m \rmi \beta \gamma^5 .
\end{eqnarray} 
This is a 1D $p_x$-wave superconductor.
Again, this is a $\mathbb{Z}_2$ state.

We now discuss the  topological character of these states in more detail.
The $\mathbb{Z}_2$ nature of the state in this 
$d=2$ example can be studied by 
the Kane-Mele invariant
\cite{KaneMele,Fu06_3Db}.
It can be expressed as
an $\mathrm{SU}(2)$ Wilson loop
\begin{eqnarray}
W_{\mathrm{SU}(2)}[L] &:=&
\frac{1}{2}
\mathrm{tr}\,
P
\exp\left[
 \oint_{L} \mathcal{A}(k) 
\right].
\label{eq: def SU2 Wilson loop}
\end{eqnarray}
Here $\mathcal{A}(k)$ is the $\mathrm{SU}(2)$ Berry connection,
$P$ represents the path-ordering. 
By definition, 
$W_{\mathrm{SU}(2)}[L]$
is a well-defined and gauge invariant 
quantity for any loop $L$ in the BZ.
For time-reversal invariant systems, it is useful to consider a loop $L$ that
satisfies time-reversal symmetry;
i.e., a loop which is mapped onto itself
(up to reparameterization)
under $k\to -k$. 
For these time-reversal invariant loops,
the SU(2) Wilson loop is quantized,
\begin{eqnarray}
W_{\mathrm{SU}(2)}[L] &=&
\pm 1,
\end{eqnarray}
and provides a way to distinguish different $\mathbb{Z}_2$ states
\cite{KaneMele,Fu06_3Db,Lee_Ryu}.
The quantization of the Wilson loop can be proved by first noting
that because of $\mathcal{T}$, 
Bloch wavefunctions at $k$ and $-k$ 
are related to each other by a unitary transformation $w(k)$, 
\begin{eqnarray}
|\Theta u^-_{\hat a}(k) \rangle 
=
w_{\hat a \hat b}(k)| u^-_{\hat a} (-k) \rangle
\end{eqnarray}
where
$\Theta$ represents 
$\mathcal{T}$
operation and 
\begin{eqnarray}
w_{\hat{a} \hat{b} } (k) 
&:=&
\langle u^-_{\hat a} (-k) |\Theta u^-_{\hat b}(k) \rangle.
\end{eqnarray}
Accordingly, the Berry connection at $k$
is gauge equivalent to $A^T_{\mu}$ at $-k$, 
\begin{eqnarray}
A_{\mu}(-k) 
&=&
+
w(k) A^T_{\mu}(k) w^{\dag}(k)
-
w(k) \partial_{\mu} w^{\dag}(k)
\nonumber \\
&=&
-
w(k) A^*_{\mu}(k) w^{\dag}(k)
-
w(k) \partial_{\mu} w^{\dag}(k).
\label{eq: u(2) sewing condition}
\end{eqnarray}
Because of this sewing condition
(\ref{eq: u(2) sewing condition}) of the gauge field, 
when plugged in 
(\ref{eq: def SU2 Wilson loop}), 
contributions at $k$ and $-k$ in the path integral
cancel pairwise, except at those momenta
which are invariant under 
$\mathcal{T}$ by themselves. 
Thus, 
\begin{eqnarray}
W_{\mathrm{SU}(2)}[L] &=&
\prod_K \mathrm{Pf}\left[w(K)\right],
\end{eqnarray}
where $K$ is a momentum which is invariant under
$k\to -k$.

We now compute the $\mathbb{Z}_2$ number
for the $d=2$ class DIII topological
superconductor (\ref{eq: d=2 DIII massive dirac}). 
From (\ref{eq: wfn 4x4 occupied}),
\begin{eqnarray}
|u^{-}_1(k)\rangle
&=&
\frac{1}{
\sqrt{2}
\lambda
}
\left(
\begin{array}{c}
-k_x + \rmi k_y \\
\rmi m  \\
0 \\
\lambda
\end{array}
\right),
\quad
|u^{-}_2(k)\rangle
=
\frac{1}{
\sqrt{2} \lambda
}
\left(
\begin{array}{c}
 \rmi m \\
-k_x - \rmi k_y \\
\lambda \\
0 \\
\end{array}
\right).
\end{eqnarray}
Then, their time-reversal counterparts are
\begin{eqnarray}
|\Theta u^{-}_1(k)\rangle
&=&
\frac{-\rmi}{
\sqrt{2}
\lambda
}
\left(
\begin{array}{c}
\lambda \\
0 \\
-\rmi m  \\
k_x + \rmi k_y \\
\end{array}
\right),
\quad
|\Theta u^{-}_2(k)\rangle
=
\frac{-\mathrm{i}}{
\sqrt{2} \lambda
}
\left(
\begin{array}{c}
0 \\
-\lambda \\
-k_x + \rmi k_y \\
 \rmi m \\
\end{array}
\right).
\end{eqnarray}
where 
$\Theta = (\sigma_y\otimes \tau_x )\mathcal{K}$
with $\mathcal{K}$ being the complex conjugation. 
The ``sewing matrix'' $w(k)$ can then be computed as 
\begin{eqnarray}
w(k) 
&:=&
\langle u_{\hat a} (-k) |\Theta u_{\hat b}(k) \rangle 
=
\frac{1}{\lambda(k)}
\left(
\begin{array}{cc}
- \rmi k_x +   k_y & m(k) \\
-m(k) & + \rmi k_x + k_y 
\end{array}
\right).
\end{eqnarray}
Here, we have regularized Dirac Hamiltonian 
properly, by making the mass $k$-dependent,
e.g.,.
$m\to m(k)= m-C k^2$. 
With this regularization, 
$k=0$ and $k=\infty$ are the two
time-reversal invariant momenta. 
One finds 
\begin{eqnarray}
\mathrm{Pf}\, w(0)
&=&
\frac{m(0)}{\lambda(0)}
\mathrm{Pf}\, (\mathrm{i}\sigma_y )
=
\frac{m}{|m|}
\mathrm{Pf}\, (\mathrm{i}\sigma_y ),
\nonumber \\
\mathrm{Pf}\, w(\infty)
&=&
\frac{m(\infty)}{\lambda(\infty)}
\mathrm{Pf}\, (\mathrm{i}\sigma_y )
=
\frac{-C}{|C|}
\mathrm{Pf}\, (\mathrm{i}\sigma_y ), 
\end{eqnarray}
and hence
\begin{eqnarray}
W_{\mathrm{SU}(2)}[L]
= 
-\mathrm{sign}(m) \mathrm{sign}(C) . 
\end{eqnarray}
That is,
when
$\mathrm{sign}(m) = \mathrm{sign}(C)$
the Hamiltonian (\ref{eq: d=3 DIII massive dirac}) 
is a $\mathbb{Z}_2$ topological superconductor.

\subsection{Topological insulators lacking chiral symmetry}
\label{sec: dimensional reduction in non-chiral classes}

Four among the 
eight (``real'') symmetry classes of
table~\ref{table: Z2 topins dimreduct} 
break chiral (``sublattice'') symmetry.
The $\mathbb{Z}$ topological insulators (superconductors)
which break chiral symmetry
are characterized by 
a Chern number [see equation~\re{Chern_number}]. 
Using this topological invariant one can derive the $\mathbb{Z}_2$
classification for the  lower dimensional descendants.
This derivation is analogous to
the discussion in section \ref{sec: dimensional reduction in chiral classes}
and has been performed previously in Ref.~\cite{Qi_Taylor_Zhang}
for the $\mathbb{Z}_2$ topological insulators which break chiral symmetry.
For these reasons we do not repeat the argument here and refer
the reader to Ref.~\cite{Qi_Taylor_Zhang}  for details.
For the $\mathbb{Z}_2$ topological insulators (superconductors)
which break chiral symmetry,
one can construct a $\mathbb{Z}_2$ index
similar to 
the one constructed by Moore and Balents \cite{Moore06},
in all (even) dimensions
\cite{Fukui}.

\section{Discussion}
\label{Discussion}

In this paper we have performed an exhaustive study of all topological insulators 
and superconductors in arbitrary dimensions.
The main part of this paper deals with dimensional reduction procedures which 
relate topological insulators (superconductors) in different dimensions and symmetry classes.
We also discussed topological field theories
in $D=d+1$ space time dimensions
describing linear responses
of topological insulators (superconductors).
Furthermore,
we studied how the presence of
inversion symmetry modifies the classification of topological insulators (superconductors)
(see appendix~\ref{secParity}).
In the following we give a brief summary of the main results of the paper.

\subsection{Dimensional reduction procedures}

We have constructed for all five symmetry classes
of topological insulators or superconductors 
a Dirac Hamiltonian representative, in all spatial dimensions. 
Using these Dirac Hamiltonians as canonical examples,
we have demonstrated that topological insulators (superconductors) in different
spatial dimensions and symmetry classes can be related to each other
by dimensional reduction procedures. 
These Dirac representatives
have been useful for constructing 
string theory realization of 
topological insulators and superconductors:
In Ref.\ \cite{RyuTakayanagi10}, 
a one-to-one correspondence between 
the ten-fold classification of
topological insulators and superconductors
and a K-theory classification of D-branes
was established, 
where open string excitations between two D-branes of
various dimensions are shown to reproduce all Dirac representatives
constructed in this paper.

\subsubsection{The ``complex'' case.}

We first studied topological insulators (superconductors)
with Hamiltonians 
which are complex (i.e., belonging to the ``complex case'' of table  \ref{periodic table},
that is, classes A and AIII from table~\ref{TableSymmetryClassesTwo}).
Here, starting from a 
topological insulator
which lacks chiral symmetry
(i.e., class A)
in even spatial dimensions $d=2n$, 
we obtain a topological insulator (superconductor)
in $d=2n-1$ 
which possesses chiral symmetry
(i.e.,  class AIII), 
by dimensional reduction. 
In order to understand how the topological characteristics of the $d=2n-1$  class AIII topological
insulator (characterized by the winding number $\nu_{2n-1}$) are inherited from its higher dimensional 
parent, the $d=2n$ class A topological insulator (characterized by the Chern number $\mathrm{Ch}_{n}$),
it is important to first point out a few properties of the momentum space topology
of Bloch wavefunctions in symmetry classes A and AIII.
First we note that the non-zero Chern number in class A leads to an obstruction
to the existence of globally defined Bloch eigenfunctions.
I.e., it is not possible to construct Bloch eigenfunctions
for the  parent class A topological insulator
globally on the $d=2n$ dimensional Brillouin zone.
Hence, the Bloch eigenfunctions can only be defined locally on 
some suitably chosen coordinate patches.
For the descendant topological insulator (superconductor)
with chiral symmetry,
on the other hand, 
since there is no Chern invariant in $d=2n-1$,
there always exists a basis (i.e., a gauge) in which 
the Bloch eigenfunctions are well-defined globally
on the entire $d=2n-1$ dimensional Brillouin zone.
Such a basis is provided by the one, in which the Hamiltonian
is block off-diagonal.
The existence of such a block off-diagonal basis, in turn, 
is implied by the chiral symmetry of the descendant class AIII topological insulator (superconductor).
As a consequence, the role played by the chiral symmetry is to
guarantee that Bloch eigenfunctions of
class AIII topological insulators (superconductors) 
can be defined globally on the entire Brillouin zone.

Now, by adopting such a block off-diagonal basis for the class AIII topological insulator,
we have shown that
the $d=2n-1$ dimensional Brillouin zone of the 
descendant class AIII
Hamiltonian defines the boundary of two local coordinate patches 
of the $d=2n$ dimensional Brillouin zone of the parent class A topological insulator.
The transition function between these two coordinate patches is given by the block 
off-diagonal projector\footnote[1]{of the type described in
equation (34)  of  the first article
in Ref.\ \cite{SchnyderRyuFurusakiLudwig2008}}
$q$ of the
descendant Hamiltonian, and the Chern number $\mathrm{Ch}_n$ of the parent Hamiltonian
can be written in terms of a winding number of this transition function 
(see section~\ref{dirac insul and dim reduc}).

\subsubsection{The ``real'' case.}

Also demonstrated is, using the Dirac Hamiltonian representatives,
the periodicity 8 shift of symmetry classes
for the topological insulators (superconductors) 
in the 8 symmetry classes
in which the Hamiltonian possesses at least one
reality condition 
(arising from $\mathcal{T}$ or $\mathcal{C}$). 
Once one has constructed a representative in terms of
a Dirac Hamiltonian,
the reality properties of the spinor
representations of the orthogonal groups $\mathrm{SO}(N)$
(which are linked to the reality properties of the
representations of the Clifford algebra formed by the gamma matrices)
leads directly to this 8-fold periodicty in the spatial dimension $d$
(see section \ref{sec: dimensional ladder: real case}).
From the Dirac Hamiltonian representatives of
topological insulators (superconductors) in all $d$ spatial
dimensions (all of which are of course massive),
one can realize a $d-1$-dimensional boundary,
by making a domain wall
at which the mass term changes sign
\cite{Callan-Harvey}.
There then appears a boundary state localized at the boundary, 
which is a defining property of topological character in the bulk. 
These boundary states are of special kind, as they are protected 
from the appearance of a gap
and from Anderson localization,
by the time-reversal and charge-conjugation symmetry properties
of the specific symmetry class.
Indeed, one of the underlying strategies of classification 
adopted in Ref.\ \cite{SchnyderRyuFurusakiLudwig2008}
was to classify the properties of these boundary Hamiltonians
in each symmetry class, including those of Dirac type,
which must be completely gapless  and delocalized when they come from a topological
bulk, but can otherwise be gapped and localized for a non-topological bulk.
By constructing massive Dirac Hamiltonian representatives in the bulk
for all topological insulators (superconductors), 
which give rise to the gapless boundary Dirac Hamiltonians,
we undertook here in some sense a complementary task 
as compared to that in Ref.\ \cite{SchnyderRyuFurusakiLudwig2008}.

\subsection{Topological field theories}

In the low-energy limit, all 
topological insulators (superconductors)
can be described by topological field theories
for the linear responses in space-time, which 
characterize universal (in principle) experimentally
accessible observables of the topological features,
such as, e.g., in transport.
For example, for symmetry class A in even spatial dimensions $d=2n$, the topological field theory
of the linear responses
in $D=d+1$ space-time dimensions
is given by the Chern-Simons
action, whose
coupling-coefficient
is the $n$th Chern number $\mathrm{Ch}_n$.
For symmetry class AIII in odd spatial dimensions $d=2n-1$, on the other hand,  the topological 
field theory in $D=2n$ space-time dimensions
is given by the $\theta$ term, whose coefficient (the $\theta$ angle) is given 
by the winding number $\nu_{2n+1}$ (see section \ref{field_theory}).
For topological singlet superconductors
(symmetry classes C and CI), 
one can couple external SU(2) gauge field to 
the conserved spin current operator of the
BdG quasiparticles. 
The spin response in topological singlet superconductors, 
can be described by 
SU(2) gauge theory 
with Chern-Simons type topological term in $d=2n$
\cite{Read00}
and with the $\theta$-type-term in $d=2n+1$
\cite{SchnyderRyuLudwig2009}.
In passing we note, that it is also possible to establish a connection among these topological field theories in various
space-time dimensions
and symmetry classes via dimensional reduction procedures.
The topological field theory formulation may be a good starting point
to explore, more generally, topological phases in interacting systems
beyond those that are presently known.

\subsection{Topological insulators with inversion symmetry and
either time reversal or charge conjugation symmetry
}

We have also studied the restrictions imposed on groundstate properties of topological 
insulators (superconductors)
by the presence of inversion symmetry. That is, we studied topological states that are protected
by a combination of spatial inversion (denoted by $\mathcal{I}$) 
and an additional discrete symmetry 
(i.e., $\mathcal{T}$ or $\mathcal{C}$).
These systems are invariant 
under the combined symmetry operations 
$\mathcal{T} \cdot \mathcal{I}$ or $\mathcal{C} \cdot \mathcal{I}$, but 
all three symmetries 
$\mathcal{T}$, $\mathcal{C}$ and $\mathcal{I}$,
are assumed to be absent.
We have determined the space of projectors describing
these topological states.
From the homotopy groups of these space of projectors follows
the classification of topological states that are protected
by $\mathcal{T} \cdot \mathcal{I}$ or $\mathcal{C} \cdot \mathcal{I}$ 
(see table~\ref{inversion and one discrete symmetry}).

\subsection{Directions for future work}

An important direction for future study 
is the search for experimental realizations of 
three-dimensional topological singlet or triplet superconductors.
Given how fast experimental realizations
of the QSHE in $d=2$ and the $\mathbb{Z}_2$ topological insulators
in $d=3$ have been found, we anticipate a similar development
for the three-dimensional topological singlet or triplet superconductors.
For example, unconventional 
superconductors in heavy fermion systems,
typically possessing strong spin-orbit effects,
have been studied extensively over the years.
They might be good candidates for topological superconductors.
Moreover, the search for non-trivial topological
quantum ground states is not restricted to free fermions, or
BCS quasiparticles,
but includes also, more generally, strongly interacting systems
other than BCS
with emergent free fermion behavior at low energies.
Furthermore, it should be noted that the classification 
scheme given by table \ref{periodic table}
is also applicable outside 
the realm of condensed matter physics.
For example, topological properties of
color superconducting phases~\cite{Rajagopal2008,Nishida10}, which
are predicted to occur in quark matter, can be discussed
in terms of the present classification scheme.

\ack

This work is supported in part by NSF Grants
No.\  PHY05-51164 (APS) and No. DMR-0706140 (AWWL),
and by a Grant-in-Aid for Scientific Research from
the Japan Society for the Promotion of Science
(Grant No.\ 21540332) (AF).
SR thanks the Center for Condensed Matter Theory at University of California,
Berkeley for its support. 
SR thanks 
A.\ M.\ Essin, J.\ E.\ Moore, and A. Vishwanath for useful discussion.

\appendix
\makeatletter
\renewcommand\thesection{\Alph{section}}
\makeatother

\section{ 
Cartan symmetric spaces: generic Hamiltonians, 
NL$\sigma$M field theories, and  classifying spaces of $K$-theory
}
\label{subsec:RMT} 

In this appendix we review the appearance of Cartan's ten-fold list of 
symmetric spaces in the context of
(i) basic quantum mechanics: where they describe the time evolution operators 
$\exp(\rmi t {\cal H} )$ of generic Hamiltonians ${\cal H}$,
(ii) NL$\sigma$M field theories:  where they describe the ``target space'' 
manifold of the NL$\sigma$M, and 
(iii) K-Theory: where they describe the ``classifying space'' (briefly discussed 
at the end of subsection \ref{Review-Classification-TopIns} of the introduction).
As reviewed in the introduction, the homotopy groups of these symmetric spaces 
play a key role in the classification of topological insulators and superconductors 
both in the approach of Ref.~\cite{SchnyderRyuFurusakiLudwig2008}, 
which makes use of results from Anderson localization physics,
as well as in the approach of Ref.~\cite{KitaevLandau100Proceedings}, 
which is based on K-theory.

\begin{table}[ht]
\begin{center}
\begin{small}
\begin{tabular}{cccc}\hline
Cartan
&time evolution op.\ 
& fermionic replica
& classifying \\
label
& 
$\exp\{ \mathrm{i} t {\cal H}\}$
& NL$\sigma$M target space
&space
           \\ \hline \hline 
A &
$\mathrm{U}(N)\times \mathrm{U}(N)/\mathrm{U}(N)$
&
$\mathrm{U}(2n)/\mathrm{U}(n)\times \mathrm{U}(n)$ 
&
$\mathrm{U}(N+M)/\mathrm{U}(N)\times \mathrm{U}(M)=C_0$ 
\\ 
{\bf AIII }& 
$\mathrm{U}(N+M)/\mathrm{U}(N)\times \mathrm{U}(M)$
& $\mathrm{U}(n)\times \mathrm{U}(n)/ \mathrm{U}(n)$
& $\mathrm{U}(N)\times \mathrm{U}(N)/ \mathrm{U}(N)=C_1$
\\ 
\hline \hline
AI &
$\mathrm{U}(N)/\mathrm{O}(N)$
&
$\mathrm{Sp}(2n)/\mathrm{Sp}(n)\times \mathrm{Sp}(n)$ 
&
$\mathrm{O}(N+M)/\mathrm{O}(N)\times \mathrm{O}(M)=R_0$
\\ 
{\bf BDI}
&
$\mathrm{O}(N+M)/\mathrm{O}(N)\times \mathrm{O}(M)$
& $\mathrm{U}(2n)/ \mathrm{Sp}(2n)$ 
&
$\mathrm{O}(N)\times \mathrm{O}(N)/\mathrm{O}(N)=R_1$
\\  
D &
$\mathrm{O}(N)\times \mathrm{O}(N)/\mathrm{O}(N)$
& $\mathrm{O}(2n)/ \mathrm{U}(n)$
&$\mathrm{O}(2N)/ \mathrm{U}(N)=R_2$
\\
{\bf DIII} &
$\mathrm{SO}(2N)/\mathrm{U}(N)$
& $\mathrm{O}(n)\times \mathrm{O}(n)/\mathrm{O}(n)$ 
& $\mathrm{U}(2N)/ \mathrm{Sp}(2N)=R_3$ 
\\ 
AII &
$\mathrm{U}(2N)/\mathrm{Sp}(2N)$
& $\mathrm{O}(2n)/\mathrm{O}(n)\times \mathrm{O}(n)$
& 
$\mathrm{Sp}(N+M)/\mathrm{Sp}(N)\times \mathrm{Sp}(M)=R_4$ 
\\ 
{\bf CII} &
$\mathrm{Sp}(N+M)/\mathrm{Sp}(N)\times \mathrm{Sp}(M)$
& $\mathrm{U}(n)/ \mathrm{O}(n)$ 
&$\mathrm{Sp}(N)\times\mathrm{Sp}(N)/\mathrm{Sp}(N)=R_5$
\\ 
C &
$\mathrm{Sp}(2N)\times \mathrm{Sp}(2N)/\mathrm{Sp}(2N)$
& $\mathrm{Sp}(2n)/ \mathrm{U}(n)$ 
& $\mathrm{Sp}(2N)/ \mathrm{U}(N)=R_6$ 
\\ 
{\bf CI} &
$\mathrm{Sp}(2N)/\mathrm{U}(N)$
& $\mathrm{Sp}(2n)\times \mathrm{Sp}(2n)/\mathrm{Sp}(2n)$
&$\mathrm{U}(N)/ \mathrm{O}(N)=R_7$ 
\\ \hline
\end{tabular}
\end{small}
\end{center}
\caption{
Three appearances of the list of Cartan's ten symmetric spaces.
First column: unitary time evolution operator,
second column: (compact) target space of NL$\sigma$Ms,
third column: classifying space. 
(We use the convention in which $m=$ even in $\mathrm{Sp}(m)$. -- Moreover,
the Cartan labels of those symmetry classes invariant under the
chiral symmetry operation 
$\mathcal{S}=\mathcal{T} \cdot \mathcal{C}$
 from table~\ref{TableSymmetryClassesTwo}
are indicated by bold face letters.)
\label{RMT, NLsM}
}
\end{table}

In table \ref{RMT, NLsM}  we list for each symmetry class denoted 
by its Cartan label  in the first column,
the time evolution operator in this symmetry class
(penultimate column of table \ref{TableSymmetryClassesTwo})
in the second column,
the target 
spaces \footnote{There are three different varieties of NL$\sigma$M field theories:
supersymmetric, fermionic replica, 
and bosonic replica models. 
In the latter two formulations, the replica limit must be taken,
where one lets the number of replicas, $N$, tend to zero
at the end of the calculations.
While the target spaces of fermionic replica models are compact, those
of bosonic replica models are non-compact.
In the supersymmetric formulation, on the other hand, the target spaces are supermanifolds
with both bosonic and fermionic coordinates.
The fermionic and bosonic replica NL$\sigma$Ms, with $N$ finite,
can be viewed as 
fermion-fermion
and 
boson-boson
subsectors, respectively,  of the corresponding 
supersymmetric NL$\sigma$Ms \cite{Efetov97}.}
of the NL$\sigma$M field theories in the third column,
and in the last column the classifying space appearing in 
K-Theory \cite{KitaevLandau100Proceedings}.

Here
we would like to re-emphasize the remarkable fact that the
{\it same} ten Cartan symmetric
spaces describe all three objects listed in the three last column
of  table~\ref{RMT, NLsM}.
Furthermore, there are remarkable relations between these 
columns.  
First, the second column (``time evolution op.'') can be obtained 
from the last column
(``classifying space'')
by shifting the entries in the last column down by one entry
(modulo eight, and
modulo two
in the ``real'' and ``complex'' cases, 
respectively).
Second, the third
column (``NL$\sigma$M target space'') is obtained
from the last column 
by performing a reflection (modulo 8) in the last column 
about the entry in the row labeled D (and an exchange for the complex case).
Consequently, there is then  
also a resulting relationship between the second and the 
third columns \cite{MirlinEtAl2009-09}.

As reviewed in 
subsection \ref{Review-Classification-TopIns} of the introduction,
the classifying spaces listed in the last column of
table \ref{RMT, NLsM} describe topological insulators (superconductors)
in zero dimensions, i.e., at one point in space\footnote{
Ref.\ \cite{KitaevLandau100Proceedings}; compare also
table III (second column, with $k=0$) of the first article of \cite{SchnyderRyuFurusakiLudwig2008}.}.
The disconnected components of this space of Hamiltonians $Q$
(which cannot be continuously deformed into each other)
are labeled by $\mathbb{Z}$ or $\mathbb{Z}_2$
appearing in the $d=0$ column of table
\ref{periodic table}, which denotes the list of zeroth homotopy
groups of the spaces in the last column of table \ref{RMT, NLsM}.
All higher homotopy groups of the same spaces can then be inferred
from table \ref{homotopy} due to the 
dimensional periodicity and shift
properties visible in that table.

\section{Spinor representations of $\mathrm{SO}(N)$}
\label{app: spinors}

In this appendix we review 
spinor representations of $\mathrm{SO}(N)$
and their properties under 
reality conditions
\cite{Polchinski,Georgi}.
It is most convenient to discuss spinor representations
in terms of Clifford algebras, i.e.,  in terms of gamma matrices 
$\{\Gamma^a_{(N)}\}_{a=1,\ldots,N}$ satisfying
$
\{ \Gamma^j_{(N)}, \Gamma^k_{(N)}  \}
= 2 \delta_{jk}, 
$
with $j,k=1, \ldots, N$. Given such a set of gamma matrices
a spinor representation of $\mathrm{SO}(N)$ can be readily obtained 
\begin{eqnarray}
M_{jk}
=
 - \frac{ \rmi }{4 } \left[ \Gamma^{j}_{(N)}, \Gamma^{k}_{(N)} \right] ,
\end{eqnarray}
with the $\mathrm{SO}(N)$ generators $M_{jk}$.

\subsection{Spinors of $\mathrm{SO}(2n+1)$}

In what follows we will focus  on $\mathrm{SO}(2n+1)$,
which has a $2^n$-dimensional irreducible spinor representation.
The gamma matrices in the Dirac representation for $N=2n+1$
are defined recursively by
\begin{eqnarray}
\Gamma^a_{(2n+1)} 
&=&
\Gamma^a_{(2n-1)} 
\otimes \sigma_3,
\quad
a=1,\cdots, 2n-2,
\nonumber \\
\Gamma^{2n-1}_{(2n+1)} 
&=&
I_{2^{n-1}}
\otimes \sigma_1,
\nonumber \\
\Gamma^{2n}_{(2n+1)} 
&=&
I_{2^{n-1}}
\otimes \sigma_2,
\nonumber \\
\Gamma^{2n+1}_{(2n+1)}
&=&
(-\rmi)^n
\Gamma^1_{(2n+1)}
\Gamma^2_{(2n+1)}
\cdots
\Gamma^{2n}_{(2n+1)},
\end{eqnarray}
where $I_{2^{n-1}}$ is the $2^{n-1}\times 2^{n-1}$ identity matrix. 
[The gamma matrices in 
the Dirac representation 
for $N=2n$ can be constructed by just
leaving out  $\Gamma^{2n+1}_{(2n+1)}$, i.e., 
$ \Gamma^{a}_{(2n)}=\Gamma^{a}_{(2n+1)}$,
with $a=1,\cdots, 2n$.]
To be more explicit,
\begin{eqnarray}
\Gamma^1_{(2n+1)} 
&=&
\sigma_1 \otimes 
\underbrace{\sigma_3 \otimes \cdots \otimes \sigma_3}_{n-1},
\nonumber \\
\Gamma^2_{(2n+1)} 
&=&
\sigma_2 \otimes 
\underbrace{\sigma_3 \otimes \cdots \otimes \sigma_3}_{n-1},
\nonumber \\
\Gamma^3_{(2n+1)} 
&=&
\sigma_0 \otimes \sigma_1 \otimes 
\underbrace{
\sigma_3 \otimes \cdots \otimes \sigma_3}_{n-2},
\nonumber \\
\Gamma^4_{(2n+1)} 
&=&
\sigma_0 \otimes \sigma_2 \otimes 
\underbrace{
\sigma_3 \otimes \cdots \otimes \sigma_3}_{n-2},
\nonumber \\
 && \vdots 
\nonumber \\
\Gamma^{2n-1}_{(2n+1)} 
&=&
\underbrace{
\sigma_0 \otimes \cdots  \otimes  \sigma_0}_{n-1}
\otimes 
\sigma_1,
\nonumber \\
\Gamma^{2n}_{(2n+1)} 
&=&
\underbrace{
\sigma_0 \otimes \cdots  \otimes  \sigma_0}_{n-1}
\otimes 
\sigma_2,
\end{eqnarray}
and
\begin{eqnarray}
\Gamma^{2n+1}_{(2n+1)} 
&=&
\underbrace{
\sigma_3 \otimes \cdots  \otimes  \sigma_3}_{n}.
\end{eqnarray}

From the explicit construction of the gamma matrices, we infer that
$\Gamma^{1,3,\cdots,2n+1}_{(2n+1)}$ 
are all real, and
$\Gamma^{2,4,\cdots,2n}_{(2n+1)}$ are purely imaginary.
In order to implement discrete symmetries on the 
space of Dirac Hamiltonians
we define the matrices
\begin{eqnarray} \label{def B1 and B2}
B^1_{(2n+1)}
&:=&
\Gamma^1_{(2n+1)}
\Gamma^3_{(2n+1)}
\cdots
\Gamma^{2n-1}_{(2n+1)},
\nonumber \\
B^2_{(2n+1)}
&:=&
\Gamma^2_{(2n+1)}
\Gamma^4_{(2n+1)}
\cdots
\Gamma^{2n}_{(2n+1)}.
\end{eqnarray}
By use of the reality and anti-commutation properties 
of the gamma matrices, one finds that
\begin{eqnarray}
\left[
B^{1}_{(2n+1)} 
\right]^*
B^{1}_{(2n+1)}
&=&
(-1)^{n(n-1)/2},
\nonumber \\
\left[
B^{2}_{(2n+1)} 
\right]^*
B^{2}_{(2n+1)}
&=&
(-1)^{n(n+1)/2}.
\label{eq: condition on B1 and B2}
\end{eqnarray}
The operator $B^2_{(2n+1)}$ is used to construct
Majorana (real) representations of 
$\mathrm{SO}(2n+1)$. 
For odd $N=2n+1$,
they are possible when \cite{Polchinski,Georgi}
\begin{eqnarray} \label{majorana condition}
&&
\left[
B^{2}_{(2n+1)}
\right]^*
B^{2}_{(2n+1)}
=
(-1)^{n(n+1)/2}
=
1
\nonumber \\
&\Longrightarrow \quad &
n=0, 3
\quad
\mathrm{mod} \quad 4 
\quad
\left(
N=2n+1=1,7
\quad
\mathrm{mod} \quad 8 
\right) .
\end{eqnarray}
Observe that when $n=\mbox{even}$
all gamma matrices can be made real and symmetric,
whereas when $n=\mbox{odd}$,
they can be made purely imaginary and skew-symmetric.

For later use, we also introduce 
\begin{eqnarray}\label{def Btilde}
B^{12}_{(2n+1)} 
&:=&
\left[
B^2_{(2n+1)} \right]^{-1}
B^{1}_{(2n+1)}
=
(- \mathrm{i} )^n 
\Gamma^{2n+1}_{(2n+1)},
\nonumber \\
\tilde{B}^{1}_{(2n+1)} &:=&  B^1_{(2n+1)}\Gamma^{2n}_{(2n+1)},
\nonumber \\
\tilde{B}^2_{(2n+1)}  &:=&  B^2_{(2n+1)} \Gamma^{2n}_{(2n+1)},
\end{eqnarray}
Note that these matrices satisfy
\begin{eqnarray}
B^{1}_{(2n+1)} \Gamma^{a}_{(2n+1)} 
\left[B^{1}_{(2n+1)}\right]^{-1}
&=&
\left\{
\begin{array}{ll}
(-1)^{n+1} \Gamma^{a *}_{(2n+1)}, & a = 1,\cdots,2n,
\\
(-1)^{n} \Gamma^{a *}_{(2n+1)},& a = 2n+1,
\end{array}
\right.
\nonumber \\
B^{2}_{(2n+1)}
 \Gamma^{a}_{(2n+1)} 
\left[
B^{2}_{(2n+1)}
\right]^{-1}
&=&
(-1)^{n} \Gamma^{a *}_{(2n+1)},
\quad
a = 1,\cdots,2n+1,
\label{eq: gamma and B1 and B2}
\nonumber \\
B^{12}_{(2n+1)} \Gamma^a_{(2n+1)} 
\left[B^{12}_{(2n+1)}\right]^{-1}
&=&
\left\{
\begin{array}{ll}
-\Gamma^{a }_{(2n+1)}, & a = 1,\cdots,2n,
\\
+\Gamma^{a }_{(2n+1)},& a = 2n+1.
\end{array}
\right.
\end{eqnarray}

In the following subsection we will use
the ``$B$-matrices'', equations \re{def B1 and B2} and \re{def Btilde},
as symmetry operators in order
to implement discrete symmetries
on the space of Dirac Hamiltonians.
To identify the character of these discrete symmetries
we first compute
the sign $\eta_{B_{(2n+1)}}$ 
picked up 
by $B_{(2n+1)}$
under transposition, 
$
[B_{(2n+1)}]^T=\eta_{B_{(2n+1)}} 
B_{(2n+1)}
$.
Note that the sign $\eta_{B_{(2n+1)}}$  is independent on the choice
of basis for the gamma matrices
(see Ref.\ \cite{Polchinski}). 
From equations~(\ref{eq: condition on B1 and B2}) and
(\ref{majorana condition})
and by use of the property
$
[B^{2}_{(2n+1)} ]^{\dag}
B^{2}_{(2n+1)} 
=
1
$,
it follows that
\begin{eqnarray}
\eta_{B^2_{(2n+1)}}
&=&
(-1)^{n(n+1)/2}.
\end{eqnarray}
Similarly,
\begin{eqnarray}
\eta_{B^1_{(2n+1)} }
&= & (-1)^{n(n-1)/2}, 
\nonumber \\
\eta_{\tilde{B}^1_{(2n+1)}}
&=&
-(-1)^{n(n+1)/2},
\nonumber \\
\eta_{\tilde{B}^2_{(2n+1)}}
&=&
(-1)^{n(n+3)/2}. 
\end{eqnarray}

\subsection{Discrete symmetries of Dirac Hamiltonians}

Let us now determine the symmetry properties
of the Dirac Hamiltonians $\mathcal{H}^{d=2n+3}_{(2n+3)}(k)$,
$\mathcal{H}^{d=2n+2}_{(2n+3)}(k)$, $\ldots$,
$\mathcal{H}^{d=2n-1}_{(2n+3)}(k,m)$,  
defined in equations (\ref{eq: top Hamiltonian 0}) through
(\ref{eq: Dirac Hamiltonian IV}). We find that these Dirac
Hamiltonians satisfy the following symmetry conditions

\begin{eqnarray}
(0): &&
B^{2}_{(2n+1)} 
\mathcal{H}^{d=2n+1}_{(2n+1)}(k) 
\left[
B^{2}_{(2n+1)} \right]^{-1}
=
(-1)^{n+1} 
\left[
\mathcal{H}^{d=2n+1}_{(2n+1)}(-k)
\right]^*, 
\end{eqnarray}
\begin{eqnarray}
(i): &
B^{1}_{(2n+1)} 
\mathcal{H}^{d=2n}_{(2n+1)}(k,m) 
\left[ B^{1}_{(2n+1)}\right]^{-1}
= (-1)^n 
\left[
\mathcal{H}^{d=2n}_{(2n+1)}(-k,m)
\right]^*,
\end{eqnarray}
\begin{eqnarray}
(ii): &&
B^{1}_{(2n+1)} 
\mathcal{H}^{d=2n-1}_{(2n+1)}(k,m) 
\left[ B^{1}_{(2n+1)}\right]^{-1}
= (-1)^n 
\left[
\mathcal{H}^{d=2n-1}_{(2n+1)}(-k,m)
\right]^*,
\nonumber \\
&&
\tilde{B}^{1}_{(2n+1)}\mathcal{H}^{d=2n-1}_{(2n+1)}(k,m)
\left[
\tilde{B}^{1}_{(2n+1)}
\right]^{-1}
=
(-1)^{n+1}\left[\mathcal{H}^{d=2n-1}_{(2n+1)}(-k,m)\right]^*.
\nonumber \\
\end{eqnarray}
\begin{eqnarray}
(iii): 
&&
B^2_{(2n+1)}
\mathcal{H}^{d=2n-2}_{(2n+1)} (k,m)
\left[B^2_{(2n+1)}\right]^{-1} 
=
(-1)^{n+1}
\left[\mathcal{H}^{d=2n-2}_{(2n+1)} (-k,m)
\right]^*, 
\nonumber \\
\end{eqnarray}
\begin{eqnarray}
(iv): &&
 B^2_{(2n+1)}
\mathcal{H}^{d=2n-3}_{(2n+1)} (k,m)
\left[B^2_{(2n+1)}\right]^{-1} 
=
(-1)^{n+1}
\left[\mathcal{H}^{d=2n-3}_{(2n+1)} (-k,m)
\right]^*, 
\nonumber \\
&&
\tilde{B}^{2}_{(2n+1)}\mathcal{H}^{d=2n-3}_{(2n+1)}(k,m)
\left[
\tilde{B}^{2}_{(2n+1)}
\right]^{-1}
=
(-1)^{n}\left[\mathcal{H}^{d=2n-3}_{(2n+1)}(-k,m)\right]^*.
\nonumber \\
\end{eqnarray}
A few remarks are in order.
\begin{itemize}
\item
The Hamiltonians
$\mathcal{H}^{d=2n-1}_{(2n+1)}(k,m)$
and
$\mathcal{H}^{d=2n-3}_{(2n+1)}(k,m)$
satisfy two different discrete symmetry conditions,
and  are  therefore also left invariant under the combination of these
two symmetries, which defines a chiral symmetry. In other words,
both $\mathcal{H}^{d=2n-1}_{(2n+1)}(k,m)$
and
$\mathcal{H}^{d=2n-3}_{(2n+1)}(k,m)$
anticommute with a unitary matrix
\begin{eqnarray}
\left\{
\mathcal{H}^{d=2n-1}_{(2n+1)}(k,m),
\Gamma^{2n}_{(2n+1)}
\right\}
=
0 ,
\nonumber \\
\left\{
\mathcal{H}^{d=2n-3}_{(2n+1)}(k,m),
\Gamma^{2n-2}_{(2n+1)}
\right\}
=
0.
\end{eqnarray}

\item
The Hamiltonians
$\mathcal{H}^{d=2n}_{(2n+3)}(k,m)$ 
and
$\mathcal{H}^{d=2n-1}_{(2n+3)}(k,m)$
can be made block diagonal because  
\begin{eqnarray}
\left[
\mathcal{H}^{d=2n}_{(2n+3)}(k,m), 
\Gamma^{a}_{(2n+3)}
\Gamma^{b}_{(2n+3)}
\right]
=0,
\nonumber \\
\left[
\mathcal{H}^{d=2n-1}_{(2n+3)}(k,m), 
\Gamma^{a}_{(2n+3)}
\Gamma^{b}_{(2n+3)}
\right]
=0, 
\end{eqnarray}
where
$
(a,b)
=
(2n+1,2n+2)$,
$(2n+2,2n+3)$,
or
$(2n+3,2n+1)$.

\item
In the case of the massive Hamiltonian $\mathcal{H}^{d=2n}_{(2n+1)}(k,m)$,
the parity transformation\footnote{
Parity here simply means $k\to -k$ symmetry.
In odd space time dimensions, it is actually not called parity. 
}
can be implemented by
$B^{12}_{(2n+1)}$ 
\begin{eqnarray}
B^{12}_{(2n+1)} 
\mathcal{H}^{d=2n}_{(2n+1)}(k,m) 
\left[B^{12}_{(2n+1)} \right]^{-1}
= \mathcal{H}^{d=2n}_{(2n+1)}(-k,m).
\end{eqnarray}
By combining the discrete symmetry
$B^1_{(2n+1)}$
and 
the parity  symmetry
$B^{12}_{(2n+1)}$, 
the Hamiltonian also satisfies 
\begin{eqnarray}
B^{2}_{(2n+1)} 
\mathcal{H}^{d=2n}_{(2n+1)}(k,m) 
\left[
B^{2}_{(2n+1)}
\right]^{-1}
= (-1)^n 
\left[
\mathcal{H}^{d=2n}_{(2n+1)}(k,m)
\right]^*.
\end{eqnarray}
This discrete symmetry is unique in that, unlike 
$\mathcal{T}$ or $\mathcal{C}$
which relates Bloch Hamiltonians
at $k$ and $-k$, 
it constrains the form of the Hamiltonian at a given $k$. 
This is nothing but the real/pseudo-real condition
for $\mathrm{SO}(2n+1)$. 
As a consequence, 
the Hamiltonian in $k$-space
can be written as a real/pseudo-real matrix.
Hence,
classifying topological insulators (superconductors)
satisfying the combination 
of $B^1_{(2n+1)}$ and $B^{12}_{(2n+1)}$ 
amounts to classifying 
{\it real} bundles. 
(It is not necessary to satisfy both, though.)
An example of 
a topological  insulator/superconductor satisfying 
the combination of $B^1_{(2n+1)}$ and $B^{12}_{(2n+1)}$
is constructed in appendix~\ref{secParity}.
\end{itemize}

\section{
Topological insulators protected by 
a combination of spatial inversion and 
either time-reversal or charge-conjugation symmetry}
\label{secParity}

In the main text, we have focused on the role of
generic symmetries such as time reversal and charge conjugation,
which are not related to any spatial symmetries. 
However, real systems often have
discrete spatial symmetries, such as 
parity, reflection, discrete rotations, etc.
Hence,  it is meaningful to study 
the restrictions imposed by these symmetries
on the Bloch wavefunctions, and thus on the  
ground state properties.
In this appendix, 
we briefly discuss
topological states that are protected by 
a combination of
spatial inversion and an additional
discrete symmetry, such as
$\mathcal{T}$ or $\mathcal{C}$.

\subsection{Inversion symmetry combined with
another discrete symmetry}

Consider a tight-binding Hamiltonian,
\begin{eqnarray}
H 
&=&
\sum_{r,r'} 
\psi^{\dag}(r)\, 
\mathcal{H}(r,r')\, 
\psi(r'), 
\quad
\mathcal{H}^{\dag}(r',r) =\mathcal{H}(r,r') ,
\end{eqnarray}
where $\psi(r)$ is an $n_f$-component 
fermion operator,
and index $r$ labels the lattice sites.
(The internal indices are suppressed.)
Each block in the single-particle Hamiltonian 
$\mathcal{H}(r,r')$ is a $n_f\times n_f$ matrix,
and we assume the total size of the 
single particle Hamiltonian is
$n_f V\times n_f V$,
where $V$ is the total number of lattice sites. 
The components in $\psi(r)$
can describe, e.g., 
orbitals or spin degrees of freedom,
as well as
different sites within a crystal unit cell
centered at $r$.

Provided the system has 
translational symmetry,
\begin{eqnarray}
\mathcal{H}(r,r') = 
\mathcal{H}(r-r'),
\end{eqnarray}
with periodic boundary conditions in each spatial direction
(i.e., the system is defined on a torus $T^d$),
we can perform the Fourier transformation and obtain 
in momentum space
\begin{eqnarray}
H 
&=&
\sum_{k \in \mathrm{BZ}} 
\psi^{\dag}(k) \,
\mathcal{H}(k)\, 
\psi (k) ,
\end{eqnarray}
where
\begin{eqnarray}
\psi(r)
=
\sqrt{V}^{-1}
\sum_{k\in \mathrm{BZ}}
e^{\rmi k\cdot r}
\psi(k),
\quad
\mathcal{H}(k)
\!\!&=&\!\!
\sum_r 
e^{- \mathrm{i} k\cdot r}
\mathcal{H}(r). 
\end{eqnarray}
[Here, if different components in $\psi(r)$
were to represent different sites
within a unit cell centered at $r$,
we could include
the phase 
$e^{ \mathrm{i}k\cdot (r-r_a)}$
in the definition of $\psi(k)$,
$\psi(k)
\to 
\mathrm{diag} 
\left(
e^{ \mathrm{i}k\cdot (r-r_a)}
\right)_{a=1,\cdots,n_f}
\psi(k)
$
where $r_a$ represents a location within a unit cell.]

In the main text of this paper, we considered topological
insulators (superconductors)
protected by discrete symmetries, 
such as $\mathcal{T}$ and $\mathcal{C}$.
We now focus on the effect of inversion symmetry. 
By definition, an inversion
$\Omega$
relates
fermion operators at $r$ and at $-r$,
\begin{eqnarray} \label{mDEFinversioin}
\Omega 
\,
\psi(r)
\,
\Omega^{-1} = U 
\psi(-r), 
\label{def: inversion}
\end{eqnarray}
where $U$ is a constant $n_f\times n_f$ unitary matrix. 
(It is important to distinguish between inversion
and parity symmetry.
The former means $r\to -r$ for 
any space-time dimension.
Parity transformation in even 
space-time dimension
agrees with inversion,
but in odd space-time dimension,
it is different from inversion.)
One can imagine, for example, 
$U$ represents
a parity eigenvalue of 
each orbital at site $r$ 
($s, p, \ldots,$ orbitals), which can be either $+1$ or $-1$.
Similarly
if different components in $\psi(r)$ represent 
different sites in the unit cell centered at $r$,
inversion transformation sends $r$ to $-r$
and at the same time can reshuffle locations of 
electron operators within a unit cell,
which can be described by 
the unitary matrix $U$.
The invariance of $H$ under $\Omega$ implies 
\begin{eqnarray}
U^{-1} \mathcal{H}(r, r') U = \mathcal{H}(-r,-r'),
\quad
U^{-1} \mathcal{H}(r) U = \mathcal{H}(-r).
\end{eqnarray}
Thus, the Bloch Hamiltonians at $k$ and
$-k$ are unitarily equivalent, 
\begin{eqnarray}
U^{-1}\mathcal{H}(k) U
=
\mathcal{H}(-k).
\end{eqnarray}
To summarize, definition (\ref{def: inversion})
implies  that (i) the spectrum at $k$ is identical to the one at $-k$,
and (ii) the unitary matrix $U$ relating the Hamiltonian at  $k$ to the one at  $-k$
is $k$-independent. 
This form of symmetry 
does not necessarily correspond to
inversion symmetry verbatim,
but might describe an inversion symmetry 
which is realized as a projective symmetry
(e.g., invariance of the Hamiltonian under
inversion followed by a gauge transformation).
The only big assumption here is that 
we have assumed $U$ is $k$-independent.

\begin{table*}[t]
\begin{center}
\begin{tabular}{c|c||c||c|c||c|cc}\hline
&Cartan & S & T$=-1$ & T$=+1$ & C$=-1$ & C$=+1$ 
 \\ \hline
$(\epsilon_V,\eta_V)$ & & & $(1,-1)$ &$(1,1)$ &$(-1,-1)$ & $(-1,1)$
\\\hline\hline
standard
&A& $\times$ & $\times$ &$\times$ & $\times$ & $\times$ \\ \cline{2-7}
&AI & $\times$ & $\times$ & $\bigcirc$ & $\times$ & $\times$ \\ \cline{2-7}
&AII& $\times$ & $\bigcirc$ & $\times$ & $\times$ &$\times$ \\ \hline\hline
chiral &AIII& $\bigcirc$ & $\times$ &$\times$ & $\times$ & $\times$ \\ \cline{2-7}
(sublattice) 
&BDI& $\bigcirc$ & $\times$ & $\bigcirc$& $\times$ & $\bigcirc$ \\ \cline{2-7}
&CII& $\bigcirc$ & $\bigcirc$& $\times$ & $\bigcirc$ & $\times$ \\\hline\hline
BdG 
& D& $\times$ & $\times$ & $\times$ & $\times$ &$\bigcirc$ \\ \cline{2-7}
& C& $\times$ & $\times$ & $\times$ & $\bigcirc$ & $\times$ \\ \cline{2-7}
& DIII& $\bigcirc$ & $\bigcirc$& $\times$ & $\times$ & $\bigcirc$ \\ \cline{2-7}
& CI& $\bigcirc$ & $\times$ & $\bigcirc$&$\bigcirc$ & $\times$ \\ \hline
\end{tabular}
\caption{
\label{tab: rmt}
The ten  generic symmetry classes of single-particle Hamiltonians
classified in terms of 
the presence ($\bigcirc$) or absence ($\times$)
of the time-reversal symmetry  ($\epsilon_V=+1$)
with T$=-1$ ($\eta_V=-1$)
and T$=+1$ ($\eta_V=+1$),
and the particle-hole symmetry ($\epsilon_V=-1$)
with C$=-1$ ($\eta_V=-1$)
and C$=+1$ ($\eta_V=+1$),
as well as chiral (or sublattice) symmetry denoted by S;
see table \ref{TableSymmetryClassesTwo}.
For the  notation of these symmetry classes in
terms of $(\epsilon_V,\eta_V)$, see the main text 
and Ref.\ \cite{SchnyderRyuFurusakiLudwig2008}.
\label{fig:epsilon-eta}
}
\end{center}
\end{table*}

We now combine 
the inversion symmetry with
another discrete symmetry,
\begin{eqnarray}
V^{-1}\mathcal{H}(k)V
=
\epsilon_{V}
\mathcal{H}(-k)^T,
\quad
\epsilon_{V}= \pm 1,
\nonumber \\
\mbox{with}
\quad
V^T = \eta_V V,
\quad
\eta_V =\pm 1.
\end{eqnarray}
As before we distinguish four different cases
[see section \ref{Review-Classification-Hamiltonian}, and in particular equations \re{DEFTimeReversal} and \re{DEFChargeConjugation}]:
$(\epsilon_V,\eta_V)=(1,1)$
corresponds to a time-reversal symmetry with 
 T$=+1$,
$(\epsilon_V,\eta_V)=(1,-1)$
denotes a time-reversal symmetry with
 T$=-1$,
$(\epsilon_V,\eta_V)=(-1,1)$
is a particle-hole symmetry with
 C$=+1$, and
$(\epsilon_V,\eta_V)=(-1,-1)$
is a particle-hole symmetry with C$=-1$.
These four cases together with the corresponding ``Cartan label'' 
are summarized in  table \ref{fig:epsilon-eta}. 
Then, for the combined transformation, 
\begin{eqnarray}
W^{-1} \mathcal{H}(k) W = \epsilon_W
\mathcal{H}(k)^T,
\quad
W := UV, 
\nonumber \\
\mbox{with}
\quad 
\epsilon_W = \epsilon_V. 
\end{eqnarray}
Iterating this transformation twice, 
\begin{eqnarray}
\mathcal{H}(k)
\to
W \mathcal{H}(k)^T W^{-1}
\to
W (W^{-1})^T \mathcal{H}(k) W^T W^{-1},
\end{eqnarray}
which suggests, from Schur's lemma, 
$W^T W^{-1} \propto I_{n_f}$.
Hence,
as before, for $W$, we distinguish the following two cases,
\begin{eqnarray}
W^{T} 
= \eta_W W,
\quad
\eta_W = \pm 1. 
\end{eqnarray}
The signature $\eta_W$ is invariant 
under unitary transformations.
It depends on the signature $\eta_V$,
the signature $\eta_U$ for 
the inversion $U^T=\eta_U U$,
and the commutation relation of $U$ and $V$. 
Here, note that 
the signature $\eta_U$ for inversion alone
is not invariant under unitary transformation,
while the signature $\eta_W$ is invariant. 
As an illustration, 
let us consider spinless fermions.
Time-reversal symmetry $\mathcal{T}$ for spinless fermions implies
\begin{eqnarray}
\mathcal{H}(i,j)^*=\mathcal{H}(i,j) .
\end{eqnarray}
Thus, 
when inversion and $\mathcal{T}$ are combined, 
\begin{eqnarray}
\mathcal{H}^*(r) = W \mathcal{H}(-r) W^{-1},
\quad
\mathcal{H}^*(k) = W^{-1} \mathcal{H}(k) W,
\quad
\mbox{where} 
\quad W=U.
\end{eqnarray}
In this example and in this basis, 
the signature of $W$ solely depends on the signature of $U$,
$\eta_W=\eta_U$.

Interesting examples are provided by Dirac Hamiltonians. 
For a massive Dirac Hamiltonian, 
the mass matrix itself can be taken as
 the unitary matrix $U$
which was introduced in equation \re{mDEFinversioin}.
Put differently,
any massive Dirac Hamiltonian
possesses a $U$-type symmetry. 
These $U$-type symmetries can be, however, different from 
the inversion symmetry which is imposed 
at the microscopic level. 
However, if one starts from a microscopic lattice model
with inversion symmetry, 
and arrives at a Dirac Hamiltonian with a mass in the continuum,
inversion symmetry in the continuum must be implemented by
a mass matrix. 

First, consider Hamiltonian \re{eq: Dirac Hamiltonian I}
\begin{eqnarray}
(i):\quad &&
\mathcal{H}^{d=2n}_{(2n+1)}(k,m)
= 
\sum_{a=1}^{d=2n} k_a \Gamma^a_{(2n+1)}
+ 
m \Gamma^{2n+1}_{(2n+1)}.
\end{eqnarray}
This Hamiltonian has a $V$-type symmetry with
\begin{eqnarray}
V=B^{1}_{(2n+1)},
\quad
\epsilon_V = (-1)^n, 
\quad
\eta_V 
=
\eta_{B^1_{(2n+1)} }
= (-1)^{n(n-1)/2}.
\end{eqnarray}
There is inversion symmetry represented by 
\begin{eqnarray}
U = \Gamma^{2n+1}_{(2n+1)},
\quad
\eta_U = +1. 
\end{eqnarray}
The combined transformation
$W$ is given by
\begin{eqnarray}
W =B^{2}_{(2n+1)} ,
\quad
\epsilon_W = (-1)^n,
\quad 
\eta_W
=
\eta_{B^{2}_{(2n+1)}}
=
(-1)^{n(n+1)/2}. 
\end{eqnarray}

Second, we consider the massive Dirac Hamiltonian \re{eq: Dirac Hamiltonian III}
\begin{eqnarray}
&&
(iii): 
\quad
\mathcal{H}^{d=2n-2}_{(2n+1)}(k,m)
=
\sum_{a=1}^{d=2n-2} k_a \Gamma^a_{(2n+1)}
+
 m 
\mathcal{M}, 
\nonumber \\
&&
\qquad
\mbox{with}
\quad 
\mathcal{M}
:=
\mathrm{i}
\Gamma^{2n+1}_{(2n+1)}
\Gamma^{2n}_{(2n+1)}
\Gamma^{2n-1}_{(2n+1)}. 
\end{eqnarray}
The $V$-type discrete symmetry for this is 
\begin{eqnarray}
V = B^2_{(2n+1)},
\quad
\epsilon_V =(-1)^{n+1},
\quad
\eta_V = 
\eta_{B^{2}_{(2n+1)}}
=
(-1)^{n(n+1)/2}. 
\end{eqnarray}
The inversion symmetry is represented by
\begin{eqnarray}
U =\mathcal{M},
\quad
\eta_U = +1.
\end{eqnarray}
Thus, the combined transformation is 
\begin{eqnarray}
W 
=
\mathcal{M}
B^2_{(2n+1)}
=
\mathrm{i}
\Gamma^{2n+1}_{(2n+1)}
\Gamma^{2n}_{(2n+1)}
\Gamma^{2n-1}_{(2n+1)}
\times 
\Gamma^2_{(2n+1)}
\Gamma^4_{(2n+1)}
\cdots
\Gamma^{2n}_{(2n+1)}.
\nonumber \\
\mbox{with}
\quad
\epsilon_W= (-1)^{n+1},
\quad
\eta_W =
(-1)^{n(n+1)/2}.
\end{eqnarray}
where we have noted 
$UV = VU$.

\subsection{
Classification of gapped Hamiltonians
in the presence of 
a combination of spatial inversion
and 
either
time-reversal
or charge-conjugation symmetry}

We now classify the ground states of gapped Hamiltonians
protected by a $W$-type symmetry. 
We distinguish four different cases, 
$(\epsilon_W, \eta_W)= (\pm 1, \pm 1)$.

Let us first 
consider the case of
$(\epsilon_W, \eta_W)= (1, 1)$,
in which case $W$ is a symmetric unitary matrix,
$W=W^T$ ($\eta_W=+1$). 
Such a matrix is an element
of 
$\mathrm{U}(n_f)/\mathrm{O}(n_f)$.
Any element in
$\mathrm{U}(n_f)/\mathrm{O}(n_f)$
can be written as 
$W=XX^T$ where $X$ is a unitary matrix. 
One can then 
take a basis in which  
the Bloch Hamiltonian 
for any given $k$
is a real symmetric matrix.
In other words, if we define
$\tilde{\mathcal{H}}:= X^{\dag}
\mathcal{H} X$,
then
$\tilde{\mathcal{H}}^*=
\tilde{\mathcal{H}}$. 
Thus, when $(\epsilon_W,\eta_W)=(+1,+1)$,
the Hamiltonian is real symmetric and hence 
Bloch wavefunctions can be taken real at each $k$. 
We assume there are $N_-$ ($N_+$)
occupied (unoccupied) Bloch wavefunctions 
for each $k$
with $N_{+} + N_{-} = N_{\mathrm{tot}} (=n_f)$.
The spectral projector 
onto the filled Bloch states
or the ``$Q$-matrix'' in this case 
can be viewed as an element of
the real Grassmannian
$G_{N_-, N_++N_-}(\mathbb{R})=
\mathrm{O}(N_++N_- )/[\mathrm{O}(N_+)\times \mathrm{O}(N_-)]$. 
For a given system,
the ``$Q$-matrix'' defines
a map from the BZ onto 
the real Grassmannian.
Hence,  classifying topological classes of
band insulators 
amounts to
counting the number of
topologically inequivalent
mappings from the BZ to
$G_{N_-, N_++N_-}(\mathbb{R})$.
Mathematically, this is given by the homotopy 
group of the space of projectors (i.e., of the real Grassmannian in the present case),
which can be read off from table~\ref{homotopy}.

The same reasoning can be repeated for the case
$(\epsilon_W,\eta_W)=(+1,-1)$. One finds that the
space of projectors is the quaternionic Grassmannian
$G_{N_-, N_++N_-}(\mathbb{H})=
\mathrm{Sp}(N_++N_- )/[\mathrm{Sp}(N_+)\times \mathrm{Sp}(N_-)]$,
whose homotopy group is again given in table~\ref{homotopy}. 

When 
$(\epsilon_W,\eta_W)=(-1,+1)$,
the Hamiltonian can be taken, in a suitable basis, 
real and skew symmetric, i.e., an element of $\mathrm{so}(n_f)$. 
Any element of $\mathrm{so}(n_f)$ 
can be transformed into a canonical form by an orthogonal transformation.
For the $Q$-matrix,
its canonical form $\tilde{Q}$ is a matrix whose matrix elements 
are $\pm 1$, 
\begin{eqnarray}
Q =
S \tilde{Q} S^T ,
\quad
S \in \mathrm{SO}(n_f), 
\nonumber \\
\mbox{where}
\quad
\tilde{Q} =
\left(
\begin{array}{cccccc}
0  & +1 &  & &  & \cdots \\
-1 & 0 &  &  & &  \\
  &  & 0 &+1 &  &  \\
 &  & -1 & 0 &  &  \\
\vdots &   & & &  & \ddots \\
\end{array}
\right),
\end{eqnarray}
where we have assumed $n_f$ is an even integer. 
We thus identified the space of projectors as 
$\mathrm{SO}(n_f)/\mathrm{U}(n_f/2)$. 
Finally, when 
$(\epsilon_W,\eta_W)=(-1,-1)$,
we identify the space of projectors as 
$\mathrm{Sp}(n_f)/\mathrm{U}(n_f)$. 

To summarize, we have listed in table \ref{inversion and one discrete symmetry},
for each of the four cases of $W$-type symmetries, 
the space of projectors ``$G/H$'' together with their homotopy
groups $\pi_d ( G/ H )$, which yields the classification of
topological insulators (superconductors) protected by
a combination of inversion and an additional discrete symmetry.
Notice, 
we focus here on the implication of
the combination of inversion and a discrete symmetries,
but we do not require each symmetry
separately.

\begin{table}
\begin{center}
\begin{tabular}{clccccccccc}\hline\hline
 & &
\multicolumn{9}{c}{$d$ }   \\  \cline{3-11}
$(\epsilon_W,\eta_W)$ & Projectors  & $0$ & $1$ & $2$ & $3$ &$4$ & $5$ & $6$ & $7$  & $ \cdots$ \\ \hline 
$(+1,-1)$ & $\mathrm{Sp}(N+M)/\mathrm{Sp}(N)\times \mathrm{Sp}(M)$ 
               &	 $\mathbb{Z}$ & 0& 0 & 0 & $\mathbb{Z}$ & $\mathbb{Z}_2$  & $\mathbb{Z}_2$ & 0   & $\cdots$\\  
$(-1,+1)$ & $\mathrm{O}(2N)/\mathrm{U}(N)$ 
               & $\mathbb{Z}_2$ & 0 & $ \mathbb{Z}$ & 0 & 0 & 0  & $\mathbb{Z}$ & $\mathbb{Z}_2$  & $\cdots$\\  
$(+1,+1)$ & $\mathrm{O}(N+M)/\mathrm{O}(N)\times \mathrm{O}(M)$ 
               & $\mathbb{Z}$ &$\mathbb{Z}_2$ & $\mathbb{Z}_2$ & 0 & $ \mathbb{Z}$ &0  & 0 & 0   & $\cdots$\\ 
$(-1,-1)$  & $\mathrm{Sp}(N)/\mathrm{U}(N)$                
              & 0 & 0 &  $\mathbb{Z}$ & $\mathbb{Z}_2$ & $\mathbb{Z}_2$ & 0   & $\mathbb{Z}$ & 0 & $\cdots$ \\  \hline \hline
\end{tabular}
\end{center}
\caption{
Classification of  topological insulators 
(superconductors) protected by
a combination of spatial inversion and an additional discrete symmetry.
\label{inversion and one discrete symmetry}
}
\end{table}

\subsubsection{Example: 
$(\epsilon_W,\eta_W) =(1,1)$ 
in $d=2$.}

We now specialize to the case 
of $(\epsilon_W,\epsilon_W)=(1,1)$.
The relevant space of the projection operators is
the real Grassmannian
$G_{N_+,N_++N_-}(\mathbb{R})$
and the projection operator 
defines a map from the BZ onto 
the real Grassmannian.
In particular,
when $d=2$,
the projector $Q(k_x,k_y)$ defines a
map from $S^2$ or $T^2$ onto 
$G_{N_+,N_-+N_+}(\mathbb{R})$.
The homotopy group 
$\pi_2[G_{N_+,N_-+N_+}(\mathbb{R})]=\mathbb{Z}_2$ tells us that
the space of quantum ground states is partitioned into two
topologically distinct sectors.  

A representative $Q$-field configurations
which is a non-trivial element of
$\pi_2[
G_{N_+,N_-+N_+}(\mathbb{R})
]=\mathbb{Z}_2$ 
can be constructed following Ref.\ \cite{Weinberg84}.
For simplicity, take the case where we have four bands in total
and two filled bands;
take $N_+=N_-=2$.
Then such a representative $Q$-matrix is given by 
\begin{eqnarray}
Q^{(l)}(k) =
\left(
\begin{array}{cccc}
n^{(l)}_z & 0 & -n^{(l)}_x & -n^{(l)}_y
\\
0 & n^{(l)}_z & n^{(l)}_y & -n^{(l)}_x
\\
-n^{(l)}_x & n^{(l)}_y & -n^{(l)}_z &  0
\\
-n^{(l)}_y &  -n^{(l)}_x &  0 & - n^{(l)}_z 
\end{array}
\right)
\nonumber \\
\mbox{with}
\quad 
n^{(l)}(k)
:= 
\left(
\begin{array}{ccc}
n^{(l)}_x, n^{(l)}_y, n^{(l)}_z
\end{array}
\right) 
\nonumber \\
\qquad \qquad \qquad =
\left(
\begin{array}{ccc}
\cos (l \phi) \sin (\theta),  &
\sin (l \phi) \sin (\theta),  &
\cos \theta
\end{array} 
\right).  
\end{eqnarray}
where $l\in \mathbb{Z}$ and
$\theta$ and $\phi$
is spherical coordinates parameterizing 
the BZ $\simeq S^2$. 
The projector is an non-trivial element of 
the second homotopy group
when $l$ is odd,
while it is trivial when $l$ is even.
Observe that the normalized vector $n^{(l)}(k)$ itself 
defines a map from $S^2$ onto $S^2$, 
and wraps $S^2$ integer ($=l$) times.

We can construct a Hamiltonian which has 
$Q^{(l)}(k)$ as a projector.
We consider 
the following four band tight-binding Hamiltonian 
on the square lattice:
\begin{eqnarray}
H &=&
\sum_{r}
\left[
\psi^{\dag}(r)\, h_0 \, \psi(r)
\right.
\nonumber \\
&&
\quad
\left. 
+
\psi^{\dag}(r) \, h_x \, \psi(r+\hat{x})
+
\mathrm{h.c.}
+
\psi^{\dag}(r) \, h_y \, \psi(r+\hat{y})
+
\mathrm{h.c.}
\right],
\end{eqnarray}
where
$\hat{x}=(1,0)$
and
$\hat{y}=(0,1)$,
respectively, and 
\begin{eqnarray}
h_0 &=&
-\mu
\left(
\begin{array}{cc}
\sigma_0  &  0 \\
0 & -\sigma_0
\end{array}
\right),
\quad
h_x
=
\left(
\begin{array}{cc}
t \sigma_0 &  \Delta \sigma_y \\
-\Delta \sigma_y & -t \sigma_0
\end{array}
\right),
\nonumber \\
h_y
&=&
\left(
\begin{array}{cc}
t \sigma_0 &  -\mathrm{i}\Delta \sigma_0 \\
-\mathrm{i}\Delta \sigma_0 & -t \sigma_0
\end{array}
\right),
\label{eq: H with BO}
\end{eqnarray}
and 
$t,\Delta,\mu\in\mathbb{R}$ are a parameter of the model.
In $k$-space,
\begin{eqnarray}
\mathcal{H}(k)
&=&
\left(
\begin{array}{cccc}
R_z & 0 & -R_x & -R_y \\
0 & R_z & R_y & -R_x \\
-R_x & R_y & -R_z & 0 \\
-R_y & -R_x & 0 & -R_z  \\
\end{array}
\right)
\label{eq: H with BO in k-space}
\end{eqnarray}
where
\begin{eqnarray}
R(k)
=
\left(
\begin{array}{ccc}
- 2\Delta \sin k_y,  &
- 2\Delta \sin k_x, &
2t \left(\cos k_x + \cos k_y\right) -\mu
\end{array}
\right). 
\end{eqnarray}

We will set $t=\Delta=1$ henceforth. 
As we change $\mu$, there are four phases separated by 
quantum phase transitions at $\mu=\pm 4$ and $\mu=0$.
When 
$|\mu| >4 $, 
the normalized vector $R(k)/|R(k)|$ does not wrap $S^2$
as we sweep the momentum 
and the system is in a trivial phase.
On the other hand, for
$|\mu| <4 $
the normalized vector $R(k)/|R(k)|$ wraps the sphere $S^2$
$n=+1$ times (or $n=-1$ times), and hence the system is in a non-trivial phase.
The projector constructed from 
(\ref{eq: H with BO in k-space})
is topologically equivalent to
$Q^{(l)}(k)$ with $l$ odd.

\begin{figure}
\begin{center} 
\includegraphics[width=0.9\textwidth,clip]{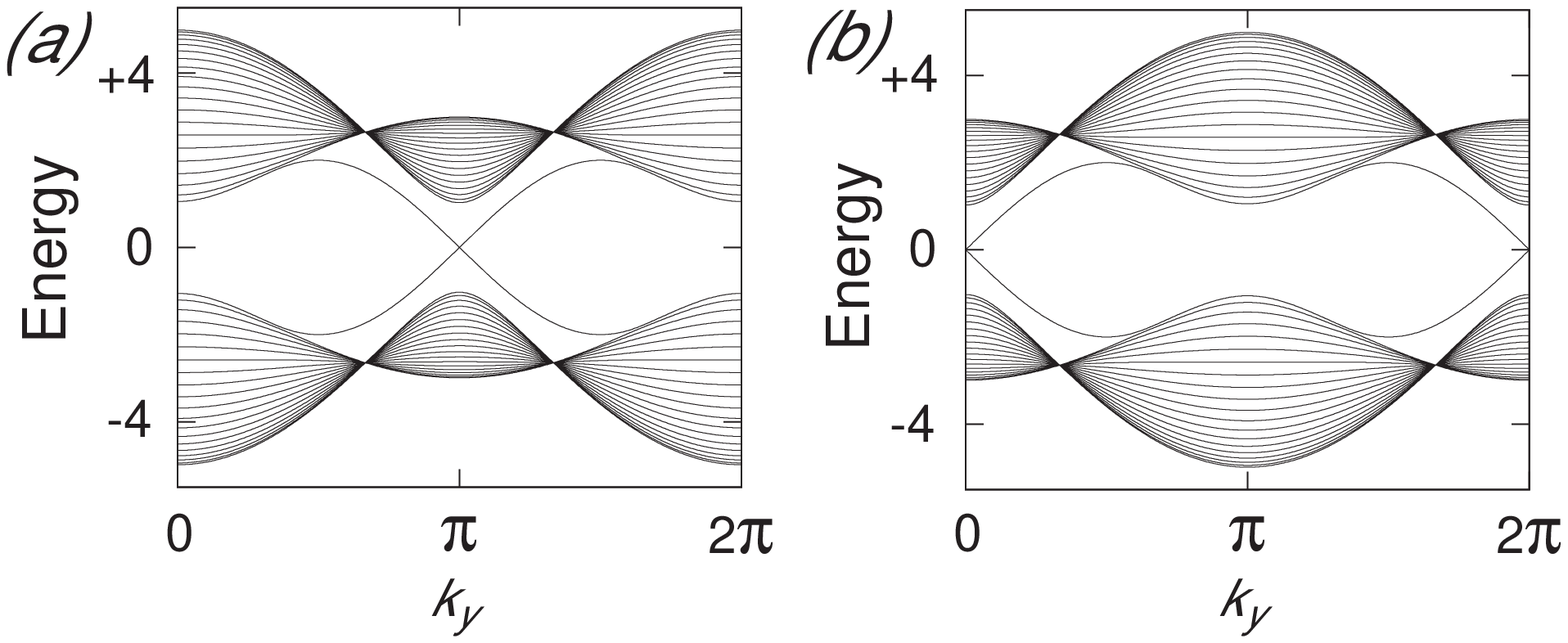}
\caption{
\label{fig: p-wave_edge.eps}
The energy spectrum with $t=\Delta=1$
and $\mu=-1$ [panel (a)] and
$\mu=+1$ [panel (b)].
}
\end{center}
\end{figure}

For topological insulators and superconductors 
protected by $\mathcal{T}$, $\mathcal{C}$,
or combination of both,
a diagnostics of bulk topological character is 
the existence of gapless edge modes. 
For insulators and superconductors 
with an inversion symmetry,
it is less clear if looking for edge modes
is useful to characterize the bulk topological character
since boundary breaks inversion symmetry. 
[Nevertheless, as discussed recently in 
Ref.\ \cite{Turner09}, 
for the entanglement entropy spectrum, 
an inversion symmetry can protect the gapless 
entanglement entropy spectrum.] 
For the present case, however, 
the Hamiltonian has several accidental symmetries (see below),
in particular $\mathcal{T}$ (odd), 
which may protect the gapless nature of edge states,
if they exist, and if the 
number of Kramers pairs  is odd. 
In Fig.\ \ref{fig: p-wave_edge.eps},
the energy spectrum for $\mu=\pm 1$
in the slab geometry is presented;
here two edges along $y$-direction 
(located at $x=0$ and $x=N_x$)
are created, and the energy eigenvalues are plotted
as a function of $k_y$.
Each eigenvalue is doubly degenerate for each $k_y$. 
For $\mu=\pm 1$, there are four edge modes,
two of which localized at $x=0$ and
the other two at $x=N_x$,
while for $|\mu|>4$ there is no edge mode.

The only symmetry we have assumed so far
for ensembles of Hamiltonians
is a $W$-type symmetry
with $(\epsilon_W,\eta_W)=(1,1)$.
The representative Hamiltonian
(\ref{eq: H with BO}-\ref{eq: H with BO in k-space}), 
however, accidentally has 
several other discrete symmetries.
The Hamiltonian is invariant under the following 
discrete symmetries: Time-reversal symmetries $\mathcal{T}$
\begin{eqnarray}
A^{-1} 
\mathcal{H}(-k)^*
A
=
\mathcal{H}(k),
\quad
A=
\tau_z\sigma_0 
\quad  \mbox{or} \quad 
\tau_z\sigma_y, 
\nonumber
\end{eqnarray}
inversion symmetries $\mathcal{I}$
\begin{eqnarray}
B^{-1} 
\mathcal{H}(-k)
B
=
\mathcal{H}(k),
\quad
B=
\tau_z\sigma_0
\quad  \mbox{or} \quad 
\tau_z\sigma_y, 
\nonumber
\end{eqnarray}
and charge-conjugation symmetries $\mathcal{C}$
\begin{eqnarray}
C^{-1} 
\mathcal{H}(-k)^*
C
=
- \mathcal{H}(k),
\quad
C=
\tau_x\sigma_x
\quad  \mbox{or} \quad  
\tau_x\sigma_z.
\nonumber
\end{eqnarray}

Although accidental,
the presence of these symmetries 
allows us to interpret the Hamiltonian 
in many different ways. 
For example, 
since the Hamiltonian is invariant under
$\mathcal{T}$,
$A=\tau_z\sigma_y$ with $A^T=-A$, 
it is a member of symmetry class AII.
Furthermore, it supports two branches of modes per edge
which are counter propagating. 
The representative Hamiltonian so constructed is thus
a $\mathbb{Z}_2$ topological insulator in class AII
(i.e., it is in the QSH phase). 
With further imposing a $\mathcal{C}$,
$C=\tau_x\sigma_z$
with $C^T=+C$, 
the single particle Hamiltonian can be 
interpreted as a member of symmetry class DIII.
Again, it is a non-trivial $\mathbb{Z}_2$ topological 
superconductor in class DIII. 
Thus, non-trivial $\mathbb{Z}_2$ topological insulators (AII) and
superconductors (DIII)
in $d=2$ dimensions
can be interpreted,
somewhat accidentally,
a non-trivial element of 
$\pi_2[
G_{N_+,N_-+N_+}(\mathbb{R})
]=\mathbb{Z}_2$. 
To implement a $W$-type symmetry,
$\mathcal{T}$
for symmetry classes AII and DIII ($A=\tau_z\sigma_y$ with $A^T=-A$)
can be combined with an inversion symmetry
represented by $B=\tau_z\sigma_y$. 
This inversion symmetry is, however, an artificial symmetry in the sense that 
it flips spin.


\section*{References}
\begin{thebibliography}{99}


\bibitem{Thouless82}
Thouless D J,
Kohmoto M,
Nightingale P,
and den Nijs M 1982
\textit{Phys.\ Rev.\ Lett.} \textbf{49} 405

\bibitem{Kohmoto}
Kohmoto M 1985
\textit{Ann.\ Phys.\ (N.Y.)} \textbf{160} 355

\bibitem{bellissard1994}
Bellissard J, van Elst A, and Schulz-Baldes H
\textit{J.\ Math.\ Phys.} \textbf{35}, 5373 (1994)


\bibitem{Panati}
Panati G 2007 
\textit{Ann.\ Henri Poincare}
\textbf{8} 995

\bibitem{Brouder}
Brounder C, Panati G, Calandra M,
Mourougane C, and Marzari N 2007
\textit{Phys.\ Rev.\ Lett.}
\textbf{98} 046402

\bibitem{Hastings}
Hastings M B 2008
\textit{J.\ Stat.\ Mech.} L01001 

\bibitem{IQHEentropy}
Ryu S and Hatsugai Y 2006
\textit{Phys.\ Rev.\ B} \textbf{73}, 245115

\bibitem{Read00}
Read N and Green D 2000
\textit{Phys.\ Rev.\ B} \textbf{61} 10267

\bibitem{KaneMele}
Kane C L and Mele E J 2005
\textit{Phys.\ Rev.\ Lett.} \textbf{95} 146802;
Kane C L and Mele E J 2005
\textit{Phys.\ Rev.\ Lett.}
\textbf{95} 226801


\bibitem{Roy06}
Roy R 2009
\textit{Phys.\ Rev.\ B} \textbf{79} 195321

\bibitem{Bernevig05}
Bernevig B A and Zhang S-C 2006
\textit{Phys.\ Rev.\ Lett.} \textbf{96} 106802


\bibitem{Moore06}
Moore J E and Balents L 2007
\textit{Phys.\ Rev.\ B} \textbf{75} 121306(R)

\bibitem{Roy3d}
Roy R 2009
\textit{Phys.\ Rev.\ B} \textbf{79} 195322

\bibitem{Fu06_3Da}
Fu L, Kane C L, and Mele E J 2007
\textit{Phys.\ Rev.\ Lett.} \textbf{98} 106803

\bibitem{Fu06_3Db}
Fu L and Kane C L 2007
\textit{Phys.\ Rev.\ B} \textbf{76} 045302

\bibitem{VolovikBook}
Volovik G E 2003 {\it The Universe in a Helium Droplet}
(Oxford: Oxford University Press);
Volovik G E 1992 \textit{Exotic Properties of Superfluid 3He}
(Singapore: World Scientific)

\bibitem{Bernevig-Taylor-Zhang06}
Bernevig B A, 
Hughes T L, 
and Zhang S-C
2006
\textit{Science} \textbf{314} 1757

\bibitem{Konig06}  
K\"onig M, Wiedmann S, Brune C, Roth A, Buhmann H, 
Mohlenkamp L, Qi X-L, and Zhang S-C 2006
\textit{Science} \textbf{314} 1757

\bibitem{Konig06b}
K\"onig M, Buhmann H, Molenkamp L W, Hughes T, Liu C-X, Qi X-L, Zhang S-C
2008 \textit{J.\ Phys.\ Soc.\ Jpn.} \textbf{77} 031007

\bibitem{Zhang09}
Zhang H, 
Liu C-X, 
Qi X-L, 
Dai X, 
Fang Z,
Zhang S-C
2009
\textit{Nature Physics} \textbf{5} 438

\bibitem{Hsieh08} Hsieh D, Qian D, Wray L, Xia Y, Hor Y, Cava R,
and Hasan M 2008 \textit{Nature} \textbf{452} 970 

\bibitem{Hsieh09} Hsieh D, Xia Y, Wray L, Qian D, Pal A, Dil J H, 
Osterwalder J, Meier R, Bihknayer G, Kane C L, Hor Y, Cava R, 
and Hasan M 2009 \textit{Science} \textbf{323} 919

\bibitem{Xia09} Xia Y, Qian D, Hsieh D, Wray L, Pal A, Lin H, Bansil A, Grauer D,
Hor Y S, Cava R J, and Hasan M Z 
2009 {\it Nature Phys.} \textbf{5} 398

\bibitem{Hsieh09b} Hsieh D, Xia Y, Qian D, Wray L, Dil J H, Meier F,
 Osterwalder J, Patthey L, Checkelsky J G, Ong N P, Fedorov A V, Lin H,
 Bansil A, Grauer D, Hor Y S, Cava R J, and Hasan M Z
 2009 \textit{Nature} \textbf{460} 1101-1105


\bibitem{Chen09}
Chen Y L,
Analytis J G,
Chu J-H,
Liu Z K, 
Mo S-K,
Qi X L,
Zhang H J, 
Lu D H, 
Dai X, 
Fang Z,
Zhang S C,
Fisher I R,
Hussain Z,
and 
Shen Z-X
2009
\textit{Science} \textbf{325}
178 - 181


\bibitem{SchnyderRyuFurusakiLudwig2008}
Schnyder A P, Ryu S,
Furusaki A, and Ludwig A W W 2008
\textit{Phys.\ Rev.\ B} \textbf{78} 195125;
Schnyder A P, Ryu S, Furusaki A, and Ludwig A W W 2009
\textit{AIP Conf.\ Proc.} \textbf{1134} 10

\bibitem{KitaevLandau100Proceedings} 
Kitaev A Yu 2009
\textit{AIP Conf.\ Proc.} \textbf{1134}  22

\bibitem{Qi_Taylor_Zhang}
Qi X-L, Hughes T, and Zhang S-C 2008 
\textit{Phys.\ Rev.\ B} \textbf{78} 195424

\bibitem{Zirnbauer96}
Zirnbauer M R 1996
\textit{J.\ Math.\ Phys.} \textbf{37} 4986

\bibitem{Altland97}
Altland A and Zirnbauer M R 1997
\textit{Phys.\ Rev.\ B} \textbf{55} 1142.

\bibitem{Huckleberry2005}
See, e.g., also a more recent discussion:
Heinzner P, Huckleberry A, and Zirnbauer M R 2005
\textit{Commun.\ Math.\ Phys.} \textbf{257} 725

\bibitem{DysonThreeFold}  Dyson F J  1962
\textit{J.\ Math.\ Phys.} \textbf{3} 1199 


\bibitem{EversMirlinRMP}
Evers F and Mirlin A D 2008
\textit{Rev.\ Mod.\ Phys.} \textbf{80} 1355



\bibitem{AndersonOriginalPaper}
Anderson P W 1958
\textit{Phys.\ Rev.} \textbf{109} 1492


\bibitem{LeeRamakrishnanRPM}
Lee P A and Ramakrishnan T V 1985
\textit{Rev.\ Mod.\ Phys.} \textbf{57}  287




\bibitem{RefToFootnoteInLandau100}
See especially footnote 22 of
 the 2nd article of 
Ref.\ \cite{SchnyderRyuFurusakiLudwig2008}.



\bibitem{Wegner}
Wegner F J 1979
\textit{Z.\ Phys.\ B} \textbf{35} 207;
Sch\"afer L and Wegner F J 1980
\textit{Z.\ Phys.\ B} \textbf{38} 113. 
 

\bibitem{Efetov97} 
Efetov K 1997 
\textit{Supersymmetry in disorder and chaos}
(Cambridge, UK: Cambridge University Press)

\bibitem{Pruisken84}
Pruisken A 1984 
\textit{Nucl.\ Phys.\ B} \textbf{235} 277

\bibitem{Fendley01}
See, e.g., Fendley P 2001
\textit{Phys.\ Rev.\ B} \textbf{63} 104429

\bibitem{WittenNPB-223-1983-422}
Witten E  1983 \textit{Nucl.\ Phys.\ B} \textbf{223} 422


\bibitem{AbanovWiegmannNPB-570-2000-685}
Abanov A G and Wiegmann P B 2000
\textit{Nucl.\ Phys.\ B} \textbf{570} 685


\bibitem{lundell_90}
Lundell A T 1992 \textit{Lect.\ Notes Math.} \textbf{1509} 250,
and references therein.



\bibitem{KitaevPeriodicity}
We were first made aware  by A. Kitaev of the existence of a 
regularity in dimensionality
in the classification of topological insulators (superconductors)
\cite{KitaevLandau100Proceedings},
which, as reviewed below \cite{SchnyderRyuFurusakiLudwig2008,RefToFootnoteInLandau100},
can be seen as a consequence of the regularity of the
present table of homotopy groups.

\bibitem{Zumino}
Zumino B 1991
\textit{Current Algebra and Anomalies}
(Singapore: World Scientific)

\bibitem{Nakahara03}
Nakahara M 2003
\textit{Geometry, Topology, and Physics}
2nd edition (Bristol: Institute of Physics Publishing)

\bibitem{Fujikawa}
Fujikawa K and Suzuki H 2004 \textit{Path Integrals and Quantum Anomalies}
(Oxford: Oxford University Press)


\bibitem{NomuraKoshinoRyu}
Nomura K, Koshino M, and Ryu S 2007
\textit{Phys.\ Rev.\ Lett.\ } \textbf{99} 146806 

\bibitem{RyuMudryObuseFurusaki}
Ryu S, Mudry C,  Obuse H, Furusaki A  2007
\textit{Phys.\ Rev.\ Lett.\ } \textbf{99} 116601

\bibitem{Ostrovsky}
Ostrovsky P M, Gornyi I V, and Mirlin A D 2007
\textit{Phys.\ Rev.\ Lett.\ } \textbf{98} 256801

\bibitem{RyuMudryFurusakiLudwig}
Ryu S, Mudry C, Furusaki A, Ludwig A W W,
\textit{unpublished}


\bibitem{KohmotoHalperionWu}
Kohmoto M, Halperin B, and Wu Y S 1992,  
\textit{Phys.\ Rev.\ B} \textbf{45} 13488

\bibitem{ran_vishwanath_09}
Ran Y,  Zhang Y and Vishwanath A 2009 
\textit{Nature Phys.} \textbf{5} 298

\bibitem{Wilczek and Zee}
Wilczek F and Zee A 1984 \textit{\PRL} \textbf{52} 2111

\bibitem{Salomaa88}
Salomaa M M and Volovik G E 1988
\textit{Phys.\ Rev.\ B} \textbf{37} 9298;
Volovik G E 2009
\textit{JETP Lett.} \textbf{90} 587


\bibitem{Ryu02}
Ryu S and Hatsugai Y 2002 \textit{\PRL} \textbf{89} 077002

\bibitem{Kingsmith93}
King-Smith R D and Vanderbilt D H 1993 \textit{\PR}\ B \textbf{47} 1651;
King-Smith R D and Vanderbilt D H 1993 \textit{\PR}\ B \textbf{48} 4442

\bibitem{Resta94}
Resta R 1994
\textit{\RMP}\ \textbf{66} 899


\bibitem{Essin08}
Essin A M, Moore J E and Vanderbilt D 2009
\textit{\PRL} \textbf{102} 146805

\bibitem{Xiao07}
Xiao D, Shi J, Clougherty D P and Niu Q 2009
\textit{\PRL} \textbf{102} 087602

\bibitem{Golterman92}
Golterman M, Jansen K, Kaplan D 1993
Phys.\ Lett.\ B \textbf{301} 219


\bibitem{Haldane88}
Haldane F D M 1988
\textit{Phys.\ Rev.\ Lett.}  \textbf{61} 2015

\bibitem{Su-Schrieffer}
Heeger A J, Kivelson S, Schrieffer J R, and Su W-P 1988 
Rev.\ Mod.\ Phys.\ \textbf{60} 781

\bibitem{Stone}
Stone M 1984 
\textit{Ann.\ Phys.\ (N.Y.)} \textbf{155} 56

\bibitem{FosterLudwigAIII-Interactions}
Foster M S, and Ludwig A W W, 2008
\textit{Phys.\ Rev. \ B} \textbf{77} 165108

\bibitem{Pavan2008}
Hosur P, Ryu S, and Vishwanath A
2010 \textit{Phys.\ Rev.\ B} \textbf{81} 045120

\bibitem{SenthilFisher2000}
Senthil T and Fisher M P A 2000 \textit{Phys.\ Rev.\ B} \textbf{61} 9690;
Read N and Ludwig A W W 2000 \textit{Phys.\ Rev.\ B}  \textbf{63} 024404;
Chalker J T, Read N, Kagalovsky V, Horovitz B, Avishai Y, and Ludwig A W W 2001
\textit{Phys. Rev. B} \textbf{65} 012506


\bibitem{FidkowskiKitaev2009}
Fidkowski L and Kitaev A 2010
\textit{Phys.\ Rev.\ B} \textbf{81} 134509

\bibitem{SenthilMarstonFisher}
Senthil T, Marston J B, and Fisher M P A 1999
\textit{Phys.\ Rev.\ B} \textbf{60} 4245;
Gruzberg I A, Ludwig A W W, and Read  1999 \textit{Phys. Rev. Lett}
\textbf{82} 4524

\bibitem{topSCstanford}
Qi X-Li, 
Hughes T L, 
Raghu S, 
Zhang S-C
2009
\textit{Phys.\ Rev.\ Lett.\ } \textbf{102}, 187001

\bibitem{topSCRoy}
Roy R 2008,
\textit{Preprint arXiv:0803.2868}


\bibitem{RyuDiamond}
Ryu S 2009
\textit{Phys.\ Rev.\ B} \textbf{79}  075124 


\bibitem{Lee_Ryu}
Lee S-S, and Ryu S 2008
\textit{Phys.\ Rev.\ Lett.\ } \textbf{100} 186807



\bibitem{Fukui}
Fukui T and Fujiwara T 2009 {\it J.\ Phys.\ A: Math.\ Theor.} \textbf{42}
362003

\bibitem{RyuTakayanagi10}
Ryu S and Takayanagi T
2010
\textit{Preprint arXiv:1001.0763}

\bibitem{Callan-Harvey}
Callan C G Jr and Harvey J A 1985
\textit{Nucl.\ Phys.} \textbf{B250} 427


\bibitem{SchnyderRyuLudwig2009}
Schnyder A P, Ryu S, and Ludwig A W W 2009
\textit{Phys.\ Rev.\ Lett.} \textbf{102} 196804


\bibitem{Rajagopal2008}
Alford G M, Schmitt A, Rajagopal K, and Schafer T
2008 \textit{Rev. Mod. Phys.} \textbf{80} 1455

\bibitem{Nishida10}
Nishida Y  2010
\textit{Phys.\ Rev.\ D} \textbf{81} 074004

\bibitem{MirlinEtAl2009-09}
Relationships of this kind have also been noted independently
in the recent preprint: Ostrovsky P M, Gornyi I V, and  Mirlin  AD 2009
\textit{arXiv:0910.1338}


\bibitem{Polchinski}
Polchinski J 1988
\textit{String theory. Vol. 1,2}
(Cambridge, UK: Cambridge University Press)

\bibitem{Georgi}
Georgi H 1999
\textit{Lie algebras in particle physics}
(Reading, MA: Perseus Books)


\bibitem{Weinberg84}
Weinberg E J,
London D, 
and  Rosner J L 1984
\textit{Nucl.\ Phys.\ B} \textbf{236} 90


\bibitem{Turner09}
Turner A M, Zang Y, and Vishwanth A,
\textit{Preprint arXiv:0909.3119}




\end{thebibliography}
\end{document}